\documentclass[reqno]{amsart}

\usepackage{amsmath,amsfonts}
\usepackage{epsfig}
\usepackage{graphicx}

\def\alg{\mathfrak{A}}
\def\amp{{\rm Amp}}

\def\bn{\bar{n}}

\def\cG{\mathcal{G}}

\def\cR{\mathcal{R}}

\def\cUe{\mathcal{U}} 
\def\delt{{\nu}}
\def\deltexp{\sigma}

\def\e{\epsilon}

\def\err{{\rm err}} 
\def\Exp{\mathbb{E}}
\def\Fo{\mathfrak{F}}

\def\Hex{H_{ex}}

\def\Hom{H_{\omega}}
\def\homega{\widehat\omega}

\def\Hzero{H_0}

\def\LL{\Lambda_L}

\def\npx{\mu}
\def\nuvar{\zeta}
\def\okappa{\overline{\kappa}}

\def\omzl{\rho_0}

\def\omzlt{\rho_t}
\def\omzlT{\rho_{T/\eta^2}}
\def\omzlTl{\rho_{T/\lambda}}

\def\talpha{\widetilde{\alpha}}

\def\tm{\widetilde{m}}

\def\tn{\widetilde{n}}
\def\ts{\widetilde{s}}

\def\Tor{\mathbb{T}}
\def\Tr{{\rm Tr}}

\def\tm{\widetilde{m}}
\def\tt{\widetilde{t}}

\def\vac{\Omega}
\def\Vom{V_\omega}

\def\bra{\big\langle}
\def\ket{\big\rangle}

\def\C{{\mathbb C}}
\def\N{{\mathbb N}}

\def\R{{\mathbb R}}
\def\Tor{\mathbb T}
\def\Z{{\mathbb Z}}

\def\um{{\underline{m}}}
\def\umq{{\underline{m}}^{(q)}}
\def\utmq{{\underline{\tm}}^{(q)}}

\def\uu{{\underline{u}}}

\def\cU{{\mathcal U}}

\def\1{{\bf 1}}

\def\eqnn{\begin{eqnarray*}}
\def\eeqnn{\end{eqnarray*}}
\def\eqn{\begin{eqnarray}}
\def\eeqn{\end{eqnarray}}

\def\bal{\begin{align}}
\def\eal{\end{align}}

\theoremstyle{plain}
\newtheorem{theorem}{Theorem}[section]
\newtheorem{definition}[theorem]{Definition}
\newtheorem{proposition}[theorem]{Proposition}

\newtheorem{lemma}[theorem]{Lemma}

\newtheorem{remark}[theorem]{Remark}

\numberwithin{equation}{section}

\def\prf{\begin{proof}}
\def\endprf{\end{proof}}

\begin{document}

\parskip=8pt

\title[Homogenous Fermi gas in a random medium]
{Boltzmann limit for a homogenous Fermi gas with dynamical 
Hartree-Fock interactions in a random medium}

\author[T. Chen]{Thomas Chen}
\address{T. Chen,  
Department of Mathematics, University of Texas at Austin.}
\email{tc@math.utexas.edu}

\author[I. Rodnianski]{Igor Rodnianski}
\address{I. Rodnianski, 
Department of Mathematics, Princeton University.}
\email{irod@math.princeton.edu}


\begin{abstract}
We study the dynamics of the thermal momentum distribution function
for an interacting, homogenous Fermi gas
on $\Z^3$ in the presence of an external weak static  random potential, 
where the pair interactions between the
fermions are modeled in dynamical Hartree-Fock theory.
We determine the Boltzmann limits associated to different scaling regimes 
defined by the size of the random potential, 
and the strength of the fermion interactions. 
\end{abstract}

\maketitle

\section{Introduction}

We study the dynamics of an interacting, homogenous Fermi gas on $\Z^3$ in
a static, weakly disordered random medium, where the pair interactions between 
the fermions are modeled in dynamical Hartree-Fock theory.
An observable of considerable importance is the momentum distribution function
at positive temperature, and we are interested in its dynamics for time scales
that are associated to kinetic scaling limits of Boltzmann type.
A main motivation is to understand the trend to equilibrium in such systems, 
and to control the interplay between the influence of the static
randomness, and nonlinear self-interactions of the particles.
We derive Boltzmann limits for the thermal momentum distribution function,
depending on different scaling ratios between the 
random potential, and the strength of the pair interactions between the fermions.

The model in discussion describes
a gas of fermions in a finite box $\LL:=[-\frac L2,\frac L2]^3\cap\Z^3$ of side
length $L\gg1$ with periodic boundary conditions, and 
associated dual lattice $\Lambda_L^*:=\LL/L\subset\Tor^3$;
we will eventually take the thermodynamic
limit $L\rightarrow\infty$.
On the fermionic Fock space of scalar electrons,
$\Fo = \C\oplus \bigoplus_{n\geq1} \, \bigwedge_1^n \, \ell^2(\LL)$,
we introduce creation- and annihilation operators $a^+_x$, $a_y$,
for $x$, $y\in\LL$,
satisfying the canonical anticommutation relations
$\{a_x^+,a_y\}:=a_x^+ \, a_y \, + \, a_y \, a_x^+ \, = \, L^3\delta_{p,q}$ where $\delta_{p,q}$ is
the Kronecker delta, and $\{a_x^\sharp,a_y^\sharp\}=0$ for $a^\sharp=a$ or $a^+$.
We denote the Fock vacuum by $\vac=(1,0,0,\dots)\in\Fo$; it is annihilated
by all annihilation operators, $a_x \, \vac=0$ for all $x\in\LL$.

Letting $\alg$ denote the $C^*$-algebra of bounded operators on $\Fo$,
we consider a time-dependent state $\omzlt$ on $\alg$ determined by
\eqn
  i\partial_t \omzlt( \, A \, ) \, = \, \omzlt( \, [ \, H(t) \, , \, A \, ] \, )
\eeqn
for $A\in\alg$,
with a translation invariant initial condition $\omzl$. 
We assume that $\omzl$ is number conserving $\omzl( [A,N] ) = 0$
for all $A\in\alg$, where $N := \sum_{x\in\LL} a_x^+ a_x$ 
denotes the particle number operator.
We study the dynamics of $\omzlt$ determined by the time-dependent Hamiltonian
\eqn\label{eq-H-def-1}
  H(t) \, = \, \Hzero \, + \, \eta \Vom \, + \, \lambda W(t) \, ,
\eeqn
where $\Hzero := \int dp  E(p) \, a_p^+  a_p$
is the second quantization of the
centered nearest neighbor Laplacian, with 
$E(p) \, = \, 2 \, \sum_{j=1}^3  \cos( \, 2\pi p_j \, )$ denoting its symbol.
The interaction of the fermion gas with the static random background potential is
described by the operator $\Vom := \sum_{x\in\LL} \omega_x \, a_x^+   a_x$
where
$\{\omega_x\}_{x\in\LL}$ is a field of i.i.d. centered, normalized, Gaussain random variables.
The small parameter $0<\eta\ll1$ determines the strength of the disorder.

The operator $W(t)$ accounts for fermion pair interactions 
in dynamical Hartree-Fock theory,  
\eqn\label{eq-Wt}
  W(t) \, := \,
  \sum_{x,y\in\LL}v(x-y)\Big[ \,
  \Exp[ \, \omzlt( \, a_x^+a_x\, ) \, ] \,   a_y^+a_y
  \, - \, \Exp [ \, \omzlt( \, a_y^+a_x \, ) \, ] \, a_x^+a_y \, \Big] \,.
\eeqn
Here, $v$ denotes a  pair potential, where
$\|  \widehat v \|_{H^{3/2+\deltexp}(\Tor^3)} <  C$ is assumed
for  $\deltexp>0$ arbitrary but fixed. 
Notably,  the unknown quantity $\Exp[ \, \omzlt( \, a_x^+a_x\, ) \, ]$ itself
appears in $W(t)$. 

Our main interest is to study the dynamics of
$\Exp [ \, \omzlt( \, a_y^+a_x \, ) \, ]$, the
average of the pair correlation function,
which is determined by
the self-consistent nonlinear evolution equation
\eqn
  i\partial_t \Exp[ \, \omzlt( \, a_x^+a_y \, ) \,] \, = \,
  \Exp[ \, \omzlt( \, [ \, H(t) \, , \, a_x^+a_y \, ] \, ) \, ]
\eeqn
with initial condition $\omzl( \, a_x^+a_y \,)$.
In particular, we derive Boltzmann equations in   kinetic scaling limits of the above 
model, for scaling regimes defined by different ratios between $\eta$ and $\lambda$.

The relevant scaling relations in the system can be understood with the help
of the following heuristics. 
The assumption of vanishing mean implies that the average effect of the random
potential on the dynamics of $\npx_t$ in a time interval $[0,t]$ is proportional to its variance,
of size $O(\eta^2 t)$.
This suggests that the strength of the pair interactions between fermions,
and the interactions of each fermion with the random potential, 
are comparable if $\lambda= O(\eta^2)$.
Accordingly, we distinguish the following scaling regimes, for which 
we derive the associated Boltzmann limits:
\begin{itemize}
\item
The scaling regime $\lambda= O(\eta^2)$,
where the interactions between pairs of fermions, and of each fermion with
the random potential are comparable.
For any  $T>0$, for  
test functions $f$, $g$, and $T/t=\eta^2$,  we prove  
\eqn\label{eq-rescfixpt-0-1}
	\lim_{\eta\rightarrow0}\lim_{L\rightarrow\infty}\Exp[ \, \omzlT( \, a^+(f)a(g) \, ) \, ]
	\, = \, \int dp\, \overline{f(p)} \, g(p) \, F_T(p) \,,
\eeqn
where $a^+(f)=\sum_xf(x)a_x^+=(a(f))^*$ (adjoint), and  
$F_T(p)$ satisfies the linear Boltzmann equation
\eqn\label{eq-linBoltz-1}
	\partial_T F_T(p) \, = \, 2 \pi \int du \, \delta( \, E(u)-E(p) \, )
	\, ( \, F_T(u) - F_T(p)\, )
\eeqn
with initial condition $F_0 =\npx_0\in H^{\frac32+\deltexp}(\Tor^3)$ for some $\deltexp>0$.
The fact that the resulting Boltzmann equation is linear follows from complicated phase cancellations
due to translation invariance.
While the microscopic dynamics is nonlinear, 
we prove that its kinetic scaling limit, as $\eta\rightarrow0$, 
is described by a {\em linear} Boltzmann equation.
The scattering kernel accounts for elastic collisions preserving the kinetic
energy. 
This result remains valid in the regime $\lambda=o(\eta^2)$.
In the case $\lambda=0$, it reduces to the one proven in \cite{ChSa}.
\\
\item
In the regime $\eta^2=o(\lambda)$ and $(T,X)= (\lambda t, \lambda x)$, we prove
that the kinetic scaling limit $\lambda\rightarrow0$ is  stationary. 
\\
\item
In the regime where $\lambda>0$ is independent of $\eta$, and for the scaling
$(T,X)=(\eta^2 t,\eta^2 x)$,
we characterize
the stationary states; those are given by solutions of an implicit equation of the form (\ref{eq-linBoltz-1}) 
with zero on the l.h.s., but where the delta distribution enforces conservation of a {\em renormalized}
energy per particle.
Accordingly, the stationary states are supported 
on level surfaces of a {\em renormalized} kinetic energy function,
determined by a nonlinear fixed point equation.
A derivation of non-stationary solutions in this kinetic scaling limit is an interesting open problem.
\end{itemize}

Our work significantly extends \cite{ChSa} which addresses the Boltzmann limit
for a homogenous free Fermi gas in a random medium.
The proofs given in \cite{ChSa} employ techniques developed
\cite{Ch1,Ch2,ErdYau,Erd} developed for the derivation of Boltzmann equations
from the quantum dynamics of a single electron in a weak random potential;
see also \cite{LukSpo,Sp}.
In the landmark works
\cite{ErdSalm,ErdSalmYau1,ErdSalmYau1-2}, this analysis has been extended 
to diffusive time scales.
We also refer to \cite{AiSiWa,Bou-1,Bou-2,DeRFrPi,Kl,RodSchl} for related works. 

For the proof of our results, we represent 
the average momentum density  $\Exp[\npx_t]$ 
in integral form, as an expansion organized in terms of Feynman diagrams. 
Our overall strategy parallels the approach in \cite{ErdYau,Ch1,ChSa,Erd,Ch2,LukSpo,Sp},
but the techniques developed in those works (for linear models) 
do not carry over directly because of the nonlinear
self-interaction of the fermion field.
In particular, resolvent methods which underlie the analysis those works are
not available here.
Instead, our approach strongly uses 
stationary phase techniques, 
in order to control the combination of
Feynman graph expansion techniques with such nonlinearities.

For the related problem of the derivation of dynamical Hartree-Fock equations from a fermion gas,
see for instance \cite{BaGoGoMa}.
We note that the derivation of macroscopic transport equations from the quantum
dynamics of Fermi gases without any simplifying assumptions on the interparticle
interactions (and without random potential) is a prominent and very challenging
open problem in this research field; see for instance 
\cite{BCEP1,ErdSalmYau2,HoLan,LukSpo-2,Spo}.
For some very interesting recent progress relevant related to this issue, see \cite{LukSpo-3}.

\newpage

\section{Definition of the model}

We consider a gas of fermions in a finite box $\LL:=[-\frac L2,\frac L2]^3\cap\Z^3$ of side
length $L\gg1$, with periodic boundary conditions.
We denote the associated dual lattice by $\Lambda_L^*:=\LL/L\subset\Tor^3$.
We assume that $L$ is much larger than any other significant length scale of the 
system, which will depend upon the case under consideration.
We will eventually take the thermodynamic
limit $L\rightarrow\infty$.

We denote the Fourier transform by
\eqn	
	\widehat f(p) \, := \, \sum_{x\in\LL} \, e^{-2\pi i p\cdot x} \, f(x) \,,
\eeqn
where $p\in \Lambda_L^*$, and the inverse transform by
\eqn
	g^\vee(x) \, = \,
	\int dp \, e^{2\pi i p\cdot x} \, g(p) \,.
\eeqn
For brevity, we are using the notation
\eqn
	\int dp \, f(p) \, \equiv \, \frac{1}{L^3}\sum_{p\in \LL^*} \, f(p) \,,
\eeqn
which recovers its usual meaning in the thermodynamic limit $L\rightarrow\infty$.

We will use the notation
\eqn
	\delta(k) \, := \, L^3\delta_k \,,
\eeqn
where
\eqn
	\delta_k \, = \,
	\left\{
	\begin{array}{ll}
	1 & {\rm if \; } k=0 \; 	\\
	0 & {\rm otherwise}
	\end{array}
	\right.
\eeqn
denotes the Kronecker delta on the momentum lattice $\LL^*$ (mod $\Tor^3$).

We denote the fermionic Fock space of scalar electrons by
\eqn
	\Fo \, = \, \bigoplus_{n\geq0} \, \Fo_n \,,
\eeqn
where
\eqn
	\Fo_0 \, = \, \C
	\; \; \; , \; \; \;
	\Fo_n \, = \, \bigwedge_1^n \, \ell^2(\LL) \; , \; n\geq1 \,.
\eeqn
We introduce creation- and annihilation operators $a^+_p$, $a_q$,
for $p$, $q\in\LL^*$,
satisfying the canonical anticommutation relations
\eqn
	a_p^+ \, a_q \, + \, a_q \, a_p^+ \, = \, \delta(p-q)
	\, := \,
	\left\{
	\begin{array}{lll}
	L^3 & & {\rm if \; \; } p \, = \, q \\
	0 & & {\rm otherwise.}
	\end{array}
	\right.
\eeqn
There is a unique unit ray  $\vac=(1,0,0,\dots)\in\Fo$, referred to as the Fock vacuum,
which is annihilated
by all annihilation operators, $a_p \, \vac=0$ for all $p\in\LL^*$.

Let $\alg$ denote the $C^*$-algebra of bounded operators on $\Fo$.
We consider a time-dependent state $\omzlt$ on $\alg$ determined by
\eqn
  i\partial_t \omzlt( \, A \, ) \, = \, \omzlt( \, [ \, H(t) \, , \, A \, ] \, )
\eeqn
for $A\in\alg$,
with a translation invariant initial condition $\omzl$. 
We assume that $\omzl$ is number conserving; that is,
\eqn
	\omzl( \, [A,N] \, ) \, = \, 0
\eeqn
for all $A\in\alg$, where
\eqn
	N \, := \, \sum_{x\in\LL} a_x^+ a_x
\eeqn
denotes the particle number operator.

The dynamics of $\omzlt$ shall be determined by the time-dependent Hamiltonian
\eqn\label{eq-H-def-1}
  H(t) \, = \, \Hzero \, + \, \eta \Vom \, + \, \lambda W(t) \, ,
\eeqn
where the right hand side is defined as follows.
The operator
\eqn
	\Hzero \, := \,
	\int dp \, E(p) \, a_p^+ \, a_p
\eeqn
is the second quantization of the
centered nearest neighbor Laplacian $(\Delta f)(x)=\sum_{|y-x|=1}f(y)$ on $\Z^3$.
The symbol of $\Delta$ is given by
\eqn
	E(p) \, = \, 2 \, \sum_{j=1}^3  \cos( \, 2\pi p_j \, ) \,,
\eeqn
corresponding to the kinetic energy of a single electron.
The interaction of the fermion gas with the static random background potential is
described by the operator
\eqn
	\Vom \, := \, \sum_{x\in\LL} \omega_x \, a_x^+ \, a_x \, .
\eeqn
We assume $\{\omega_x\}_{x\in\LL}$ to be a field of i.i.d. random variables which
is centered, normalized, and Gaussian,
\eqn
	\Exp[ \, \omega_x \, ] \, = \, 0
	\; \; , \; \;
	\Exp[ \, \omega_x^2 \, ] \, = \, 1 \,,
\eeqn
for $x\in\LL$.
The small parameter $0<\eta\ll1$ controls the strength of the disorder.

The operator $W(t)$ accounts for fermion pair interactions 
in dynamical Hartree-Fock theory,
\eqn
  W(t) \, := \,
  \sum_{x,y\in\LL}v(x-y)\Big[ \,
  \Exp[ \, \omzlt( \, a_x^+a_x\, ) \, ] \,   a_y^+a_y
  \, - \, \Exp [ \, \omzlt( \, a_y^+a_x \, ) \, ] \, a_x^+a_y \, \Big] \,.
\eeqn
Here, $v$ denotes a $\LL$-periodic pair potential, for which we assume that
\eqn\label{eq-v-bd-1}
	\| \, \widehat v \, \|_{H^{3/2+\deltexp}(\Tor^3)} \, < \, C \, 
\eeqn
for  $\deltexp>0$ arbitrary but fixed. 
Notably, we make no assumption on the sign of $v$.
We note that the unknown quantity $\Exp[ \, \omzlt( \, a_x^+a_x\, ) \, ]$
appears in $W(t)$.

We are interested in the dynamics of the
average of the pair correlation function,
\eqn
	\Exp [ \, \omzlt( \, a_y^+a_x \, ) \, ] \,,
\eeqn
which is determined by
the self-consistent nonlinear evolution equation
\eqn
  i\partial_t \Exp[ \, \omzlt( \, a_x^+a_y \, ) \,] \, = \,
  \Exp[ \, \omzlt( \, [ \, H(t) \, , \, a_x^+a_y \, ] \, ) \, ]
\eeqn
with initial condition $\omzl( \, a_x^+a_y \,)$.
Its Fourier transform is diagonal in momentum
space,
\eqn
  \Exp[ \, \rho_t( \, a_p^+a_q \, )\, ] \, = \, \delta(p-q) \, \frac{1}{L^3} \,
  \Exp[ \, \rho_t( \, a_p^+a_p \, )\, ] \,,
\eeqn
because $\Exp [ \, \omzlt( \, a_y^+a_x \, ) \, ]$ is translation invariant.
This follows from the homogeneity of the randomness.

We remark that for fermions,
\eqn\label{eq-density-bd-1}
	0 \, \leq \, \frac{1}{L^3} \, \omzl( \, a_p^+ a_p \, )  \, \leq \, 1 \,,
\eeqn
since $\|a_p^{(+)}\|=L^{d/2}$ in operator norm, $\forall p\in\LL^*$. 
The expected occupation density
of the momentum $p$ in the lattice
$\LL^*$ is given by
\eqn\label{eq-mut-def-1}
  \npx_t(p) \, := \,
  \frac{1}{L^3} \, \Exp[ \, \rho_t( \, a_p^+a_p \, )\, ] \,.
\eeqn
The dynamical Hartree-Fock interaction can be written as
\eqn
  W(t) & = & \frac{1}{L^3} \, \rho_0(N) \Big( \, \sum_x \, v(x)\, \Big) \, N
  \, - \, \int_{\LL^*} dp \, ( \, \widehat v \, * \, \npx_t \, )(p) \, a_p^+a_p
  \nonumber\\
  & =: & W_{dir}(t) \, + \, W_{ex}(t) \,
\eeqn
where, following standard terminology,
$W_{dir}(t)$ denotes the direct, and $W_{ex}(t)$ the exchange term.
It is clear that $\Hom(t)$ is particle number conserving, 
\eqn
	[ \, \Hom(t),N \, ] \, = \, 0
	\; \; \; , \; \; \;
	\forall \; t \in \R \,.
\eeqn 
Since $\omzl$ is translation invariant and number conserving, 
we conclude that whenever $[A,N]=0$ holds for an operator $A$, 
it follows that
\eqn
	i\partial_t\omzlt( \, A \, ) \, = \, \omzlt( \, [\Hex(t),A] \, ) \, 
\eeqn
where the operator
\eqn
	\Hex(t) \, := \, H_0 \, + \, \eta \Vom \, + \, \lambda W_{ex}(t) \,
\eeqn
contains only the exchange term of $W(t)$.
\\
\\
\\

\newpage

\section{Statement of the main results}

In order to determine the dynamics of
the average  momentum distribution function
$\npx_t$ defined in (\ref{eq-mut-def-1}), we consider   
\eqn\label{eq-fixpt-1}
	\int dp \, \overline{f(p)} \, g(p) \, \npx_t(p)
	& = & 
	\Exp[ \, \omzlt( \, a^+(f)a(g) \, ) \, ] \,
	\nonumber\\
	& = & 
	\Exp[ \, \omzl( \, \cU_t^* \, a^+(f)a(g) \, \cU_t \, ) \, ]
\eeqn
for a translation invariant and particle number conserving initial state $\omzl$,
where $f$ and $g$ are test functions.
The linear operator $\cU_t$ denotes the unitary flow generated by $H_{ex}(t)$.
It satisfies $\cU_0=\1$,
and notably depends on $\mu_s$, $s\in[0,t]$. 
 
Accordingly, we make the key observation that (\ref{eq-fixpt-1}) is a 
{\em fixed point equation} for $\npx_t$,  tested against  $f$, $g$.
The right hand side of  (\ref{eq-fixpt-1}) is a complicated nonlinear functional of $\npx_s$ 
which we will discuss in detail in Section \ref{ssec-Duh-1}.

We introduce macroscopic variables $(T,X)$, 
related to the microscopic variables $(t,x)$ by
\eqn
	(T,X) \, = \,  (\zeta t, \zeta x ) \,,
\eeqn
with $\zeta>0$ a real parameter.
We will study kinetic scaling limits associated to different 
scaling ratios between $\zeta$, $\eta$ and $\lambda$.

As stated in the introduction, the random potential 
has an average effect on the dynamics of $\npx_t$ by an amount 
proportional to its variance, $O(\eta^2t)$, in the time interval $[0,t]$. 
Since the strength of the fermion pair interactions is $O(\lambda)$,
both effects are comparable if $\lambda = O(\eta^2)$.
This implies that the relevant scaling regimes of the system are determined
by those addressed below, in 
Theorems \ref{prop-approx-fixpt-1}, \ref{prop-approx-fixpt-2}, and
Theorem \ref{prop-approx-fixpt-3}.

\begin{theorem}\label{thm-main-1}\label{prop-approx-fixpt-1}
Assume that $\lambda\leq O(\eta^2)$. Then, for any fixed, finite $T>0$,
and any choice of test functions $f$, $g$,
\eqn\label{eq-rescfixpt-0-1}
	\lim_{\eta\rightarrow0}\lim_{L\rightarrow\infty}\Exp[ \, \omzlT( \, a^+(f)a(g) \, ) \, ]
	\, = \, \int dp\, \overline{f(p)} \, g(p) \, F_T(p)
\eeqn
holds, where
$F_T(p)$ satisfies the linear Boltzmann equation
\eqn\label{eq-linBoltz-1}
	\partial_T F_T(p) \, = \, 2 \pi \int du \, \delta( \, E(u)-E(p) \, )
	\, ( \, F_T(u) - F_T(p)\, )
\eeqn
with initial condition $F_0 =\npx_0\in H^{\frac32+\deltexp}(\Tor^3)$ for some $\deltexp>0$.
\end{theorem}

We note that the linear Boltzmann equation (\ref{eq-linBoltz-1}) can be explicitly
solved. The solution 
is given by
\eqn
	F_T(p) \, = \, F_\infty(p) \, + \, e^{-Tm(p)}\frac{2\pi}{m(p)} \, \int du \, \delta( \, E(u)-E(p) \, )
	\, ( \, F_0(u) - F_0(p) \, ) \,,
\eeqn
where
\eqn
	m(p) \, := \, 2\pi \int du \, \delta( \, E(u)-E(p) \, )
\eeqn
and
\eqn\label{eq-statsolsimple-1}
	F_\infty(p) \, := \,  \frac{2\pi}{m(p)} \, \int du \, \delta( \, E(u)-E(p) \, ) \, F_0(u) \,.
\eeqn
As an important example, we note that the following is obtained if the initial state $\omzl$ is given by 
the Gibbs state for a non-interacting fermion gas
(with inverse temperature $\beta$ and chemical potential $\mu$),
\eqn\label{eq-free-Gibbs}
	\omzl(A) \, = \, \frac{1}{Z_{\beta,\mu}} \, \Tr ( \, e^{-\beta( T -\mu N)} A\, ) \,,
\eeqn
where $Z_{\beta,\mu}:=\Tr( \, e^{-\beta( T -\mu N)} \, )$. 
The associated momentum distribution function is given by the Fermi-Dirac distribution
\eqn
	\lim_{L\rightarrow\infty} \frac{1}{L^3} \, \omzl( \, a_p^+ a_p \, )
	\, = \, \frac{1}{1+e^{\beta( E(p)-\mu)}} \,,
\eeqn
which is a {\em stationary solution} of the linear Boltzmann equation (\ref{eq-linBoltz-1}),
for all $\beta>0$.
This result remains true in the zero temperature limit $\beta\rightarrow\infty$  
where, in the weak sense,
\eqn
	\frac{1}{1+e^{\beta( E(p)-\mu)}} \, \rightarrow \, \chi[E(p)<\mu] \,
\eeqn
(see also \cite{ChSa}).

In the case $\eta^2=o_\lambda(1)$ and $T=\lambda t$, we find the following kinetic
scaling limit.

\begin{theorem}\label{thm-main-1-1}\label{prop-approx-fixpt-2}
Assume that $\eta^2=O(\lambda^{1+\delta})$ for $\delta>0$ arbitrary.
Then, for any fixed, finite $T>0$, and any choice of test functions $f$, $g$,
\eqn
	\lim_{\lambda\rightarrow0}\lim_{L\rightarrow\infty}\Exp[ \, \omzlTl( \, a^+(f) \, a(g) \, ) \, ]
	\, = \, \int dp\, \overline{f(p)} \, g(p) \, F_T(p) \,,
\eeqn
and
\eqn\label{eq-lBoltz-2}
	\partial_T F_T(p) \, = \, 0 \,,
\eeqn
for $F_0 =\npx_0\in H^{\frac32+\deltexp}(\Tor^3)$ for some $\deltexp>0$
Accordingly, $F_T =F_0 $ is stationary.
\end{theorem}

Finally, we prove a partial result that highlights some interesting 
aspects about the problem of determining the kinetic scaling limit determined by
$T=\eta^2 t$ and $\eta\rightarrow0$, with $\lambda$ small but independent of $\eta$.
That is, we are considering, for $\lambda=O(1)$, the rescaled, formal fixed point equation
\eqn\label{eq-rescfixpt-0} 
	\int dp \, \overline{f(p)} \, g(p) \, \npx_{T/\eta^2}(p)
	& = & 
	\cG^{(L)}[ \, \npx_{\bullet}( \, \bullet \, )
	 ; \eta ; \lambda ; T ; f , g  \, ]
	\nonumber\\
	& := & \Exp[ \, \omzlT( \, a^+(f)a(g)\, ) \, ]   \, 
\eeqn
for $\npx_{\bullet}( \, \bullet \, )$. 
The existence and uniqueness of solutions for this fixed point equation is
currently an open problem. 
Below, we will make the assumption that there exist limiting stationary solutions,
and determine a their form under this hypothesis.

We base our discussion on the following hypotheses for the case $\lambda=O(1)$:
\begin{itemize}
\item[{\em (H1)}]
There exist solutions $F^{(\eta)}(T):=\lim_{L\rightarrow\infty}\npx_{T/\eta^2}$
of \eqref{eq-rescfixpt-0}, such that the limit $w-\lim_{\eta\rightarrow0}F^{(\eta)}(T)=: F(T)=F(0)$
exists and is stationary. 
\\
\item[{\em (H2)}]
The stationary fixed point solution in  {\rm (H1)} satisfies
\eqn 
	F(T)
	& = & \lim_{\eta\rightarrow0}\lim_{L\rightarrow\infty}\cG^{(L)} [ \, F^{(\eta)} 
	 ; \eta ; \lambda ; T ; f , g \, ] 
	 \nonumber\\
	 & = & \lim_{\eta\rightarrow0}\lim_{L\rightarrow\infty}\cG^{(L)} 
	 [ \, F  ; \eta ; \lambda ; T ; f , g \, ] \,.
\eeqn
The first equality sign here is equivalent to {\em (H1)}, 
while the second equality sign
accounts for the assumption that $ F^{(\eta)} $ can be replaced by 
the limiting fixed point $F$ before letting $\eta\rightarrow0$,
to produce the same result.
\end{itemize}

We remark that based on the analysis given in this paper, 
we are able to prove hypothesis {\em (H2)} if $F^{(\eta)}=F+O(\eta^2)$.
Error bounds of order $O(\eta^2)$ require more precise estimates of 
"crossing" and "nesting" terms in the Feynman graph expansion than
considered in this paper, but are available from 
\cite{ErdSalm, ErdSalmYau1,ErdSalmYau1-2,ErdSalmYau2}. 
We will not further pursue this issue in the work at hand.

\begin{proposition}
Let $\lambda$ be small but independent of $\eta$, 
and assume that $F\in L^\infty(\Tor^3)$ independent of $t$. 
Then, the thermodynamic limit 
\eqn 
	\cG[ \, F \,
	 ; \eta ; \lambda ; T ; f , g \, ]
	\, := \, \lim_{L\rightarrow\infty}\cG^{(L)} [ \, F 
	 ; \eta ; \lambda ; T ; f , g \, ] 
\eeqn
exists. 
\end{proposition}

The proof of this proposition follows directly from results established 
in \cite{Ch1,Ch2,ChSa,ErdYau}, and will not be reiterated here.


\begin{theorem}\label{thm-main-2}\label{prop-approx-fixpt-3}
Assume that $\lambda\leq O_\eta(1)$, and let
\eqn\label{eq-rescfixpt-5}
	\widetilde E_\lambda(u) \, := \, E(u) \, - \, \lambda( \, \widehat v*F \, )(u) \,.
\eeqn 
We assume that $F\in L^\infty(\Tor^3)$ admits the bounds
\eqn\label{eq-intcond-1}
  \sup_\alpha\int dp \, \frac{1}{|\widetilde E_\lambda(q)-\alpha-i\e|}
  \, , \,
  \sup_q\int d\alpha \, \frac{1}{|\widetilde E_\lambda(q)-\alpha-i\e|}
  \, \leq \, C \, \log\frac1\e \,,
\eeqn
and
\eqn\label{eq-intcond-2}
  &&\sup_{\alpha_i}\sup_{u\in\Tor^3}\int dq \, dp \,
  \frac{1}{|\widetilde E_\lambda(q)-\alpha_1-i\e|}\,
  \frac{1}{|\widetilde E_\lambda(p)-\alpha_2-i\e|} \,
  \frac{1}{|\widetilde E_\lambda(p\pm q + u)-\alpha_3-i\e|}
  \nonumber\\
  &&\quad\quad\quad\quad\quad\quad\quad\quad\quad\quad\quad\quad
  \quad\quad\quad\quad\quad\quad\quad\quad\quad\quad\quad\quad \,
  \, \leq \, \e^{-b}
\eeqn
for some $0<b<1$. 

Then, $F$ satisfies
\eqn 
	\int dp\, \overline{f(p)} \, g(p) \, F(p)   
	\, = \, \lim_{\eta\rightarrow0} \cG[ \, F 
	 ; \eta ; \lambda ; T ; f , g \, ]  \,,
\eeqn
independent of $T$, if and only if it satisfies
\eqn\label{eq-rescfixpt-4}
	F(p) \, = \, \npx_0(p) \, = \, \frac{1}{\widetilde m_\lambda(p)} \int du \,
	\delta( \, \widetilde E_\lambda(u) - \widetilde E_\lambda(p) \, )
	\, F(u)  \, ,
\eeqn
where 
\eqn
	\widetilde m_\lambda(p) \, := \, 2\pi \int du \,
	\delta( \, \widetilde E_\lambda(u) - \widetilde E_\lambda(p) \, ) \, 
\eeqn
is the (normalized) measure of the level surface of $\widetilde E_\lambda$ 
for the value $\widetilde E_\lambda(p)$. 
\end{theorem}

$\;$  

\begin{remark}
The following comments refer to Theorem \ref{thm-main-2}.
\begin{enumerate}
\item 
The solution of  \eqref{eq-rescfixpt-5} corresponds to
a renormalized kinetic energy which is shifted by the average interaction 
energy for fermion pairs. 
\\
\item
The fixed point equation
(\ref{eq-rescfixpt-4}) for $F$ shows that the stationary kinetic limits of $\npx_t$ are concentrated and
equidistributed on level surfaces of the
renormalized kinetic energy function
$\widetilde E_\lambda(\, \cdot \,)$.
\\
\item
The bounds (\ref{eq-intcond-1}) and (\ref{eq-intcond-2})
correspond to the ``crossing estimates'' in 
\cite{Ch1,ErdYau,ErdSalm,Luk}.
They ensure sufficient non-degeneracy of the renormalized energy
level surfaces so that the Feynman graph expansions 
introduced below are convergent. 
However, they do not seem sufficient to prove hypothesis
{\em (H2)} under the assumption that {\em (H1)} holds.
\\
\item
We note that if $\lambda\leq o_\eta(1)$, the stationary solutions
found in  Theorem \ref{thm-main-2} reduce to those of
the linear Boltzmann equation derived in Theorem \ref{thm-main-1},
see (\ref{eq-statsolsimple-1}).
\end{enumerate}
\end{remark}

\newpage

\section{Feynman graphs and amplitudes}

In this section, we set up the Feynman graph expansions underlying our
proofs of Theorems \ref{prop-approx-fixpt-1},  \ref{prop-approx-fixpt-2}, and
Theorem \ref{prop-approx-fixpt-3}.

\subsection{Duhamel expansion}
\label{ssec-Duh-1}
 
Let $\cUe_t$ denote the unitary flow generated by $\Hex(t)$,
determined by
\eqn
	i\partial_t \, \cUe_t \, = \, \Hex(t) \, \cUe_t\, 
	\; \; \; \; {\rm and} \; \; \; \; 
	\cUe_0 \, = \, 1 \,.
\eeqn
It then follows that
\eqn
	\omzlt( \, A \, ) \, = \, \omzl( \, \cUe_t^* \, A \, \cUe_t \, ) \,,
\eeqn
and that
\eqn
	\omzlt( \, a^+(f) \, a(g)\, ) \, = \, \omzl( \, a^+(f,t) \, a(g,t) \, ) \,.
\eeqn
The Heisenberg evolution of the creation- and annihilation
operators is determined by
\eqn
	a(f,t) \, := \,  \cUe_t^*  \, a(f) \,  \cUe_t \,,
\eeqn
with 
\eqn
	a(f) \, = \, \int dp \, f(p) \, a_p
	\; \; \; \; , \; \; \; \; 
	a^+(f,t) \, = \, (a(f,t))^* \,.
\eeqn
It suffices to discuss the annihilation operators $a(f,t)$.
Because $\Hex(t)$ is bilinear in $a^+$ and $a$, it follows that $a(f,t)$
is a linear superposition of annihilation operators.
Therefore, there exists a function $f_t$ such that
\eqn
	a(f,t) \, = \, a(f_t) \,,
\eeqn
satisfying
\eqn
	i\partial_t a(f_t) & = &
	[ \, \Hex(t) \, , \, a(f_t) \, ]
	\nonumber\\
	& = & \int dp \, f_t(p) \, E(p) \, a_p
	\, + \, \eta \int dp \int du \, f_t(p) \, \homega(u-p) \, a_u
	\nonumber\\
	&&\quad\quad\quad\quad\quad\quad\quad\quad\;
	\, - \, \lambda \, \int dp \, ( \, \widehat v* \npx_t \, )(p) \, f_t(p) \, a_p \,,
\eeqn
with initial condition  
\eqn
	a(f,0) \, = \, a(f_0) \, = \, a(f) \,.
\eeqn
We conclude that $f_t$ is the solution of the 1-particle
random Schr\"odinger equation
\eqn\label{eq-onepart-RSE-1}
	i\partial_t f_t(p)
	\, = \, E(p) f_t(p)
	\, + \, \eta \,  ( \, \homega*f_t \, )(p)
	\, - \, \lambda \,  ( \, \widehat v *\npx_t \, )(p) f_t(p)
\eeqn
with initial condition
\eqn\label{eq-onepart-RSE-2}
	f_0 \, = \, f \,.
\eeqn
Here, $\homega(u)=\sum_x e^{2\pi i ux}\omega_x$, and
$v$ is the fermion pair interaction potential. 

Noting that the Hamiltonian $\Hex(t)$ itself depends on the unknown quantity
$\npx_t$, we determine $\npx_t$ by writing the fixed point equation (\ref{eq-fixpt-1})
in integral form, as an expansion in powers of $\eta$.  

For arbitrary test functions $f$ and $g$, 
we consider the pair correlation function
\eqn\label{eq-omzlt-weak-def-1}
	\omzlt( \, a^+(f) \, a(g) \, ) & = & \omzl( \, a^+(f_t) \, a(g_t )\, )
	\nonumber\\
	& = & \int dp \, dq \, \omzl( \, a_p^+ \, a_q) \, \overline{f_t(p)} \, g_t(q)
	\nonumber\\
	& = & \int dp \, J(p) \, \overline{f_t(p)} \, g_t(p) \,,
\eeqn
where the state $\omzlt$ equals the one in the definition of $\npx_t$, (\ref{eq-mut-def-1}).
Passing to the last line, we have used the momentum conservation condition
\eqn\label{eq-J-def-1}
	\omzl( \, a_p^+ \, a_q \, ) \, = \, J(p) \, \delta(p-q) \,
\eeqn
obtained from the translation invariance of the initial state $\omzl$, where 
\eqn
	0 \, \leq \, J(p) \, = \, \frac{1}{L^3}\omzl( \, a_p^+ a_p \, ) \, \leq \, 1 \,,
\eeqn
as noted before, in (\ref{eq-density-bd-1}).

The solution $f_t$ of (\ref{eq-onepart-RSE-1}),
(\ref{eq-onepart-RSE-2}), satisfies the Duhamel (respectively, variation of constants) formula 
\eqn\label{eq-Duh1step-1}
	f_t(p) \, = \, U_{0,t}(p) \, f(p)
	\, + \, i \, \eta \, \int_0^t ds \, U_{s,t}(p)
	\, ( \, \homega*f_s \, )(p) \,
\eeqn	
with
\eqn\label{eq-def-Uprop}
	U_{s,t}(p) \, := \, e^{i\int_s^t ds' \, ( \, E(p)-\lambda \kappa_{s'}(p) \, ) } \,,
\eeqn
where we treat
\eqn\label{eq-def-kappa}
	\kappa_s(u) \, := \, ( \, \widehat v*\npx_s \, )(u) \, 
\eeqn
as an external (a priori bounded) source term.
We note that $U_{0,t}(p)f(p)$ solves (\ref{eq-onepart-RSE-1}) for $\eta=0$
(no random potential)
with initial condition (\ref{eq-onepart-RSE-2}).

Let $N\in\N$, which remains to be optimized.
The $N$-fold iterate of (\ref{eq-Duh1step-1}) produces the truncated Duhamel expansion
with remainder term,
\eqn
	f_t \, = \,   f_t^{(\leq N)} \, + \, f_t^{(>N)}  \,,
\eeqn
where
\eqn
	f_t^{(\leq N)} \, := \, \sum_{n=0}^N f_t^{(n)}  \,,
\eeqn
and $f_t^{(>N)}$ is the Duhamel remainder term of order $N$.
We define
\eqn
	t_{-1} \, := \, 0
	\;  \; , \; \;
	t_j \, = \, s_0 \, + \, \cdots \, + \, s_j   \,,
\eeqn
for $j=0,\dots,n$, and
\eqn\label{eq-cR-def-1}
	\cR(k_0,\dots,k_n;z) \, := \, \int_{\R_+^{n+1}} ds_0 \cdots ds_n \,
	\Big( \, \prod_{j=0}^n
	e^{-is_j(E(k_j)-z)}e^{i\lambda\int_{t_{j-1}}^{t_j} ds' \, \kappa_{s'}(k_j)} \, \Big) \,,
\eeqn
for $z\in\C$.

The $n$-th order term in the Duhamel expansion is given by
\eqn\label{eq-ftn-def-1}
	f_t^{(n)}(p) & := & (i\eta)^n
	\int_0^t dt_n \cdots \int_0^{t_{2}}dt_1 \,
	\int dk_0\cdots dk_n \, \delta(p-k_0) \,
	\\
	&&\quad
	\Big[ \, \prod_{j=0}^n U_{t_{j-1},t_j}(k_j) \, \Big]
	\Big[ \, \prod_{\ell=1}^n\homega(k_\ell-k_{\ell-1}) \, \Big] \, f(k_n) \,.
	\nonumber
\eeqn
Expressed in terms of the time increments $s_j:=t_j-t_{j-1}$,  
\eqn\label{eq-ftn-def-2}
	f_t^{(n)}(p) & = & (i\eta)^n
	\int ds_0 \cdots ds_n \, \delta(t-\sum_{j=0}^n s_j)
	\, \int dk_0\cdots dk_n \, \delta(p-k_0) \,
	\\
	&&\quad
	\Big[ \, \prod_{j=0}^n e^{-i\int_{t_{j-1}}^{t_j} ds'(E(k_j)-\lambda\kappa_{s'}(k_j))}  \, \Big]
	\Big[ \, \prod_{\ell=1}^n\homega(k_\ell-k_{\ell-1}) \, \Big] \, f(k_n) \,.
	\nonumber
\eeqn
Expressing the delta distribution $\delta(t-\sum_{j=0}^n s_j)$ in terms
of its Fourier transform, we find
\eqn\label{eq-ftn-def-3}
	f_t^{(n)}(p) & = & (i\eta)^n \, e^{\e t} \, \int d\alpha \, e^{-it\alpha} \, 
	\, \int dk_0\cdots dk_n \, \delta(p-k_0) \,
	\nonumber\\
	&&\quad
	\cR(k_0,\dots,k_n;\alpha+i\e)
	\, \Big[ \, \prod_{j=1}^n\homega(k_j-k_{j-1}) \, \Big] \, f(k_n) \,.
\eeqn
The above three equivalent expressions for $f_t^{(n)}(p)$ 
have different advantages which we will make use of.

The Duhamel remainder term of order $N$ is given by
\eqn
	f_t^{(>N)}  \, = \,  i\eta
	\int_0^t ds \, \, \cUe_{s,t} \,  \Vom^{(1)} \, f_s^{(N)}  \,.
\eeqn
We choose
\eqn
	\e \, = \, \frac1t
\eeqn
so that the factor $e^{\e t}$ in \eqref{eq-ftn-def-3} remains bounded for all $t$.

Substituting the truncated Duhamel expansion for $a^+(f_t)$,
$a(g_t)$ in (\ref{eq-omzlt-weak-def-1}), one obtains
\eqn\label{eq-Duh-exp-1}
	\omzlt( \, a^+(f)  \, a(g) \, ) \, = \,
	\omzl( \,  a^+(f_t) \, a(g_t)   \, ) 
	\, = \, \sum_{n,\tn=0}^{N+1} \omzlt^{(n,\tn)}(f,g)
\eeqn
where
\eqn
	\omzlt^{(n,\tn)}(f,g) \, := \,
	\omzl( \, a^{+}(f_t^{(n)}) \, a(g_t^{(\tn)}) \, )
\eeqn
if $n,\tn\leq N$, and
\eqn
	\omzlt^{(n,N+1)}(f,g) \, := \,
	\omzl( \, a^{+}(f_t^{(n)}) \, a(g_t^{(>N)}) \, )  \, ,
\eeqn 
\eqn 
	\omzlt^{(N+1,\tn)}(f,g) \, := \,
	\omzl( \, a^{+}(f_t^{(>N)}) \, a(g_t^{(\tn)}) \, )
\eeqn
if $n,\tn\leq N$.
Moreover,
\eqn
	\omzlt^{(N+1,N+1)}(f,g) \, := \,
	\omzl( \, a^{+}(f_t^{(>N)}) \, a(g_t^{(>N)}) \, ) \,.
\eeqn
In particular, all Duhamel terms indexed by $n,\tn\leq N$
depend on $\homega$ like polynomials. 
Accordingly,
\eqn
	\lefteqn{
	\Exp[\omzlt^{(n,\tn)}(f,g)] 
	\, = \,   
	\eta^{2\bn} \, \sum_{\pi\in\Gamma_{n,\tn}}
	\int_0^t dt_q \, \cdots\, \int_0^{t_{2}}dt_1
	\int_0^{t} d\tt_q \, \cdots\, \int_0^{\tt_{2}}d\tt_1 
	}
	\nonumber\\
	&&\int du_0\cdots du_{2\bn+1}
	\, \overline{f(u_0)} \, g(u_{2\bn+1})
	\, J(u_n) \, \delta(u_n-u_{n+1})
	\nonumber\\
	&&
	\quad\quad\quad\quad\quad\quad
	\Big[\prod_{j=0}^n U_{t_{j-1},t_j}(u_j)\Big]
	\, 
	\Big[\prod_{j=n+1}^{2\bn+1} \overline{U_{\tt_{j-1},\tt_j}(u_j)} \Big] 
  	\label{eq-rhontn-1}\\
	&&
	\quad\quad\quad\quad\quad\quad\quad\quad\quad\quad
	\Exp\Big[ \, \prod_{j=1}^{n}\homega(u_j-u_{j-1})
	\, \prod_{j=n+2}^{2\bn+1}\homega(u_j-u_{j-1}) \, \Big]
	\nonumber
\eeqn
and using (\ref{eq-ftn-def-3}), this is equivalent to
\eqn
	\lefteqn{
	\Exp[\omzlt^{(n,\tn)}(f,g)] 
	\, = \,   
	\eta^{2\bn} \, e^{2 \e t} \, \sum_{\pi\in\Gamma_{n,\tn}}
	\int d\alpha \, d\talpha \, e^{it(\alpha-\talpha)} 
	}
	\nonumber\\
	&&\int du_0\cdots du_{2\bn+1}
	\, \overline{f(u_0)} \, g(u_{2\bn+1})
	\, J(u_n) \, \delta(u_n-u_{n+1})
	\nonumber\\
	&&
	\quad\quad\quad\quad\quad\quad
	\cR(u_0,\dots,u_n;\alpha+i\e) \, \cR(u_{n+1},\dots,u_{2\bn+1};\talpha-i\e)
  \label{eq-rhontn-2}\\
	&&
	\quad\quad\quad\quad\quad\quad\quad\quad\quad\quad
	\Exp\Big[ \, \prod_{j=1}^{n}\homega(u_j-u_{j-1})
	\, \prod_{j=n+2}^{2\bn+1}\homega(u_j-u_{j-1}) \, \Big]
	\nonumber
\eeqn
where $t_{-1},\tt_{-1}:=0$ in (\ref{eq-rhontn-1}).

\subsection{Graph expansion}
\label{ssec-Feynman-graphs-1}

By assumption, $\{\omega_x\}$ is  a centered, i.i.d., Gaussian random field.
Accordingly, we may explicitly determine the correlations of the 
random potential in the expressions (\ref{eq-rhontn-1}), (\ref{eq-rhontn-2}).
The expectation of any product of even degree $n\in2\N$ is equal to the sum of all 
possible products of pair correlations of the same degree,
\eqn 
	\lefteqn{
	\Exp\big[ \, \prod_{j=1}^n \, \homega( \, u_j - u_{j-1}\, ) \, \big]
	}
	\\
	&&
	\, = \, \sum_{{\rm pairings} \, (\ell_i,\ell'_i)}
	\prod_{i=1}^{\frac n2} \, \Exp\big[ \,
	\homega( \, u_{\ell_i} - u_{\ell_i-1}\, ) \,
	\homega( \, u_{\ell'_i} - u_{\ell'_i-1} \, )
	\, \big] \,.
	\nonumber
\eeqn
The sum extends over all possible pairings $(\ell_i,\ell'_i) \in\{1,\dots,n\}^2$ 
with $\ell_i\neq\ell'_i$, $i=1,\dots,\frac n2$,
where every element of $\{1,\dots,n\}$ appears in precisely one pairing.
This expansion is often referred to as {\em Wick's theorem}.
Expectations of products of $\homega$ of odd degree are identically zero.

To organize the terms in these sums of products of pair correlations, 
we introduce {\em Feynman graphs}.
The set of Feynman graphs $\Gamma_{n,\tn}$, with $\bn=n+\tn\in2\N$,
is given as follows; see also \cite{Ch1,ChSa,ErdYau}:
\begin{itemize}
\item[-]
We consider two horizontal solid lines, which we refer to as
{\em particle lines}, joined by a distinguished vertex
which we refer to as the $\omzl$-vertex
(corresponding to the term $\omzl( \, a^+_{u_n} a_{u_{n+1}} \, )$.
See Figure 1 for an example.
\item[-]
On the line on its left, we introduce $n$ vertices, and on the line on its right, 
we insert $\tn$ vertices.
We refer to those vertices as {\em interaction vertices}, and enumerate them
from 1 to $2\bn$ starting from the left.
\item[-]
The edges between the interaction vertices are referred to as {\em propagator lines}.
We label them by the momentum variables $u_0$, ..., $u_{2\bn+1}$, increasingly indexed
starting from the left. To the $j$-th propagator line, we associate the propagator
$U_{t_{j-1},t_j}(u_j)$ if $0\leq j \leq n$, 
and $\overline{U_{\tt_{j-1},\tt_j}(u_j)}$ if $n+1\leq j \leq 2\bn+1$
(with reference to the expression (\ref{eq-rhontn-1})).
\item[-]
To the $\ell$-th interaction vertex (adjacent to the edges labeled by
$u_{\ell-1}$ and $u_{\ell}$), we associate the random potential $\homega(u_\ell-u_{\ell-1})$,
where $1\leq \ell\leq 2\bn+1$.
\end{itemize}

\centerline{\epsffile{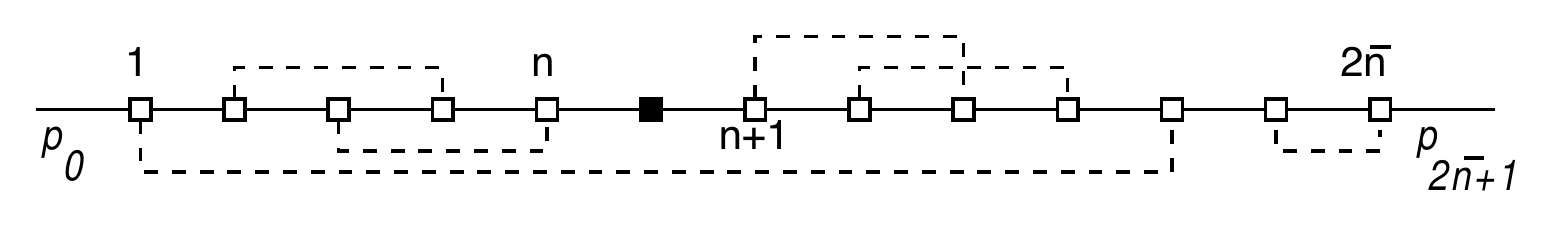} }

\noindent
Figure 1. A contraction graph.
\\

A {\em contraction graph} associated to the above pair of  particle lines
joined by the $\omzl$-vertex, and decorated by $n+\tn$ interaction vertices,
is the graph obtained by pairwise connecting interaction vertices by
{\em contraction lines}.
We denote the set of all such contraction graphs
by $\Gamma_{n,\tn}$; it contains
\eqn
	|\Gamma_{n,\tn}| \, = \, (2\bn-1)(2\bn-3)\cdots 3\cdot 1
	\, = \, \frac{(2\bn)!}{\bn!2^{\bn}} \, = \, O(\bn!)
\eeqn
elements.

If in a given graph $\pi\in\Gamma_{n,\tn}$,
the $\ell$-th and the $\ell'$-th vertex are joined by a contraction line,
we write
\eqn
	\ell \, \sim_\pi \, \ell' \,,
\eeqn
and we associate the delta distribution
\eqn
	\delta(u_{\ell}-u_{\ell-1}-(u_{\ell'}-u_{\ell'-1}))
	\, = \,
	\Exp[ \, \homega(\, u_{\ell}-u_{\ell-1} \,) \, \homega(\, u_{\ell'}-u_{\ell'-1} \,) \, ]
\eeqn
to this contraction line.

We classify Feynman graphs as follows; see also \cite{Ch1,ErdYau}:
\begin{itemize}
\item[-]
A subgraph consisting of one propagator line
adjacent to a pair of vertices $\ell$ and $\ell+1$,
and a contraction line connecting them, i.e., $\ell\sim_\pi\ell+1$,
where both $\ell$, $\ell+1$ are either $\leq n$
or $\geq n+1$, is called an {\em immediate recollision}.

\item[-]
The graph $\pi\in\Gamma_{n,n}$ (i.e., $n=\tn=\bn$) with
$\ell\sim_\pi 2n-\ell$ for all $\ell=1,\dots,n$,
is called a {\em basic ladder} diagram.
The contraction lines are called {\em rungs} of the ladder.
We note that a rung contraction always has the form
$\ell\sim_\pi\ell'$ with $\ell\leq n$ and $\ell'\geq n+1$.
Moreover, in a basic ladder diagram one always has that if
$\ell_1\sim_\pi\ell_1'$ and $\ell_2\sim_\pi\ell_2'$
with $\ell_1<\ell_2$, then $\ell_2'<\ell_1'$.

\item[-]
A diagram $\pi\in\Gamma_{n,\tn}$ is called a
{\em decorated ladder} if
any contraction is either an immediate recollision,
or a rung contraction $\ell_j\sim_\pi\ell_j'$
with $\ell_j\leq n$ and $\ell_j'\geq n$ for $j=1,\dots,k$,
and $\ell_1<\cdots<\ell_k$,
$\ell'_1>\cdots>\ell'_k$.
Evidently, a basic ladder diagram is the special case of a decorated
ladder which contains no immediate recollisions
(so that necessarily, $n=\tn$).

\item[-]
A diagram $\pi\in\Gamma_{n,\tn}$ is called {\em crossing}
if there is a pair of contractions $\ell\sim_\pi\ell'$,
$j\sim_\pi j'$, with $\ell<\ell'$ and $j<j'$, such that $\ell<j$.

\item[-]
A diagram $\pi\in\Gamma_{n,\tn}$ is called {\em nesting}
if there is a subdiagram with $\ell\sim_\pi\ell+2k$, with $k\geq1$,
and either $\ell\geq n+1$ or $\ell+2k\leq n$,
with $j\sim_\pi j+1$ for $j=\ell+1,\ell+3,\dots,\ell+2k-1$.
The latter corresponds to a progression of $k-1$ immediate recollisions.
\end{itemize}
We note that any diagram that is not a decorated ladder contains at least a crossing
or a nesting subdiagram.



\subsection{Feynman amplitudes}

To every Feynman graph $\pi\in\Gamma_{n,\tn}$ 
we associtate its {\em Feynman amplitude}, as follows.

To start with, we recall from (\ref{eq-cR-def-1})  
\eqn\label{eq-cR-def-1-1}
  &&
  \cR(u_0,\dots,u_n;\alpha+i\e)
  \\
  && \quad\quad\quad
  \, := \, \int_{\R_+^{n+1}} ds_0 \cdots ds_n \,
  \Big( \, \prod_{j=0}^n
  e^{-is_j(E(k_j)-\alpha-i\e)}e^{i\lambda\int_{t_{j-1}}^{t_j} ds' \, 
  \kappa_{s'}(k_j)} \, \Big) \,,
  \nonumber
\eeqn
where $t_j=s_0+\dots+s_{j}$ for $j>0$, and $t_{-1}=0$.
Moreover, we recall that
\eqn
  \kappa_s(p) \, = \, ( \, \widehat v*\npx_s \, )(p) \,
\eeqn
where
\eqn
  0 \, \leq \, \npx_s(p) \, \leq \, 1
\eeqn
holds uniformly in $s$ and $p$.
As a consequence,
\eqn\label{eq-kapp-bd-1}
	\| \, \kappa_s \, \|_{L^\infty(\Tor^3)} & \leq & \| \, \widehat v \, \|_{L^1(\Tor^3)} 
	\, = \, \int_{\Tor^3} dp \, \Big| \, \sum_x \, v(x) \, e^{-2\pi i px} \, \Big|
	\nonumber\\
	&\leq&{\rm Vol}\{\Tor^3\} \, \| \, \langle \, \cdot \, \rangle^{3/2+\deltexp} \, v \, \|_{\ell^2(\Z^3)}
	\, \| \, \langle \, \cdot \, \rangle^{-3/2-\deltexp} \, \|_{\ell^2(\Z^3)}
	\nonumber\\
	& < & C \, \| \, \widehat v \, \|_{H^{3/2+\deltexp}} 
	\nonumber\\
	& < &  C'
\eeqn
uniformly in $s\in\R_+$.
Here, we have recalled the property (\ref{eq-v-bd-1}) 
satisfied by the fermion pair interaction potential $v$,
for constants $C$, $C'$ that depend on $\deltexp>0$.

Given $\pi\in\Gamma_{n,\tn}$, we define
\eqn
	\delta_\pi( \, \{u_j\}_{j=0}^{2\bn+1} \, )
	\, := \, \prod_{ \ell \sim_\pi \ell' }
	\delta( \, u_{\ell}-u_{\ell-1}-(u_{\ell'}-u_{\ell'-1}) \, ) \,,
\eeqn
where every contraction line in $\pi$ corresponds to one of the factors on the rhs,
without any repetitions.
Clearly, $\delta_\pi$ determines the momentum conservation conditions 
on the graph $\pi$.

\begin{definition}
The Feynman amplitude associated to the graph $\pi\in\Gamma_{n,\tn}$ is defined by
\eqn\label{eq-rhontn-3}
	\lefteqn{
	\amp_\pi(f,g;\e;\eta)
	}
	\nonumber\\
	& := &
	\eta^{2\bn} \, e^{2 \e t} \,
	\int d\alpha \, d\talpha \, e^{it(\alpha-\talpha)}
	\\
	&&\int du_0\cdots du_{2\bn+1}  \, \overline{f(u_0)} \, g(u_{2\bn+1})
	\, J(u_n) \, \delta(u_n-u_{n+1}) \,
	\nonumber\\
	&&
	\quad\quad
	\delta_\pi( \, \{u_j\}_{j=0}^{2\bn+1} \, ) \,
	\cR(u_0,\dots,u_n;\alpha+i\e)
	\,\cR(u_{n+1},\dots,u_{2\bn+1};\talpha-i\e) \,.
	\nonumber
\eeqn
\end{definition}

Our choice of $\e$   will be $\e=\frac1t$.

Since $\{\omega_x\}$ are i.i.d. centered Gaussian,
\eqn
  \Exp\big[ \, \prod\homega(u_\ell-u_{\ell-1}) \, \big]
  \, = \,
  \sum_{\pi\in\Gamma_{n,\tn}} \prod_{i\sim_{\pi}j}
  \Exp\big[ \, \homega( \, u_{i}-u_{i-1} \, ) \, \homega( \, u_j-u_{j-1} \, ) \, \big]
\eeqn
equals the sum of all possible
products of pair correlations
\eqn
  \Exp[\, \homega( \, u \, ) \, \homega( \, u' \, ) \,] \, = \, \delta( \, u+u' \, )
\eeqn
(Wick's theorem).
Accordingly,  
\eqn
	\Exp[ \, \omzlt^{(n,\tn)}(f,g) \, ] \, = \, \sum_{\pi\in\Gamma_{n,\tn}}
	\amp_\pi(f,g;\e;\eta)
\eeqn
is the sum of Feynman amplitudes of all Feynman graphs $\pi\in\Gamma_{n,\tn}$.

As a consequence of translation invariance of $\omzl$, we have that
\eqn\label{eq-omzl-delta-1}	
	\omzl( \, a^+_{u_n} a_{u_{n+1}} \, )
	\, = \, J(u_n) \,
	\delta(\, u_n - u_{n+1} \, ) \,,
\eeqn
as we recall from (\ref{eq-J-def-1}).
Moreover, translation invariance also implies overall momentum conservation,
that is,
\eqn\label{eq-mom-cons-1}
	u_0 \, - \, u_{2\bn+1} \, = \, 0 \,,
\eeqn
which one easily verifies by summing up the arguments of all delta distributions.

Accordingly, we arrive at the expansion 
\eqn\label{eq-Expomzlt-Feyn-1}
	\Exp[ \, \omzlt(f,g) \, ] & = & \sum_{n,\tn=0}^{N+1}\sum_{\pi\in\Gamma_{n,\tn}}
	\amp_\pi(f,g;\e;\eta)
	\nonumber\\
	&+&\sum_{n=0}^{N}( \, \Exp[ \, \omzlt^{(n,N+1)}(f,g) \, ] 
	\, + \, \Exp[ \, \omzlt^{(N+1,n)}(f,g) \, ] \, )
	\nonumber\\
	&+&\Exp[ \, \omzlt^{(N+1,N+1)}(f,g) \, ]
\eeqn
where the first term on the rhs is entirely expressed in terms of Feynman graphs and Feynman amplitudes.
The terms on the second and third line on the r.h.s. involve the Duhamel remainder term, 
and will be shown only to contribute to a small error.

\newpage

\section{Proof of Theorem {\ref{thm-main-1}}: I. Bounds on error terms}
  
In order to prove the Boltzmann limit stated in Theorem {\ref{thm-main-1}}, 
we separate the main terms in the expression \eqref{eq-Expomzlt-Feyn-1} from the error terms.
We subsequently show that the main terms converge to a solution of the Boltzmann equation, 
while the error terms tend to zero. 
The Feynman graphs associated to the main term 
correspond to those appearing in the works \cite{ErdYau,Ch1,ChSa}.
However, many aspects of the approach developed in those works 
(for the weakly disordered Anderson model, which is linear) 
are not suitable for the problem at hand.
The main issue is the presence of the 
phase $\lambda\int_\tau^{\tau+s}\kappa_{s'}(u)ds'$ in \eqref{eq-cR-def-1-1}, 
which depends on the unknown quantity $\mu_t$ itself. 
In this section, we introduce a main tool, given in Lemmata \ref{lm-resolvexp-1}
and \ref{lm-phasecancel-1},   that enables us to control the
nonlinear self-interaction of the fermion field.

In a first step, we prove an estimate that will serve as a substitute for
resolvent estimates. The latter were abundantly used in \cite{ErdYau,Ch1},
but due to the nonlinear self-interactions of the fermion field, they are
not available here.
 
\begin{lemma}\label{lm-resolvexp-1}
Let $\e=\frac1t\ll1$.
Then, there exists a constant $C<\infty$ independent of $\eta$, $\lambda$, $\e$
such that
\eqn
	| \, \cR(u_0,\dots,u_n;\alpha+i\e) \, |
	\, \leq \, C^{n+1}  \,  \Big( \, 1+\frac\lambda\e \, \Big)^{n+1}
	\prod_{j=0}^n  \, \frac{1}{|E(u_j)-\alpha|+\e} \,
\eeqn
for all $n\in\N$.
\end{lemma}

Lemma \ref{lm-resolvexp-1} follows from Lemma \ref{lm-phasecancel-1} below.
We note that in our proof below, we allow for at most one integration by parts
with respect to the time variable $s$, which requires no smoothness
assumption $\kappa_s(u)$.
Iterating integration by parts with respect to $s$ would easily produce the asserted result,
but under the assumption that $\kappa_s$ is $C^n$-smooth in $s$.
However, we will only rely on the a priori boundedness of $\kappa_s(u)$, which
is a consequence of the Fermi statistics satisfied by the quantum field,
and will not assume any smoothness with respect to $s$.

\begin{lemma}\label{lm-phasecancel-1} 
Assume (\ref{eq-kapp-bd-1}). 
Then, uniformly in $\tau\geq0$, 
\eqn\label{eq-phaseint-0}
	\Big| \, \int_{\R^+} ds \, e^{-is(E(u)-\alpha-i\e)}
	\, e^{-i\lambda\int_\tau^{\tau+s}\kappa_{s'}(u) ds' } \, \Big|
	\, < \, \big( \, 1+\frac\lambda\e \, \big) \, \frac{C}{|E(u)-\alpha|+\e} \,,
\eeqn
where $E(u)$ is the
symbol of the nearest neighbor Laplacian on $\Z^3$.
\end{lemma}

\prf
We define
\eqn
	\okappa_{t,t+s}(u) \, := \, \frac{1}{s} \, \int_t^{t+s} \, ds' \, \kappa_{s'}(u) \,.
\eeqn
Clearly, $|\okappa_{t,t+s}(u)|<C_0$, uniformly in $t$ and $s\geq0$.

The integral on the left hand side of (\ref{eq-phaseint-0}) can be written as
\eqn\label{eq-phaseint-1}
	\int_{\R^+} ds \, e^{-is(E(u)-\alpha+\lambda\okappa_{t,t+s}(u) )}e^{-\e s} \,.
\eeqn
To estimate it, we split $\R_+$ into disjoint intervals
\eqn
	I_j \, := \, [ \, j\nuvar \, , \,  (j+1) \nuvar \, )
	\; \; \; , \; \; j\in\N_0 \,
\eeqn
of length
\eqn
	\nuvar \, := \, \frac{\pi}{|E(u)-\alpha|} \,.
\eeqn
We find
\eqn\label{eq-phaseint-2}
	(\ref{eq-phaseint-1}) & = & \sum_{j\in 2\N_0}
	\int_{I_j} ds \, \Big( \,
	e^{-is(E(u)-\alpha+\lambda\okappa_{t,t+s}(u) )}e^{-\e s}
	\nonumber\\
	&& \quad\quad\quad\quad\quad\quad
	\, + \,
	e^{-i(s+\nuvar)(E(u)-\alpha+\lambda\okappa_{t,t+s+\nuvar}(u) )}
	e^{-\e (s+\nuvar) }
	\, \Big) \,,
\eeqn
where the second term in the bracket accounts for the integrals over $I_j$ with $j$ odd.

Evidently,
$e^{-i\nuvar(E(u)-\alpha)} \, = \,  e^{\mp i\pi} \, = \, -1$.
Therefore, we get, for $j$ fixed,
\eqn
	\lefteqn{
	\int_{I_j} ds \, \Big( \,
	e^{-is(E(u)-\alpha+\lambda\okappa_{t,t+s}(u) )}e^{-\e s}
	}
	\nonumber\\
	&& \quad\quad\quad\quad
	\, + \,
	e^{-i(s+\nuvar)(E(u)-\alpha+\lambda\okappa_{t,t+s+\nuvar}(u) )}
	e^{-\e (s+\nuvar) }
	\, \Big)
	\nonumber\\
	\label{eq-aux-term-1}
	&=&
	\int_{I_j} ds \, e^{-is(E(u)-\alpha+\lambda\okappa_{t,t+s}(u))}
	\big( \, e^{-\e s} - e^{-\e (s+\nuvar )}  \, \big)
	\\
	\label{eq-aux-term-1-1}
	&+&
	\int_{I_j} ds \, e^{-is(E(u)-\alpha) }
	e^{-\e(s+\nuvar)} \,
	\big( \, e^{-i\lambda s \okappa_{t,t+s}(u)} \, - \, e^{-i\lambda(s+\nuvar)\okappa_{t,t+s }(u)}\, \big)
	\\
	\label{eq-aux-term-2}
	&+&
	\int_{I_j} ds \, e^{-is(E(u)-\alpha) }
	e^{-\e(s+\nuvar)} \,
	\big( \, e^{-i\lambda(s+\nuvar)\okappa_{t,t+s}(u)}
  \, - \, e^{-i\lambda(s+\nuvar)\okappa_{t,t+s+\nuvar}(u)}\, \big) \,.
	\; \; \; \; \; \; \; \; \; \;
\eeqn
Clearly,
\eqn
	\sum_{j\in2\N_0}|(\ref{eq-aux-term-1})| \, < \,  \int_{\R_+}ds \, e^{-\e s} \, \e \, \nuvar \,
	= \, \frac{\pi}{|E(u)-\alpha|} \,,
\eeqn
and
\eqn
	\sum_{j\in2\N_0}|(\ref{eq-aux-term-1-1})| \, < \,  \int_{\R_+}ds \, e^{-\e s} \, \lambda \, \nuvar \,
	= \,  \frac{\lambda}{\e} \, \frac{\pi}{|E(u)-\alpha|} \,.
\eeqn
On the other hand, we observe that for $s_1<s_2$,
\eqn
	\okappa_{t,t+s_2}(u)-\okappa_{t,t+s_1}(u)
	& = &
	(\frac{1}{s_2}-\frac{1}{s_1}) \int_t^{t+s_2} ds' \, \kappa_{s'}(u)
	\\
	& + &
	\frac{1}{s_1} \Big(\int_t^{t+s_2}-\int_t^{t+s_1}\Big) \, ds' \, \kappa_{s'}(u) \,.
\eeqn
Since $|\kappa_{s'}(u)|<C_0$ uniformly in $s'$, we immediately obtain
\eqn
	|\okappa_{t,t+s_2}(u)-\okappa_{t,t+s_1}(u)| \, < \, C \, \frac{s_2-s_1}{s_1} \,,
\eeqn
so that in particular,
\eqn\label{eq-okappa-bd-1}
	|\okappa_{t,t+\nuvar+s}(u)-\okappa_{t,t+s}(u)| \, < \, C \, \frac{\nuvar}{s} \,.
\eeqn
Thus, we conclude that
\eqn
	\sum_{j\in2\N_0}(\ref{eq-aux-term-2}) & \leq &
	 C \, \int_{\R_+} ds \, \lambda \, \nuvar \, \frac{ (s+\nuvar)}{s} \, e^{-\e (s+\nuvar)}
	\nonumber\\
	& \leq & C \, \frac{\lambda}{\e} \, \frac{\pi}{|E(u)-\alpha |} \,.
\eeqn
This proves that for $|E(u)-\alpha|>0$,
\eqn
	|(\ref{eq-phaseint-2})| \, < \, \frac{C}{|E(u)-\alpha|} \,,
\eeqn
under the assumption that $\lambda=O(\e)$.

If $|E(u)-\alpha|\leq\e$, then the trivial bound
\eqn
	|(\ref{eq-phaseint-2})| \, < \, \int_{\R_+} ds \, e^{-\e s} \, < \, \frac{C}{\e} \, 
\eeqn
is better, which ignores phase cancellations,
so that in conclusion,
\eqn
	|(\ref{eq-phaseint-2})| \, < \, \frac{C}{|E(u)-\alpha|+\e} \,,
\eeqn
as claimed.
\endprf

We may now prove Lemma \ref{lm-resolvexp-1}.

\prf
We consider
\eqn\label{eq-phaseint-manybody-1}
	\int_{\R_+^{n+1}} ds_0\cdots ds_n \, \prod_{j=0}^n \, e^{-i\Phi(s_j,u_j,t_{j-1})}
	e^{-\e s_j}
\eeqn
where
\eqn
	\Phi(s,u,t) \, := \, s \, ( \, E(u) - \alpha + \lambda \okappa_{t,t+s}(u) \, )
\eeqn
and
\eqn
	t_j \, := \, s_0 + \, \cdots \, + s_j
	\; \; \; , \; \; \; t_{-1} \, := \, 0  \,.
\eeqn
Then, subdividing $\R_+$ into intervals $I_\ell$ as before,
\eqn\label{eq-phaseint-manybody-2}
	|(\ref{eq-phaseint-manybody-1})|
	& \leq &\sum_{\ell_0,\dots,\ell_n\in2\N_0}\int_{I_{\ell_0}}ds_0 \,
	\Big| \, e^{-i\Phi(s_0,u_0,t_{-1})}e^{-\e s_j} -  e^{-i\Phi(s_0+\nuvar,u_0,t_{-1})}e^{-\e (s_0+\nuvar)} \, \Big|
	\nonumber\\
	&&\Big\{ \, \cdots \int_{I_{\ell_n}}ds_n \,
	\Big| \, e^{-i\Phi(s_n,u_n,t_{n-1})}e^{-\e s_n} \, -
	\,   e^{-i\Phi(s_n+\nuvar,u_n,t_{n-1})}e^{-\e (s_n+\nuvar)} \, \Big| \, \Big\}
	\nonumber\\
\eeqn
we may use $L^1-L^\infty$ bounds in $s_j$ to get
\eqn
	(\ref{eq-phaseint-manybody-2}) & \leq & \sum_{\ell_0,\dots,\ell_n\in2\N_0}
	\prod_{j=0}^n \sup_{s_0,\dots,s_{j-1}}\int_{I_{\ell_j}}ds_j \,
	\Big| \, e^{-i\Phi(s_j,u_j,t_{j-1})}e^{-\e s_j}
	\nonumber\\
	&&\quad\quad\quad\quad\quad\quad\quad\quad\quad\quad\quad\quad
	- \, e^{-i\Phi(s_j+\nuvar,u_j,t_{j-1})}e^{-\e (s_j+\nuvar)} \, \Big|
	\nonumber\\
	&\leq&
	\prod_{j=0}^n  C \, ( \, \e+\lambda\, ) \, \frac{1}{|E(u_j)-\alpha|+\e} \,
	\int_{\R_+}ds_j \, e^{-\e s_j}
	\nonumber\\
	&\leq& \prod_{j=0}^n  \, \frac{C}{|E(u_j)-\alpha|+\e}
\eeqn
by application of Lemma \ref{lm-phasecancel-1}, recalling the assumption that $\lambda\leq O(\e)$.
In particular, the fact has been proven here that the bound proven in  Lemma \ref{lm-phasecancel-1}
holds uniformly with respect to $t$.
\endprf

Moreover, we prove the following stationary phase estimate.

\begin{lemma}\label{lm-statphest-1}
Assume that $ f$, $\widehat v \in H^{3/2+\deltexp}(\Tor^3)$ for some
$\deltexp>0$, and that $\lambda\leq O(\eta^2)$. 
Then, for $0<s<t=O(\eta^{-2})$, and uniformly in $\tau\in\R$,
\eqn\label{eq-proplamb-uint-bd-1}
	\lefteqn{
	\Big| \, \int_{\Tor^3} du \, e^{-is(E(u)-\alpha-i\e)} \, e^{i\lambda\int_\tau^{\tau+s}\kappa_s(u)} \, 
	f(u) \, \Big|
	}
	\nonumber\\
	&&\quad\quad\quad
	\, < \, C(\deltexp) \, \langle s\rangle^{-3/2}
	\, \| \widehat v\|_{ H^{3/2+\deltexp}(\Tor^3)} \, \| \, f \, \|_{H^{3/2+\deltexp}(\Tor^3)}
\eeqn
where $\kappa_s=\npx_s*\widetilde v$.
\end{lemma}

\prf
Let 
\eqn 
	g_s(u) & := & e^{-is(E(u)-\alpha-i\e)} \,,
	\nonumber\\
	h_s(u) & := & e^{i\lambda\int_t^{t+s}ds' \, \kappa_{s'}(u)} \,.
\eeqn
Clearly, the left hand side of (\ref{eq-proplamb-uint-bd-1}) satisfies
\eqn  
	\Big| \, \int_{\Tor^3} du \, g_s(u) \, h_s(u) \, \Big| 
	&\leq&
	\| \langle x\rangle^{-3/2-\deltexp}g_s^\vee\|_{\ell^2(\Z^3)}
	\, \| \langle x\rangle^{3/2+\deltexp}(f h_s)^\vee\|_{\ell^2(\Z^3)}
	\nonumber\\
	&<& \frac{C'}{\deltexp} \, \|g_s^\vee\|_{\ell^\infty(\Z^3)}
	\,   \| \, f h_s \, \|_{H^{3/2+\deltexp}(\Tor^3)}
	\label{eq-basic-statphase-1}\\
	&<& \frac{C}{\deltexp} \, \langle s\rangle^{-3/2}
	\, \, \, \| \, f  \, \|_{H^{3/2+\deltexp}(\Tor^3)}
	\| \, h_s \, \|_{H^{3/2+\deltexp}(\Tor^3)}\,,
\eeqn
where we used the fact that $H^{\alpha}(\Tor^d)$, with $\alpha>\frac d2$, is a Banach algebra.

With $\lambda s \leq O(1)$,
\eqn\label{eq-kappader-bd-1}
	\| h_s\|_{ H^{3/2+\deltexp}(\Tor^3)}
	\, < \, C \, \|\kappa_s\|_{ H^{3/2+\deltexp}(\Tor^3)} \, \leq \, C \, \|\npx_s\|_{L^1(\Tor^3)} 
	\, \|\widehat v\|_{ H^{3/2+\deltexp}(\Tor^3)}
\eeqn
where $\|\npx_s\|_{L^1(\Tor^3)} \leq\|\npx_s\|_{L^\infty(\Tor^3)} <c$ uniformly in $s$.
Accordingly, \eqref{eq-proplamb-uint-bd-1} follows.
\endprf


\subsection{Feynman diagrams contributing to the error term}
\label{ssec-error-terms-1}

In this section, we combine Lemmata \ref{lm-resolvexp-1} and \ref{lm-phasecancel-1},
and the results of  Section \ref{ssec-nested-diagrams-1} below,
to prove that the amplitudes of
all Feynman graphs that include a crossing or nesting subdiagram
contribute only to a small error.
In addition, we show that all terms involving the Duhamel remainder term
likewise contribute to a small error.

We point out that in our approach, the estimates provided 
in Lemmata \ref{lm-resolvexp-1} and \ref{lm-phasecancel-1},
and in  Section \ref{ssec-nested-diagrams-1}, require
first integrating out all time variables $s_j$, and subsequently
integrating out the momenta $u_j$ in \eqref{eq-rhontn-3} and \eqref{eq-cR-def-1-1}.

This makes it possible to straightforwardly adopt estimates on crossing, nesting,
and remainder terms from \cite{Ch1,Ch2,ChSa,ErdYau}. 
Accordingly, our discussion in this section can be kept short.

\subsubsection{Crossing and nesting diagrams}

We defer the discussion of nesting diagrams to Section \ref{ssec-nested-diagrams-1} below
because the necessary ingredients are introduced only in later sections.
We shall here anticipate the result from the analysis given there.

Using Lemmata \ref{lm-resolvexp-1} and \ref{lm-phasecancel-1},
we may infer from the analysis presented in \cite{Ch1,Ch2,ChSa,ErdYau} 
that the Feynman amplitude of 
any graph $\pi\in\Gamma_{n,\tn}$ containing a crossing or nesting subdiagram
is bounded by
\eqn\label{eq-crossnest-aux-1}
	\lefteqn{
	\lim_{L\rightarrow\infty} |\amp_\pi(f,g;\e;\eta)| 
	}
	\\
	&&\quad\quad\quad 
	\, \leq \, \| \, f \, \|_2 \,
	\| \, g \, \|_2 \, \| \, J \, \|_\infty \,
	\e^{1/5} \, ( \, \log\frac1\e \, )^4 ( \, c\eta^2\e^{-1} \log\frac1\e \, )^{\bn} \,,
	\nonumber
\eeqn
where $2\bn=n+\tn\in2\N$.
The thermodynamic limit $L\rightarrow\infty$ and the upper bound
in (\ref{eq-crossnest-aux-1}) are
obtained in the same manner as in \cite{Ch1,Ch2,ChSa,ErdYau}
where we refer for details.
We do not repeat the discussion here, and instead refer to those references.

Let
\eqn
	\Gamma_{n,\tn}^{c-n} \, \subset \, \Gamma_{n,\tn}
\eeqn
denote the subset of Feynman graphs of crossing or nesting type, and
\eqn
	\Gamma_{2\bn}^{c-n} \, := \, \bigcup_{n+\tn=2\bn} \, \Gamma_{n,\tn}^{c-n} \,,
\eeqn
with cardinality $|\Gamma_{2\bn}^{c-n}| \, \leq \, 2^{\bn}\bn!$.

Accordingly, summing over all graphs with a crossing or nesting subdiagram,
\eqn\label{eq-cnsum-bd-1}
	\lefteqn{
	\sum_{1\leq \bn \leq N} \sum_{\pi\in\Gamma_{2\bn}^{c-n}}
	\lim_{L\rightarrow\infty} |\amp_\pi(f,g;\e;\eta)|
	}
	\\
	&&\quad\quad\quad
	\, < \, \|f\|_2 \, \|g\|_2 \, \| \, J \, \|_\infty
  \, \e^{1/5} \, ( \, \log\frac1\e \, )^4 \, ( \, c \, \eta^2\e^{-1}N\log\frac1\e \, )^{N} \,,
	\nonumber
\eeqn
where we replaced the factorials by $N!<N^N$.
Since $f,g$ are of Schwartz class,  $\|f\|_2,\|g\|_2<C$.

Moreover,
\eqn\label{eq-density-bd-2}
	\| \, J \, \|_\infty \, \leq \, 1 \,,
\eeqn
from (\ref{eq-density-bd-1}). 
As a concrete example, 
\eqn
	0 \, \leq \, J(p) \, = \, (1+e^{\beta(E(p)-\mu)})^{-1} \, \leq \, 1 \,,
\eeqn 
for the Gibbs state of a free Fermi gas, at inverse temperature $0\leq\beta\leq\infty$.

\subsubsection{Remainder term}
\label{ssec-remainder-1}

Using Lemmata \ref{lm-resolvexp-1} and \ref{lm-phasecancel-1}, 
and following \cite{ChSa} (which uses results in \cite{Ch1,Ch2,ErdYau}),
we straightforwardly find the following bounds on the contributions of the
Duhamel remainder term.

If at least one of the indices $n$, $\tn$ equals $N+1$, one obtains
\eqn\label{eq-rem-est-aux-1}
	\lefteqn{
	\lim_{L\rightarrow\infty} |\Exp[\rho^{(n,\tn)}(f,g)]| \,
	}
	\nonumber\\
	& \leq & \| \, f \, \|_2 \, \| \, g \, \|_2 \, \| \, J \, \|_\infty
	\, \Big[ \,
	\frac{N^2 \, \kappa^2 \, }{(N!)^{1/2}} \, \max\{c \, \e^{-1} \, \eta^2 \, , \, (c \, \e^{-1} \, \eta^2)^N\}
	\\
	&&
	+ \, ( \, N^2 \kappa^2 \e^{1/5} \, + \, \e^{-2} \kappa^{-N} \, ) \,
	( \, \log\frac1\e \, )^4 ( \, c\eta^2\e^{-1}N\log\frac1\e \, )^{8N} \, \Big]\,,
  \nonumber
\eeqn
(again using $(4N)!<(4N)^{4N}$)
where the constant $1\ll\kappa\ll t$ remains to be optimized.
The first term on the right hand side of (\ref{eq-rem-est-aux-1})
bounds the contribution from all basic ladder diagrams contained in the
Duhamel expanded remainder term.
A detailed discussion of an analogous result, and more details, can
be found in \cite{Ch1,Ch2,ChSa,ErdYau}.

\subsubsection{Choice of parameters and error bounds}
\label{ssec-constants-1}

The kinetic scaling limit asserted in Theorem {\ref{thm-main-1}} is determined by 
\eqn 
	t \, = \, \frac1\e \, = \, \frac{T}{\eta^2} \,
\eeqn
where $t$ denotes the microscopic, and $T$ the macroscopic time variable.
Similarly as in \cite{Ch1,Ch2,ChSa,ErdYau}, we choose
\eqn\label{eq-param-choice-1} 
	N \, = \, N(\e) & = & \frac{\log\frac1\e}{10 \log\log\frac1\e}
	\nonumber\\
	\kappa & = & (\log\frac1\e)^{15 } \,,
\eeqn
so that
\eqn
	\e^{-1/11} \; < \;  N! & < & \e^{-1/10}
	\nonumber\\
	\kappa^N & > & \e^{-3/2} \,.
\eeqn
One can easily verify that, accordingly, for any choice of $T>0$ finite and fixed,
\eqn\label{eq-rem-est-aux-1-1}
	(\ref{eq-cnsum-bd-1}) 
  \, < \, \eta^{1/15}
\eeqn 
\eqn\label{eq-rem-est-aux-1-2}
	(\ref{eq-rem-est-aux-1}) \, < \, \eta^{1/4} \,,
\eeqn
for $\eta$ sufficiently small.

In conclusion, we have proven the following overall bound on the
amplitudes of all Feynman graphs that we attribute to 
the error term in the expansion (\ref{eq-Expomzlt-Feyn-1}).

\begin{proposition}\label{prop-totalerr-bd-1}
For the choice of parameters (\ref{eq-param-choice-1}),  
\eqn\label{eqn-totalerr-bd-1}
	(\ref{eq-cnsum-bd-1}) \, + \, (\ref{eq-rem-est-aux-1})
	\, < \, \eta^{\frac{1}{20}} \, 
\eeqn
holds, for any choice of $T>0$ finite and fixed.
The l.h.s. of \eqref{eqn-totalerr-bd-1} contains the sum of Feynman amplitudes
associated to all diagrams containing crossing and/or nesting
subgraphs. It also contains the sum of all terms that depend on the remainder term of the
Duhamel expansion.
\end{proposition}

\newpage

\section{Proof of Theorem {\ref{thm-main-1}}: II. Resummation of main terms}
\label{sec-prfmainthm-1-ii}

In this section, we discuss the main terms in the
expression \eqref{eq-Expomzlt-Feyn-1}, which are associated to
the set of decorated ladder diagrams.
This is the complement of the set of Feynman graphs containing crossing and/or nesting
subgraphs studied in the previous section.
We prove that the sum of amplitudes of all decorated ladders
converges to a solution of the linear Boltzmann equation
(\ref{eq-linBoltz-1}), for $\lambda\leq O(\eta^2)$.
We remark that because of the presence of the nonlinear self-interaction 
of the fermion field, our analysis here
is much more involved than the analogous part 
in \cite{Ch1,ErdYau}.

We let $\Gamma_{n,\tn}^{(lad)}\subset\Gamma_{n,\tn}$ denote the subset of all
decorated ladders based on $n+\tn$ vertices.
Our goal is to prove that in the given kinetic scaling limit, the sum of amplitudes
associated to decorated ladder graphs converges to a solution of 
the linear Boltzmann equation
(\ref{eq-linBoltz-1}).  

\subsection{Outline of the proof}

Our discussion is organized as follows: 
\begin{itemize}
\item[\underline{\em Step 1}.]
In Sections \ref{ssec-immed-recoll-1} and \ref{ssec-renprop-1}, we 
consider the Feynman amplitudes associated to propagator lines decorated with an 
arbitrary number of immediate recollisions. The sum of all such terms produces
a renormalized propagator (renormalization in one-loop approximation,
according to standard terminology 
in quantum field theory). We prove that the leading term corresponds to a 
multiplicative renormalization of the basic propagator $U_{t_a,t_b}(u)$; 
see (\ref{eq-def-Uprop}) for its definition.
\item[\underline{\em Step 2}.]
In Section \ref{sssec-basiclad-1}, we discuss the Feynman amplitudes
of basic ladder diagrams (that is, the propagators between rungs   
do not contain any immediate recollisions). 
We prove that, in the kinetic scaling limit, the Feynman amplitudes
of these graphs determine  a transport equation
obtained from omitting the loss term on the rhs of
the linear Boltzmann equation (\ref{eq-linBoltz-1}). To prove that the 
error terms thus obtained are small, we apply various stationary phase
estimates to the propagators $U_{t_a,t_b}(u)$.
\item[\underline{\em Step 3}.]
In Section \ref{ssec-decladders-1}, we combine the results obtained in the 
previous two steps, and show that by decorating the basic ladder diagrams with immediate
recollisions, the full Boltzmann
equation  (\ref{eq-linBoltz-1}) is obtained in the kinetic scaling limit.
\end{itemize}

As a key result of Step {\em 2}, we obtain that the dominant part of the Feynman amplitudes is
in fact independent of $\lambda$ in the regime $\lambda=O(\eta^2)$.
This is a consequence of cancellations due to the translation invariance
of the system.

\subsection{Immediate recollisions}
\label{ssec-immed-recoll-1}

According to step 1 outlined above, we study the amplitude of a progression of $m$ 
immediate recollisions.  

\begin{lemma}
\label{lm-lambdazero-1}
For any connected interval $I\subset\R_+$ of length $|I|\leq O(t)$,
\eqn\label{eq-aux-immedrec-1}
  &&\Big| \, \int_{\Tor^3}du \, \int_{I} ds 
  \, e^{-i\int_{t}^{t+s}ds' \, (E(u)-\alpha-i\e\lambda\kappa_{s'}(u))} 
  \\
  &&\quad\quad\quad\quad\quad\quad\quad
  \, - \, \int_{\Tor^3}du \, \int_{I} ds 
  \, e^{-is (E(u)-\alpha-i\e)}  \, 
  \Big| \, < \, C \, \lambda |I|^{1/2} \,,
  \nonumber
\eeqn
and
\eqn\label{eq-aux-immedrec-2}
   \, \int_{I} ds
   \, \Big| \, \int_{\Tor^3}du 
   \, e^{-i\int_{t}^{t+s}ds' \, (E(u)-\lambda\kappa_{s'}(u))} \, \Big|
   \, \leq \, C \, ( \, 1 + \lambda |I|^{1/2} \, )
\eeqn
uniformly in $t\geq0$.
\end{lemma}

\prf
To prove (\ref{eq-aux-immedrec-1}), we write
\eqn
  \int_{I} ds \, \int_{\Tor^3}du \,  e^{-is(E(u)-\alpha-i\e)} 
  \, e^{i\lambda\int_t^{t+s}ds' \, \kappa_{s'}(u)}
  \, = \,
  (I) \, + \, (II)
\eeqn
where
\eqn
  (I) & := & \int_{I} ds \, \int_{\Tor^3}du \, e^{-is(E(u)-\alpha-i\e)} \,
\eeqn
and
\begin{equation}
  (II) \, := \,
  i \int_0^\lambda d\lambda' \,  \int_{I} ds \, \int_{\Tor^3}du \, e^{-is(E(u)-\alpha-i\e)} \,
  \int_t^{t+s}ds' \, \kappa_{s'}(u) \, e^{i\lambda'\int_t^{t+s'}ds''  \kappa_{s''}(u)} \, ,
\end{equation}
from differentiating and integrating with respect to the parameter $\lambda$.
To estimate $(II)$, we introduce a smooth partition of unity,
\eqn
  \int_{\Tor^3} du \, = \, \sum_{j=1}^{8} \int_{\Tor^3} du \, \chi_j(u)
\eeqn
where the smooth, $\Tor^3$-periodic bump function
$\chi_j$ is centered around the $j$-th critical point of $E(u)$, which
is a real analytic, perfect Morse function.
The properties (\ref{eq-v-bd-1}) satisfied by $\widehat v$ also hold for
$\kappa_s=\widehat v*\npx_s$ (see (\ref{eq-kappader-bd-1})).
Lemma \ref{lm-statphest-1}, applied on the support of each $\chi_j$, implies that
\eqn
  \lefteqn{
  \Big| \, \int_{\Tor^3}du \, e^{-is(E(u)-\alpha-i\e)} \,
  \int_t^{t+s}ds' \, \kappa_{s'}(u) \, e^{i\lambda'\int_t^{t+s'}ds''  \kappa_{s''}(u)} \, \Big|
  }
  \nonumber\\
  &\leq&C \,  \langle s \rangle^{-3/2} \,   e^{-\e s}
  \Big\| \, \int_t^{t+s}ds' \, \chi_j(u) \, \kappa_{s'}(u) \,
  e^{i\lambda'\int_t^{t+s'}ds''  \kappa_{s''}(u)} \, \Big\|_{H^{3/2+\deltexp}(\Tor^3)}
  \nonumber\\
  &\leq& C \, \langle s \rangle^{-3/2} \, s \, e^{-\e s}
\eeqn
for any finite $\deltexp>0$, as long as $\lambda\leq O(\e)$, and $s\leq\e^{-1}$ 
(where $\langle s \rangle = (1+s^2)^{1/2}$).
It thus follows  that
\eqn
	| \, (II) \, |
	\, \leq \, \lambda \,  \int_{I}ds \, \langle s \rangle^{-3/2} \, s \, e^{-\e s}
	\, = \, O( \, \lambda \, |I|^{1/2} \, )  \,,
\eeqn
which proves (\ref{eq-aux-immedrec-1}).
In the same manner,
\eqn
	\Big| \, \int_{\Tor^3}du \, 
	\, e^{-i\int_{t}^{t+s}ds' \, (E(u)-\lambda\kappa_{s'}(u))} \, \Big|
	\, < \, C \, \langle s \rangle^{-3/2} \, ( \, 1 + \lambda s \, ) \,,
\eeqn
implies  (\ref{eq-aux-immedrec-2}).
\endprf

\subsection{Renormalized propagators}
\label{ssec-renprop-1}

Next, we resum progressions of immediate recollisions of arbitrary length,
leading to a propagator renormalization.

Let $m\in \N$, and  $S:=\{1,\dots,2m+1\}$.
Moreover, let $0\leq t_a <t_b \leq t$.
We consider the integral
\eqn\label{eq-Umtatb-def-1}
  \lefteqn{
  U^{(m)}_{t_a,t_b}(u) \, := \, (-\eta^2)^{m}
  \int_{\R_+^{2m+1}} ds_{1}\cdots ds_{2m+1}
  \, \delta( \, t_{b}-t_a-\sum s_j \, )
  }
  \nonumber\\
  & &
  \int_{(\Tor^3)^{2m}} du_{2} \cdots du_{2m+1}
  \, \prod_{j=1}^{m}\delta\big( \, u_{ 2j+1} - u_{ 2j-1} \, \big)
  \nonumber\\
  & &
  \quad\quad\quad\quad\quad\quad
  \, \prod_{\ell=1}^{2m+1}
  \, e^{-i \int_{t_{\ell-1}}^{t_\ell}ds'(E(u_\ell)-\lambda\kappa_{s'}(u_\ell))}
\eeqn
corresponding to the Feynman graph given by a progression of $m$ immediate recollisions.
Integrating out the product of $m$ delta distributions,
\eqn\label{eq-Umtatb-def-2}
  \lefteqn{
  U^{(m)}_{t_a,t_b}(u) 
  \, = \,
  U^{(0)}_{t_a,t_b}(u) \,
  }
  \nonumber\\
  && \;  
  (-\eta^2)^{m} \,
  \int_{\R_+^{2m+1}} ds_{1}\cdots ds_{2m+1}
  \, \delta( \, t_{b}-t_a-\sum s_j \, ) \,
  \label{eq-Ubmtt-def-1}\\
  & &
  \quad\quad
  \, \prod_{j\in  S\cap 2\N} \int_{\Tor^3} du' \,
  \, e^{-i \int_{t_{j-1}}^{t_j}ds'(E(u')-E(u)-\lambda(\kappa_{s'}(u')-\kappa_{s'}(u)))}
  \nonumber 
\eeqn
where  
\eqn
	t_j \, =  
	\, t_a+s_{1}+\cdots+s_j \,,
\eeqn
for $j=1,\dots,2m+1$, and where
\eqn\label{eq-Uzero-def-1}
	U^{(0)}_{t_a,t_b}(u) \, = \, e^{-i \int_{t_{a}}^{t_{b}}ds'(E(u)-\lambda\kappa_{s'}(u))}
\eeqn
by definition of $U^{(m)}_{t_a,t_b}(u)$.

In the following lemma, we determine the Feynman amplitude of
a progression of $m$ immediate recollisions, and extract the dominant part.
We identify the latter
as a multiplicative renormalization of the free evolution term (\ref{eq-Uzero-def-1}). 

\begin{lemma}
\label{lm-mimmrec-1}
Assume that $\lambda\leq O(\eta^2)$ and $0\leq t_a<t_b\leq t$. Then, for 
all $m\in\N$ and $\delt=O(\eta^2)$,
\eqn 
	U^{(m)}_{t_a,t_b}(u) 
	\, = \,  U^{(main;m)}_{t_a,t_b}(u) 
	\, + \, \Delta U^{(m)}_{t_a,t_b}(u)
\eeqn
where
\eqn\label{eq-Umain-m-def-1} 
	U^{(main;m)}_{t_a,t_b}(u) 
	\, := \, \frac{1}{m!} 
	\; \Big( -\eta^2 \, (t_b - t_a) \, \int du' \frac{1}{E(u')-E(u)-i\delt} \, \Big)^m 
	\; 
	U^{(0)}_{t_a,t_b}(u) 
	\; 
\eeqn
and
\eqn\label{eq-DeltU-bd-lm-1}
	\| \, \Delta U^{(m)}_{t_a,t_b} \, \|_{L^\infty(\Tor^3)}
	\, \leq 
	\, \big( \, m \, \lambda \, \delt^{-1/2} 
	\, + \,  \delt^{1/2} \, \big) \, \big( \, c \, \eta^2 \, \delt^{-1} \, \big)^m \,.
\eeqn
In particular, $\Delta U^{(0)}_{t_a,t_b}(u)=0$ in the case $m=0$.
\end{lemma}

\prf
We represent the delta distribution in (\ref{eq-Umtatb-def-2}) by
\eqn\label{eq-delta-id-1}
	\delta(t_b-t_a-\sum s_j) & = & 
	e^{\delt(t_b-t_a)}e^{-\delt\sum s_j}\delta(t_b-t_a-\sum s_j)
	\nonumber\\
	& = &  e^{\delt(t_b-t_a)}e^{-\delt\sum s_j} \frac{1}{2\pi }\int_{\R } d\gamma \, 
	e^{-i(t_b-t_a-\sum s_j) \gamma }
	\nonumber\\ 
	& = & e^{\delt(t_b-t_a)}
	\, \frac{1}{2\pi }\int_{\R} d\gamma \, 
	e^{-i(t_b-t_a)\gamma}e^{ i(\sum s_j)(\gamma+i\delt)} \,,
\eeqn
where we will pick $\delt=\e=\frac1t=\frac{\eta^2}{T}$ so that 
$(t_b-t_a) \delt\leq \delt t =1$ and in particular, $e^{\delt(t_b-t_a)}\leq e$ uniformly in $\eta$.
In order to keep track of the origin of these two parameters, 
we will continue to notationally distinguish $\delt$ and $\e$.

From (\ref{eq-Umtatb-def-2}) and (\ref{eq-delta-id-1}),
\eqn 
  \lefteqn{
  U^{(m)}_{t_a,t_b}(u)  \, = \, 
  e^{-i \int_{t_{a}}^{t_{b}}ds'(E(u)-\lambda\kappa_{s'}(u)+i\delt)} \,
  }
  \nonumber\\
  & &
  \frac{1}{2\pi}\int_{\R} d\gamma \, e^{-i \gamma (t_b-t_a)}
  (-\eta^2)^{m} \,
  \int_{\R_+^{2m+1}} ds_{1}\cdots ds_{2m+1} \prod_{\ell\in  S\cap 2\N-1} e^{is_\ell (\gamma+i\delt)}
  \label{eq-Ubmtt-def-3} 
  \nonumber\\
  & &
  \quad\quad\quad
  \, \prod_{j\in  S\cap 2\N} \int_{\Tor^3} du' \,
  \, e^{-i \int_{t_{j-1}}^{t_j}ds'(E(u')-E(u)-\gamma-i\delt-\lambda(\kappa_{s'}(u')-\kappa_{s'}(u)))} \,.
  \; \;
\eeqn
where according to the above definitions, $t_j=t_{j-1}+s_j$. 
Our goal is to prove that
\eqn
	\lefteqn{
	U^{(m)}_{t_a,t_b}(u)  \,  =  \,  
	e^{-i \int_{t_{a}}^{t_{b}}ds'(E(u)-\lambda\kappa_{s'}(u)+i\delt)} \,  
	}
	\nonumber\\
  	& &
  	\Big\{ \,\frac{1}{2\pi}\int d\gamma \, e^{-i\gamma (t_b - t_a)} \,
  	\Big( \, \frac{i}{\gamma+i\delt} \, \Big)^{m+1} \,
  	\Big( \, \int du' \, \frac{-i\eta^2}{E(u')-E(u)-\gamma-i\delt}  \, \Big)^m
	\nonumber\\
	& &
	\quad\quad\quad\quad\quad\quad\quad\quad\quad\quad\quad\quad\quad\quad\quad\quad
	\quad\quad\quad\quad
	\, + \, \err_1 \, \Big\}
	\nonumber\\
	& = &
	e^{-i \int_{t_{a}}^{t_{b}}ds'(E(u)-\lambda\kappa_{s'}(u))} \,
	\Big\{ \, e^{\delt (t_b-t_a)} \, 
	\nonumber\\
  	& &
  	\frac{1}{2\pi}\int d\gamma \, e^{-i\gamma (t_b - t_a)} \,
  	\Big( \, \frac{i}{\gamma+i\delt} \, \Big)^{m+1} \,
  	\Big( \, \int du' \, \frac{-i\eta^2}{E(u')-E(u)-i\delt}  \, \Big)^m
  	\nonumber\\
  	& &
  	\quad\quad\quad\quad\quad\quad\quad\quad\quad\quad\quad\quad\quad\quad\quad\quad
	\quad\quad\quad\quad
  	\, + \, \err_1 \, + \, \err_2 \, \Big\}
  	\nonumber\\
	& = &  
	i e^{-i \int_{t_{a}}^{t_{b}}ds'(E(u)-\lambda\kappa_{s'}(u))} 
	\, \Big\{ \, \frac{(t_b - t_a)^m}{m!} 
	\, \Big( -\eta^2 \int du' \frac{1}{E(u')-E(u)-i\delt} \Big)^m
	\nonumber\\
	& &
	\quad\quad\quad\quad\quad\quad\quad\quad\quad\quad\quad\quad\quad\quad\quad\quad
	\quad\quad\quad\quad
	\, + \, \err_1 \, + \, \err_2 \, \Big\} \,,
	\label{eq-Utt-main-err-id-1}
\eeqn
where we claim that  
\eqn\label{eq-err1-bd-1}
	| \, \err_1 \, | \, < \,  m \, \lambda \, \delt^{-1/2} \, \big( \, c \, \eta^2 \, \delt^{-1} \, \big)^m 
\eeqn
and
\eqn
	| \, \err_2 \, | \, < \,  \delt^{1/2} \, \big( \, c \, \eta^2 \, \delt^{-1} \, \big)^m \,.
\eeqn
Clearly, this implies the asserted bound.

We first prove (\ref{eq-err1-bd-1}). To this end, we will use that
$\|\kappa_s\|_{H^{3/2+}(\Tor^3)}<c$ uniformly in $s$, see (\ref{eq-kappader-bd-1}), and the fact that 
the kinetic energy $E(u)$ is a real analytic, perfect
Morse function on $\Tor^3$. A stationary phase estimate similar to the one applied
in the proof of Lemma \ref{lm-statphest-1} yields
\eqn
	\lefteqn{
	\Big| \, \int_{\Tor^3} du' \,
	\, e^{-i \int_{t_{j-1}}^{t_j}ds'(E(u')-E(u)-\gamma-i\delt-\lambda(\kappa_{s'}(u')-\kappa_{s'}(u)))} 
	}
	\nonumber\\
	& & \quad\quad\quad\quad\quad\quad
	\, - \, 
	\int_{\Tor^3} du' \,
	\, e^{-i (t_j-t_{j-1})(E(u')-E(u)-\gamma-i\delt) } 
	\, \Big|
	\nonumber\\
	& \leq & \int_0^\lambda d\lambda' \, \Big| \, \int du' 
	\, \Big( \int_{t_{j-1}}^{t_j}ds'' \, \kappa_{s''}(u') \, \Big)
	\nonumber\\
	& & \quad\quad\quad\quad
	 e^{-i \int_{t_{j-1}}^{t_j}ds'(E(u')-E(u)-\gamma-i\delt-\lambda'(\kappa_{s'}(u')-\kappa_{s'}(u)))} \, \Big|
	\nonumber\\
	& < & C \,  \lambda \, s_j \, \langle s_j \rangle^{-3/2} \,,
\eeqn
where $t_j=t_{j-1}+s_j$.
Thus,
\eqn
	| \, \err_1 \, | 
	& < & 
  	C \, \eta^{2m} \,
  	\int_{\R_+^{2m+1}} ds_{1}\cdots ds_{2m+1}
  	\, \delta( \, t_{b}-t_a-\sum s_j \, ) \,
  	\nonumber\\
  	& &
  	\quad\quad
  \,\sum_{\ell\in  S\cap 2\N}  \lambda \,\langle s_\ell\rangle^{-1/2}
  \prod_{j\in  S\cap 2\N \, ; \, j\neq\ell} \langle s_j\rangle^{-3/2}
  \nonumber\\
  & < & C \, m \, \lambda \, \delt^{-1/2} \, ( \, C \, \delt^{-1} \, \eta^2 \, )^{m} \,.
\eeqn
This implies (\ref{eq-err1-bd-1}).

To estimate $\err_2$, we use  
\eqn
	\lefteqn{
	|\err_2| \, = \, \Big| \, \frac{1}{2\pi }\Big(\int_{|\gamma|<\delt } 
	+ \int_{|\gamma|\geq \delt } \Big) \, d\gamma \, e^{-i\gamma (t_b - t_a)} \,
	\Big( \, \frac{i}{\gamma+i\delt} \, \Big)^{m+1} \,
	}
	\nonumber\\
	& & \quad\quad\quad \Big[ \, \Big( \, \int du' \, \frac{i\eta^2}{E(u')-E(u)-\gamma-i\delt}  \, \Big)^m 
	\nonumber\\
	& & \quad\quad\quad\quad\quad\quad\quad\quad\quad\quad\quad
	- \, \Big( \, \int du' \, \frac{i\eta^2}{E(u')-E(u)-i\delt}  \, \Big)^m
	\, \Big] \, \Big|
	\nonumber\\
	& \leq & 
	C \, \eta^{2m} \int_{|\gamma|\geq\delt } d\gamma \, \Big( \, \frac{1}{|\gamma|+\delt} \, \Big)^{m+1}
	\nonumber\\
	& &
	\quad\quad\quad\quad\quad\quad\quad\quad
	\Big(2\sup_{\gamma\in\R}\sup_{E(u)\in[-6,6]}
	 \Big| \, \int du' \, \frac{1}{E(u')-E(u)-\gamma-i\delt} \, \Big|\Big)^m
	\nonumber\\
	& + & 
	C \, \eta^{2m} \int_{|\gamma|<\delt } d\gamma \, \Big( \, \frac{1}{|\gamma|+\delt} \, \Big)^{m+1} 
	\label{eq-aux-immedcolldiff-3}
	\\
	& & \quad\quad\quad\quad
	\sum_{j=1}^m 
	{m\choose j}
	(\Delta(\delt))^j \Big| \, \int du' \, \frac{1}{E(u')-E(u)-i\delt} \, \Big|^{m-j} 
	\nonumber
\eeqn
where
\eqn
	 \Delta(\delt) & := & \sup_{|\gamma|<\delt }\sup_{E(u)\in[-6,6]}
	 \Big| \, \int du' \, \frac{1}{E(u')-E(u)-\gamma-i\delt} 
	\\
	 & & \quad\quad\quad\quad\quad\quad\quad\quad\quad\quad\quad\quad
	 - \int du' \, \frac{1}{E(u')-E(u)-i\delt} \, \Big| \,.
	 \nonumber
\eeqn
We prove below that
\eqn\label{eq-aux-immedcolldiff-1}
	\Delta(\delt) \, < \, C \, \delt^{1/2} \, 
\eeqn 
and 
\eqn\label{eq-aux-immedcolldiff-2}
	\sup_{\gamma\in\R}\sup_{E(u)\in[-6,6]}
	\Big| \, \int du' \, \frac{1}{E(u')-E(u)-i\delt} \, \Big| \, < \, C \,.
\eeqn
Recalling that $\sum_{j=1}^m {m\choose j}=2^m-1$, this implies that
\eqn
	(\ref{eq-aux-immedcolldiff-3}) 
	\, < \, ( \, C \, \delt^{-1/2} \eta^{2} \, )^m
	\, + \, \delt^{1/2} \, ( \, C \, \delt^{-1} \, \eta^2 \, )^m \,.
\eeqn
Choosing $\delt=\e=O(\eta^2)$, we arrive at the asserted bound for $|\err_2|$.

To prove (\ref{eq-aux-immedcolldiff-1}), we use that, from a stationary phase estimate,
\eqn
	\lefteqn{
	\Big| \, \int du' \, \frac{1}{E(u')-E(u)-\gamma-i\delt} 
	- \int du' \, \frac{1}{E(u')-E(u)-i\delt} \, \Big|
	}
	\nonumber\\
	& = & \Big| \, \int_0^\gamma d\gamma' \int du' 
	\, \Big(\frac{1}{E(u')-E(u)-\gamma-i\delt}\Big)^2 \, \Big|
	\nonumber\\
	& = & \Big| \, \int_0^\gamma d\gamma' \int_0^\infty ds_1 \int_{s_1}^\infty ds_2
	\int du' \, e^{-is_2(E(u')-E(u)-\gamma-i\delt)}  \, \Big|
	\nonumber\\
	& \leq & C \, |\gamma| \, \int_0^\infty ds_1 \int_{s_1}^\infty ds_2 \,
	\langle s_2 \rangle^{-3/2} \, e^{-\delt s_2}
	\nonumber\\
	& \leq & C \, \delt \, \delt^{- 1/2} \,,
\eeqn
since $|\gamma|<\delt$, and since $E( \, \cdot \, )$ is a real analytic, 
perfect Morse function on $\Tor^3$,
as noted before.
This implies  (\ref{eq-aux-immedcolldiff-1}).

On the other hand,  
\eqn
	\lefteqn{
	\Big| \, \int du' \, \frac{1}{E(u')-E(u)-\gamma-i\delt} \, \Big| 
	}
	\nonumber\\
	& = & \Big| \, \int_0^\infty ds \, \int du' \, e^{-is(E(u')-E(u)-\gamma-i\delt)} \, \Big|
	\nonumber\\
	& < &
	C' \, \int_0^\infty ds \, \langle s \rangle^{-3/2} \, e^{-\delt s} 
	\nonumber\\
	& < & C 
\eeqn
for all $\gamma\in\R$, and in particular for $|\gamma|\geq\delt$. 
This implies (\ref{eq-aux-immedcolldiff-2}).
\endprf

\subsection{Sum of basic ladders}
\label{sssec-basiclad-1}

Following step 2 described at the beginning of Section \ref{sec-prfmainthm-1-ii}, 
we now determine the kinetic scaling limit of 
the sum of all Feynman amplitudes associated to basic ladder graphs.
In combination with 
the propagator renormalization addressed in the previous section,
this will allow us to 
complete the proof of the Boltzmann limit asserted in Theorem \ref{thm-main-1}.

The Feynman amplitude of a single basic ladder graph with $q$ rungs 
is given by
\eqn
	\lefteqn{
	\cU_{t}^{(basic;q)}(J;f,g)
	\, := \,
	(-\eta^2)^q\int_0^t dt_q \, \cdots\, \int_0^{t_{2}}dt_1
	\int_0^{t} d\tt_q \, \cdots\, \int_0^{\tt_{2}}d\tt_1 
	}
	\nonumber\\
	&&
	\int du_0 \, \cdots \, du_q \, J(u_q) \, \overline{f(u_0)} \, g(u_0) 
	\\
	&& \quad\quad\quad
	U_{t_{q},t}^{(0)}(u_q)
	\, \overline{U_{\tt_{q},t}^{(0)}(u_q)} \, \cdots \, 
	U_{t_{1},t_2}^{(0)}(u_1) \, \overline{U_{\tt_{1},\tt_2}^{(0)}(u_1)}\, 
	U_{0,t_1}^{(0)}(u_0) \, \overline{U_{0,\tt_1}^{(0)}(u_0)}
	\nonumber
\eeqn
where the definition of $U_{t_{j-1},t_j}^{(0)}(u)$ is given in (\ref{eq-Uzero-def-1}).
 
The following intermediate result will be important for the derivation of the full transport
equations in the next section.

\begin{proposition}
\label{prp-simplad-1}
Assume that $\lambda=o(\eta)$, and $M(\eta)\in\N$ with 
$M(\eta)\leq O(\frac{\log\frac1\eta}{\log\log\frac1\eta})$ and 
$\lim_{\eta\rightarrow0}M(\eta)=\infty$. Then, for any fixed, finite $T>0$, 
\eqn
	F_T^{(basic)}(J;f,g) \, := \,
	\lim_{\eta\rightarrow0}\sum_{q=0}^{M(\eta)}\cU_{T/\eta^2}^{(basic;q)}(J;f,g)
\eeqn
exists, and 
\eqn
	F_T^{(basic)}(J;f,g) \, = \, \int du \,  \overline{f(u)} \, g(u) \, F_T^{(basic)}(u)
\eeqn
where $F_T^{(basic)}(u)$ satisfies
\eqn
	\partial_T F_T^{(basic)}(u) \, = \, \int du' \, \delta( \, E(u) - E(u')\, )
	\, F_T^{(basic)}(u') \, ,
\eeqn 
with initial condition $F_0(u)=J(u)$.
\end{proposition}

This proposition is an immediate consequence of the following lemma.

\begin{lemma}
\label{lm-derUbmtt-1}
Assume that $\lambda=o(\eta)$. Then,
\eqn
	\lefteqn{
	\Big| \, \cU_{t}^{(basic;q)}(J;f,g) 
	\, - \, 
	\cU_{t}^{(basic-main;q)}(J;f,g) \, \Big|
	}
	\\
	& &
	\, < \, C \, q \, \lambda \, t^{1/2} \, ( \, \eta^2 t \log\frac1\eta \, )^{q-1}  
	\, + \, C \, q \, \eta \, ( \, \eta^2 t \log\frac1\eta \, )^{q-1} \,,
	\nonumber
\eeqn
where $\cU_{t}^{(basic-main;q)}(J;f,g)$ is defined in (\ref{eq-Ubasic-main-q-def-1}) below, and
\eqn
	F_T^{(basic;q)}(J;f,g) \, := \, \lim_{\eta\rightarrow0} \cU_{T/\eta^2}^{(basic-main;q)}(J;f,g)
\eeqn
exists for any fixed, finite $T>0$. 
In particular,
\eqn
	F_T^{(basic;q)}(J;f,g) \, = \, \int du \,  \overline{f(u)} \, g(u) \, F_T^{(basic;q)}(u)
\eeqn
where $F_T^{(basic;q)}(u)$ satisfies
\eqn
	\partial_T F_T^{(basic;q)}(u) \, = \, \int du' \, \delta( \, E(u) - E(u')\, )
	\, F_T^{(basic;q-1)}(u') 
\eeqn
and $F_0^{(basic;q)}(u)=0$ if $q\geq1$, and $ F_0^{(basic;0)}(u)=J(u)$.
\end{lemma}

\prf
First, we write
\eqn
	U_{t_{j-1},t_j}^{(0)}(u_j) \, = \, 
	U_{0,t_j}^{(0)}(u_j) \, \overline{ U_{0,t_{j-1}}^{(0)}(u_j) } \,, 
\eeqn
so that
\eqn
	\lefteqn{
	U_{t_{j-1},t_j}^{(0)}(u_j) \, U_{t_{j-2},t_{j-1}}^{(0)}(u_{j-1}) 
	}
	\nonumber\\
	&&
	\, = \, 
	U_{0,t_j}^{(0)}(u_j) \, 
	\Big( \, U_{0,t_{j-1}}^{(0)}(u_{j-1}) \, \overline{U_{0,t_{j-1}}^{(0)}(u_{j})} \, \Big) 
	\, \overline{ U_{0,t_{j-2}}^{(0)}(u_{j-1}) } 
\eeqn
and
\eqn
	\lefteqn{
	U_{t_{j-1},t_j}^{(0)}(u_j) \, U_{t_{j-2},t_{j-1}}^{(0)}(u_{j-1}) 
	\, 
	\overline{U_{\tt_{j-1},\tt_j}^{(0)}(u_j) } \, 
	\overline{U_{\tt_{j-2},\tt_{j-1}}^{(0)}(u_{j-1}) }
	}
	\nonumber\\
	&&
	\, = \, 
	U_{\tt_{j},t_j}^{(0)}(u_j) \, 
	\Big( \, U_{\tt_{j-1},t_{j-1}}^{(0)}(u_{j-1}) \, \overline{U_{\tt_{j-1},t_{j-1}}^{(0)}(u_{j})} \, \Big) 
	\, \overline{U_{\tt_{j-2},t_{j-2}}^{(0)}(u_{j-1}) } 
\eeqn
where $U_{t,t'}^{(0)}(u)=\overline{U_{t',t}^{(0)}(u)}$
(due to $\int_t^{t'}=-\int_{t'}^t$ on the rhs of  (\ref{eq-Uzero-def-1})) 
independently of $t>t'$ or $t<t'$.
Accordingly, we find
\eqn
	\lefteqn{
	\cU_{t}^{(basic;q)}(J;f,g)
	}
	\nonumber\\
	& := &
	(-\eta^2)^q\int_0^t dt_q \int_0^{t_{q}}dt_{q-1} \, \cdots\, \int_0^{t_{2}}dt_1
	\int_0^{t} d\tt_q \int_0^{\tt_q}d\tt_{q-1} \, \cdots\, \int_0^{\tt_{2}}d\tt_1 
	\nonumber\\
	&&
	\int du_0 \, \cdots \, du_q \, J(u_q) \, \overline{f(u_0)} \, g(u_0) 
	\label{eq-Utbasic-diffeq-0}\\
	&& \quad\quad\quad 
	\, U_{t,t}^{(0)}(u_q) \,   
	\Big( \, \overline{U_{\tt_q,t_q}^{(0)}(u_q)} \, U_{\tt_q,t_{q}}^{(0)}(u_{q-1}) \, \Big)
	\, \cdots \, 
	\Big( \, \overline{ U_{\tt_1,t_{1}}^{(0)}(u_1) } \, U_{\tt_{1},t_1}^{(0)}(u_0) \, \Big) \, 
	\nonumber
\eeqn
where, evidently, $U_{t,t}^{(0)}(u_q)=1$. 
Let $\delta_u(u'):=\delta(u'-u)$ so that  
\eqn
	\cU_{t}^{(basic;q)}(J;f,g) \, = \, \int du \, J(u) \, \cU_{t}^{(basic;q)}(\delta_u;f,g) \,.
\eeqn
Our proof comprises the following main steps.
\begin{itemize}
\item[\underline{\em Step (1)}]
First, we verify that $\cU_{t}^{(basic;q)}(J;f,g)$ satisfies the following approximate recursive identity,
\eqn
	\lefteqn{
	\cU_{t}^{(basic;q)}(J;f,g)  
	}
	\nonumber\\
	& = &
	- \eta^2 \, \int du_q \, J(u_q)\, \int du_{q-1} \, \int_0^t dt_q \int_0^t d\tt_q  \,
	\Big( \, \overline{U_{\tt_q,t_q}^{(0)}(u_q)} \, U_{\tt_q,t_{q}}^{(0)}(u_{q-1}) \, \Big)
	\nonumber\\
	&& \quad\quad\quad\quad\quad\quad\quad\quad\quad\quad\quad\quad\quad\quad\quad\quad\quad\quad\quad\quad
	\cU_{t_{q}}^{(basic;q-1)}(\delta_{u_{q-1}};f,g) 
	\nonumber\\
	&& 
	\, + \, O\Big( \, t^{-1/2} \, \frac{( \, C \, \eta^2 t  \, )^{q }}{(q-2)!} \, \Big) \,.
	\label{eq-Utbasic-diffeq-1}
\eeqn
We note that the
main term in (\ref{eq-Utbasic-diffeq-1}) differs from (\ref{eq-Utbasic-diffeq-0}) 
only by the 
upper integration boundary for the variable
$\tt_{q-1}$, which is replaced by $\tt_{q-1}\rightarrow t_q$. 
\\

\item[\underline{\em Step (2)}]
Next, we prove that for $\lambda=o(\eta)$, the nonlinear self-interaction of the
fermion field only contributes to a small error,
\eqn
	\lefteqn{
	\cU_{t}^{(basic;q)}(J;f,g)  
	}
	\nonumber\\
	& = &
	- \eta^2 \, \int du_q \, J(u_q)\, \int du_{q-1} \, \int_{0}^t dt_q \int_{-t}^t ds  \,
	\exp\Big( \, -i s\big( \, E(u_q)-E(u_{q-1}) \, \big) \, \Big)
	\nonumber\\
	&& \quad\quad\quad\quad\quad\quad\quad\quad\quad\quad\quad\quad\quad\quad\quad\quad\quad\quad\quad\quad
	\cU_{t_{q}}^{(basic;q-1)}(\delta_{u_{q-1}};f,g) 
	\nonumber\\
	&& 
	\, + \, O( \, \lambda \, t^{1/2} \, ( \, \eta^2 t \log\frac1\eta \, )^{q-1} \, )
	\, + \, O( \, \eta \, ( \, \eta^2 t \log\frac1\eta \, )^{q-1} \, ) \,.
	\label{eq-Utbasic-diffeq-2}
\eeqn

\item[\underline{\em Step (3)}]
Iterating (\ref{eq-Utbasic-diffeq-2}), one gets
\eqn 
	\lefteqn{
	\cU_{t}^{(basic;q)}(J;f,g)   \, = \, \cU_{t}^{(basic-main;q)}(J;f,g)
	}
	\nonumber\\
	&&
	\, + \, O( \, q \, \lambda \, t^{1/2} \, ( \, \eta^2 t \log\frac1\eta \, )^{q-1} \, )
	\, + \, O( \, q \, \eta \, ( \, \eta^2 t \log\frac1\eta \, )^{q-1} \, )
	\label{eq-Utbasic-diffeq-3}
\eeqn
where
\eqn\label{eq-Ubasic-main-q-def-1}
	\lefteqn{
	\cU_{t}^{(basic-main;q)}(J;f,g) 
	}
	\nonumber\\
	&& := \,
	(- \eta^2)^q \, \int du_q \, J(u_q)\, \int du_{q-1} \, \cdots \, du_0 \, 
	\overline{f(u_0)} \, g(u_0)
	\label{eq-Utbasic-diffeq-4}\\
	&&
	\, \int_{0}^t dt_q \, \cdots \, \int_0^{t_2} dt_1 \,
	\int_{-t}^t ds_q \, \cdots \, \int_{-t_2}^{t_2} ds_1  \,
	\prod_{j=1}^q\exp\Big( \, -i s_j\big( \, E(u_j)-E(u_{j-1}) \, \big) \, \Big) \,.
	\nonumber
\eeqn
Introducing the variables $T=\eta^2 t$ and $T_j=\eta^2 t_j$, 
\eqn
	\lefteqn{
	\cU_{T/\eta^2}^{(basic-main;q)}(J;f,g) 
	}
	\nonumber\\
	&& = \,
	(- 1)^q \, \int du_q \, J(u_q)\, \int du_{q-1} \, \cdots \, du_0 \, 
	\overline{f(u_0)} \, g(u_0)
	\, \int_{0}^T dT_q \, \cdots \, \int_0^{T_2} dT_1 \,
	\nonumber\\
	&&
	\int_{-T/\eta^2}^{T/\eta^2} ds_q \, \cdots \, \int_{t_2/\eta^2}^{t_2/\eta^2} ds_1  \,
	\prod_{j=1}^q\exp\Big( \, -i s_j\big( \, E(u_j)-E(u_{j-1}) \, \big) \, \Big) \,.
	\label{eq-Utbasic-diffeq-5}
\eeqn
For every $T_{j+1}>0$,  
\eqn 
	\int_{T_{j+1}/\eta^2}^{T_{j+1}/\eta^2} ds_j
	\exp\Big( \, -i s_j\big( \, E(u_j)-E(u_{j-1}) \, \big) \, \Big)
	\, \longrightarrow \, \delta\big( \, E(u_j)-E(u_{j-1}) \, \big)
	\label{eq-Utbasic-diffeq-6}
\eeqn
weakly in the limit $\eta\rightarrow0$. Therefore,
\eqn 
	\lefteqn{
	\lim_{\eta\rightarrow0}\cU_{T/\eta^2}^{(basic-main;q)}(J;f,g) 
	}
	\label{eq-Utbasic-diffeq-7} \\
	&&  = \,
	\frac{(- T)^q}{q!} \, \int du_0 \, \cdots \, du_q \, J(u_q)\,  
	\overline{f(u_0)} \, g(u_0)
	\, \prod_{j=1}^q\delta\big( \, E(u_j)-E(u_{j-1}) \, \big) \,.
	\nonumber
\eeqn
It is easy to check that (\ref{eq-Utbasic-diffeq-7}) has the properties
asserted in the lemma.  
For more details on the limit $\eta\rightarrow0$, we refer to \cite{ErdYau}.
\\
\end{itemize}

It thus remains to prove steps {\em (1)} and {\em (2)}.
\\

\noindent{\underline{\em Proof of step (1)}.}
The difference between  $\cU_{t}^{(basic;q)}(J;f,g)$
and the first term on the rhs of (\ref{eq-Utbasic-diffeq-1}) 
is given by
\eqn
	\lefteqn{
	\Delta \cU_{t}^{(basic;q)}(J;f,g) 
	}
	\nonumber\\
	& := & 
	\big( \, {\rm First \; term \; on \; rhs \; of \;  (\ref{eq-Utbasic-diffeq-1})} \, \big)
	\, - \, \cU_{t}^{(basic;q)}(J;f,g) 
	\nonumber\\
	& = &
	(-\eta^2)^q\int_0^t dt_q \int_0^{t_{q}}dt_{q-1} \, \cdots\, \int_0^{t_{2}}dt_1
	\int_0^{t} d\tt_q \int_{\tt_q}^{t_q}d\tt_{q-1} \, \cdots\, \int_0^{\tt_{2}}d\tt_1 
	\nonumber\\
	&&
	\int du_0 \, \cdots \, du_q \, J(u_q) \, \overline{f(u_0)} \, g(u_0) 
	\label{eq-Utbasic-diffeq-8}\\
	&& \quad\quad\quad 
	\, U_{t,t}^{(0)}(u_q) \,   
	\Big( \, \overline{U_{\tt_q,t_q}^{(0)}(u_q)} \, U_{\tt_q,t_{q}}^{(0)}(u_{q-1}) \, \Big)
	\, \cdots \, 
	\Big( \, \overline{ U_{\tt_1,t_{1}}^{(0)}(u_1) } \, U_{\tt_{1},t_1}^{(0)}(u_0) \, \Big) \,.
	\nonumber
\eeqn
The only difference between this expression and the expression (\ref{eq-Utbasic-diffeq-0})  
consists of the integration boundaries for the variable $\tt_{q-1}$.

We use a stationary phase argument similarly as in Lemma \ref{lm-statphest-1}
to bound the integral in $u_q$, 
which yields
\eqn 
	\Big| \, \int du_q \, J(u_q) \, \overline{U_{\tt_q,t_q}^{(0)}(u_q)} \, \Big| 
	\, < \, C \, \| \, J \, \|_{H^{\frac32+\sigma}(\Tor^3)} \, \langle t_q-\tt_q \rangle^{-3/2} \,,
\eeqn
for some $\sigma>0$, and similarly,
\eqn 
	\Big| \, \int du_0 \, \overline{f(u_0)} \, g(u_0) \,
	 \, U_{\tt_{1},t_1}^{(0)}(u_0) \, \Big| 
	\, < \, C \, \| \, f \, \|_{H^{\frac32+\sigma}(\Tor^3)}
	\, \| \, g \, \|_{H^{\frac32+\sigma}(\Tor^3)} 
	\, \langle t_1-\tt_1 \rangle^{-3/2} \,,
\eeqn 
recalling that $H^{\frac32+\sigma}(\Tor^3)$ is an algebra. 

For the integrals in $u_j$ with $j=2,\dots,q-1$,
\eqn 
	\Big| \, \int du_j \, U_{\tt_j,t_{j}}^{(0)}(u_{j})
	\, \overline{U_{\tt_{j-1},t_{j-1}}^{(0)}(u_j)} \, \Big| 
	\, < \, C \, \langle t_j-\tt_j -(t_{j-1}-\tt_{j-1})\rangle^{-3/2} \,.
\eeqn
Accordingly, writing
\eqn\label{eq-BJfg-def-1}
	B_{J,f,g} \, := \, \| \, J \, \|_{H^{\frac32+\sigma}(\Tor^3)} 
	\, \| \, f \, \|_{H^{\frac32+\sigma}(\Tor^3)}
	\, \| \, g \, \|_{H^{\frac32+\sigma}(\Tor^3)}
\eeqn
we find 
\eqn 
	\lefteqn{
	\Big| \, \Delta \cU_{t}^{(basic;q)}(J;f,g) \, \Big|
	}
	\nonumber\\
	& \leq & C^q \, B_{J,f,g}
	\, \eta^{2q}\int_0^t dt_q \int_0^{t} d\tt_q \, \frac{1}{\langle t_q-\tt_q \rangle^{3/2}}
	\nonumber\\
	&&\int_0^{t_{q}}dt_{q-1} 
	\int_{\tt_q}^{t_q}d\tt_{q-1} 
	\frac{1}{ \langle t_q-\tt_q -(t_{q-1}-\tt_{q-1})\rangle^{3/2} }
	\nonumber\\
	&&
	\int_0^{t_{q-1}}dt_{q-2} \, \cdots\, \int_0^{t_{2}}dt_1
	\int_{\tt_q}^{t_{q-1}}d\tt_{q-2} \, \cdots\, \int_0^{\tt_{2}}d\tt_1 
	\nonumber\\
	&&\quad\quad\quad\quad
	\frac{1}{\langle t_1-\tt_1 \rangle^{3/2}}
	\prod_{j=2}^{q-1}\frac{1}{\langle t_j-\tt_j -(t_{j-1}-\tt_{j-1})\rangle^{3/2}} \,.
\eeqn
To bound the integrals in $\tt_j$ with $j=1,\dots,q-1$, we use
\eqn 
	\lefteqn{
	\int_0^{\tt_{q-1}}d\tt_{q-2}
	\, \frac{1}{\langle t_{q-1}-\tt_{q-1} -(t_{q-2}-\tt_{q-2})\rangle^{3/2}} 
	\cdots \cdots
	}
	\nonumber\\
	&&\quad\quad\quad\quad
	\int_0^{\tt_{2}}d\tt_1 \,
	\frac{1}{ \langle t_2-\tt_2 -(t_{1}-\tt_{1})\rangle^{3/2} } \,
	\frac{1}{\langle t_1-\tt_1 \rangle^{3/2}}
	\nonumber\\
	&<&\int_{\R}d\tt_{q-2}
	\, \frac{1}{\langle t_{q-1}-\tt_{q-1} -(t_{q-2}-\tt_{q-2})\rangle^{3/2}} 
	\cdots \cdots 
	\nonumber\\
	&&\quad\quad\quad\quad
	\int_{\R}d\tt_1 \,
	\frac{1}{ \langle t_2-\tt_2 -(t_{1}-\tt_{1})\rangle^{3/2} } \,
	\frac{1}{\langle t_1-\tt_1 \rangle^{3/2}}
	\nonumber\\
	&<&\Big(\int_{\R}d\tt'
	\, \frac{1}{\langle \tt'\rangle^{3/2}} \Big)^{q-3} 
	\, \sup_{t'\in\R}\int_{\R}d\tt_1 \,
	\frac{1}{ \langle \tt_{1}-t'\rangle^{3/2} } \,
	\frac{1}{\langle \tt_1 \rangle^{3/2}}
	\nonumber\\
	&<&C^{q-2} \,,
\eeqn 
where we first majorized the expression by extending all
integration intervals to $\R$, and subsequently translated all variables $\tt_j\rightarrow\tt_j+t_j$.
The integrals in $t_j$ with $j=1,\dots,q-2$ are easily seen to be bounded by $\frac{t^{q-2}}{(q-2)!}$.
We thus find
\eqn 
	\lefteqn{
	\Big| \, \Delta \cU_{t}^{(basic;q)}(J;f,g) \, \Big|
	}
	\nonumber\\
	&\leq& C^q \, B_{J,f,g}
	\, \eta^4\int_0^t dt_q 
	\sup_{t_q}\Big\{ \, \int_0^{t} d\tt_q \, \frac{1}{\langle t_q-\tt_q \rangle^{3/2}}
	\nonumber\\
	&&
	\int_{\tt_q}^{t_q}d\tt_{q-1} 
	\sup_{t_q,\tt_q,\tt_{q-1}}\Big[\int_0^{t_{q}}dt_{q-1} 
	\frac{1}{ \langle t_q-\tt_q -(t_{q-1}-\tt_{q-1})\rangle^{3/2} }\Big] \, \Big\}
	\frac{(\eta^2 t)^{q-2}}{(q-2)!} 
	\nonumber\\
	&\leq& C^q \, B_{J,f,g}
	\, \eta^4\int_0^t dt_q 
	\sup_{t_q}\Big\{ \, \int_0^{t} d\tt_q \, \frac{1}{\langle t_q-\tt_q \rangle^{1/2}} \, \Big\}
	\frac{(\eta^2t)^{q-2}}{(q-2)!} 
	\nonumber\\
	&\leq& C^q \, B_{J,f,g} 
	\, \eta^4 t \, t^{1/2} 
	\frac{(\eta^2t)^{q-2}}{(q-2)!} \,, 
\eeqn
using that
\eqn
	\int_{\tt_q}^{t_q}d\tt_{q-1} 
	\sup_{t_q,\tt_q,\tt_{q-1}}\Big[\int_0^{t_{q}}dt_{q-1} 
	\frac{1}{ \langle t_q-\tt_q -(t_{q-1}-\tt_{q-1})\rangle^{3/2} }\Big]
	\, \leq \, 
	\bra \, t_q-\tt_q \, \ket \,.
\eeqn
Consequently, 
\eqn 
	\Big| \, \Delta \cU_{t}^{(basic;q)}(J;f,g) \, \Big| 
	\, < \, \, B_{J,f,g}
	\, t^{-1/2} \, \frac{(C \, \eta^2 t)^q}{(q-2)!} \, 
\eeqn
follows, as claimed. 
\\

\noindent{\underline{\em Proof of step (2)}.}
Similarly as in  Lemma \ref{lm-lambdazero-1},
\eqn  
	\Big( \, \overline{U_{\tt_q,t_q}^{(0)}(u_q)} \, U_{\tt_q,t_{q}}^{(0)}(u_{q-1}) \, \Big)
	\, = \, 
	\exp\big( \, -i(t_q-\tt_q) \, ( \, E(u_q)-E(u_{q-1}) \, )\, \big) 
	\nonumber\\
	\times \,
	\Big(\, 1  \, + \,   \, g(u_q,u_{q-1};t_{q},\tt_{q};\lambda) \, \Big) \, ,
\eeqn
for $g(u_q,u_{q-1};t_{q},\tt_{q};\lambda)=
e^{i\lambda\int_{\tt_q}^{t_q}ds (\kappa_s(u_{q-1})-\kappa_s(u_q))}-1$ satisfying 
\eqn
	\|g(\,\bullet\,,u_{q-1};t_{q},\tt_{q};\lambda)\|_{H^{3/2+\sigma}} 
	\, < \, C_0 \, \lambda \, | \, t_q-\tt_q \, |
\eeqn 
as a function of $u_q$, with $C_0$ dependent on $\sigma>0$, 
but independent of $u_{q-1}$, $t_{q}$, $\tt_{q}$,  $\lambda$.
The regularity of $g$ is inherited from the pair interaction potential, 
$\widehat v\in H^{3/2+\sigma}(\Tor^3)$ (see (\ref{eq-v-bd-1})), 
and proven as in (\ref{eq-kappader-bd-1}).

Similarly as in the proof of Lemma \ref{lm-lambdazero-1}, 
a stationary phase argument yields
\eqn 
	&&\Big| \, \int du_q \, \exp\big( \, -i(t_q-\tt_q) \, ( \, E(u_q)-E(u_{q-1}) \, )\, \big)\,
	g(u_q,u_{q-1};t_{q-1},t_{q};\lambda) \,  \Big|
	\nonumber\\
 	&&\quad\quad\quad\quad\quad\quad\quad\quad
	\, < \, C \, \lambda \, \langle t_q-\tt_q \rangle^{-1/2} \,.
\eeqn
For the integral in $\tt_q$, we thus find
\eqn 
	&&\int_0^t d\tt_q \, \Big| \, \int du_q \, 
	\Big( \, \overline{U_{\tt_q,t_q}^{(0)}(u_q)} \, U_{\tt_q,t_{q}}^{(0)}(u_{q-1})  
	\, - \, 
	e^{ -i(t_q-\tt_q) (  E(u_q)-E(u_{q-1}) ) } \, \Big) \, \Big|
	\nonumber\\
	&&\quad\quad\quad\quad\quad\quad\quad\quad
	\, < \, C \, \lambda \, t^{1/2} \,.
\eeqn
It is then straightforward to arrive at (\ref{eq-Utbasic-diffeq-2}).
\endprf
 
We may now prove Proposition \ref{prp-simplad-1}.

\prf
To establish Proposition \ref{prp-simplad-1}, it suffices to verify that
\eqn\label{eq-Utbasic-diffeq-7-0}
	\lim_{\eta\rightarrow0}
	\Big| \, \sum_{q=0}^{M(\eta)}\cU_{T/\eta^2}^{(basic;q)}(J;f,g)
	\, - \, \sum_{q=0}^{M(\eta)}\cU_{T/\eta^2}^{(basic-main;q)}(J;f,g) \, \Big|
	\, = \, 0 \,.
\eeqn
For the left hand side, we obtain the bound
\eqn\label{eq-Utbasic-diffeq-7-1}
	\lefteqn{ 
	\sum_{q=0}^{M(\eta)} 
	\Big| \, \cU_{T/\eta^2}^{(basic;q)}(J;f,g)
	\, - \, \cU_{T/\eta^2}^{(basic-main;q)}(J;f,g) \, \Big|
	}
	\nonumber\\
	&& \leq \, \sum_{q=1}^{M(\eta)}          
	\Big[ \,  \lambda \, t^{1/2} \, ( \, C \, \eta^2 t \log\frac1\eta \, )^{q-1} \,  
	\, + \, \eta \, ( \, C \, \eta^2 t \log\frac1\eta \, )^{q-1} \, \Big] \,,
\eeqn
using the estimates on the error terms in (\ref{eq-Utbasic-diffeq-3}).
For the choice of parameters
\eqn 
  	\lambda \, = \, O(\eta^2)
	\; \; \; , \; \; \;
	t=O(\eta^{-2})
	\; \; \; , \; \; \;
	M(\eta) \, = \, \frac{\log\frac1\eta}{c_0\log\log\frac1\eta} \,  
\eeqn 
where we assume that $c_0>2$, this is bounded by
\eqn 
	(\ref{eq-Utbasic-diffeq-7-1}) &<& 
	\eta \, ( \, C_0 \, \log\frac1\eta \, )^{M(\eta)} \, M(\eta) \, 
	\nonumber\\
	&\leq& \eta \, \cdot \, 
	\eta^{-(\frac{\log C_0}{\log\log\frac1\eta}+\frac1{c_0}) } \,  \log\frac1\eta
	\nonumber\\
	&\leq& \eta^{\frac1{10}}
\eeqn
if $\eta$ is sufficiently small.  
This implies (\ref{eq-Utbasic-diffeq-7-0}).
\endprf

\subsection{Remark on bounds related to nested diagrams}
\label{ssec-nested-diagrams-1}
The analysis given above enables us to control the Feynman amplitudes
associated to nested diagrams,
as was noted in Section \ref{ssec-error-terms-1}.
To this end, we recall that for the  Feynman amplitudes
belonging to all diagrams containing crossings or
nestings are evaluated by first integrating over
all time variables $s_j$, and subsequently integrating over momentum variables $u_j$.
We note that our argument below exhibits the same ordering of integration steps.

Following the definition of nested diagrams in Section \ref{ssec-Feynman-graphs-1}, 
a nesting subgraph of length $m$ is a progression of $m$ consecutive 
immediate recollisions connected via a propagator line to two outermost vertices 
which are mutually contracted.
The contribution to the Feynman amplitude associated to this segment of the graph is
proportional to
\eqn
	 \eta^2\int_{\Tor^3} du \int_{0}^\tau ds \, U^{(m)}_{t_a,t_a+s}(u)  
\eeqn
for some $\tau\leq 1=T/\eta^2$, where the integration variable $u$ appears 
nowhere else in the full expression of the Feynman amplitude (for the entire graph). 
The factor $\eta^2$ accounts for the 
contraction of the two outermost vertices.

Using Lemma \ref{lm-mimmrec-1}, one straightforwardly verifies that 
\eqn
	\lefteqn{
	 \eta^2\int_{\Tor^3} du \Big| \, \int_{0}^\tau ds \, U^{(m)}_{t_a,t_a+s}(u) \, \Big|
	}
	\nonumber\\
	&\leq&
	\eta^2\int_{\Tor^3} du \Big| \, \int_{0}^\tau ds \, U^{(main;m)}_{t_a,t_a+s}(u) \, \Big|
	\, + \, \eta^2 \tau \sup_{s,u}|\Delta U^{(m)}_{t_a,t_a+s}(u)|
	\nonumber\\
	&\leq&
	C^m \, \eta^2 \int_{\Tor^3} du \, 
	\Big|\int_{0}^\tau ds \, \frac{(\eta^2 s)^m}{m!} \, U^{(0)}_{t_a,t_a+s}(u) \, \Big|
	\, + \, C^m \, \eta \,,
	\label{eq-nesting-aux-1}
\eeqn
where we have recalled that in  Lemma \ref{lm-mimmrec-1}, the values of the
parameters are given by $\lambda,\delt=O(\eta^2)$.

We claim that
\eqn\label{eq-savprop-bd-1}
	\Big| \, \int_0^\tau ds \,  \frac{ s^m}{m!} \, U_{t_a,t_a+s}^{(0)}(u) \, \Big|
	\, < \, C^m  \, \frac{1}{|E(u) |+\tau^{-1}} \, \frac{\tau^m}{m!} \,.
\eeqn 
The proof of (\ref{eq-savprop-bd-1}) is similar to the one of Lemma \ref{lm-resolvexp-1}.
If $|E(u) |\leq \tau^{-1}$, the bound is trivially fulfilled. If $|E(u)|>\tau^{-1}$,
we define 
\eqn 
	\nuvar \, := \, \frac{\pi}{|E(u) |} \,,
\eeqn 
and the intervals $I_j:=[j \nuvar \, , \, (j+1)\nuvar)$
with $j\in J:=\N\cap [0,\frac{\tau}{\nuvar}]$.
Then, the left hand side of (\ref{eq-savprop-bd-1}) can be bounded by
\eqn 
	\lefteqn{
	\int_0^\tau ds\, \frac{s^m}{m!} \, U_{t_a,t_a+s}^{(0)}(u) 
	}
	\nonumber\\
	&=&
	\sum_{j\in 2\N_0\cap J}  \int_{I_j} ds \, \Big( \, U_{t_a,t_a+s}^{(0)}(u) 
	\, \frac{(s+\nuvar)^m}{m!}
	\, + \, U_{t_a,t_a+s}^{(0)}(u)  \, \frac{s^m}{m!} \, \Big)
	\nonumber\\
	&=&\sum_{j\in 2\N_0\cap J}  \int_{[0,\nuvar]} ds \, e^{-is E(u) }
	\, \Big( \, e^{is\lambda\okappa_{t_a+j\nuvar,t'+j\nuvar+s}(u)}\, \frac{s^m}{m!}
	\\
	&&\quad\quad\quad\quad\quad\quad\quad\quad\quad
	\, - \, e^{i(s+\nuvar)\lambda\okappa_{t_a+(j+1)\nuvar,t'+(j+1)\nuvar+s}(u)}\, 
	\frac{(s+\nuvar)^m}{m!} \, \Big) \,.
	\nonumber
\eeqn
Similarly as in the proof of Lemma \ref{lm-resolvexp-1},
\eqn 
	|e^{is\lambda\okappa_{t_a+j\nuvar,t_a+j\nuvar+s}(u)}
	 \, \frac{s^m}{m!} 
	\, - \, e^{i(s+\nuvar)\lambda\okappa_{t_a+(j+1)\nuvar,t_a+(j+1)\nuvar+s}(u)}|
	\, < \, C \, \lambda \, \nuvar \, ,
\eeqn
and evidently,
\eqn 
	0 \, \leq \,\frac{(s+\nuvar)^m}{m!}-\frac{ s^m}{m!} \, \leq \, \nuvar \, \frac{(s+\nuvar)^{m-1}}{(m-1)!} \,.
\eeqn
Accordingly,
\eqn 
	\lefteqn{
	\Big| \, \int_0^\tau ds \, U_{t_a,t_a+s}^{(0)}(u) \, \frac{s^m}{m!} \, \Big|
	}
	\nonumber\\
	& \leq& C \, \Big[ 
	\, \lambda \, \nuvar \, \int_0^\tau ds \,  \frac{(s+\nuvar)^{m}}{m!} 
	\, + \, \nuvar \, \int_0^\tau ds \,  \frac{(s+\nuvar)^{m-1}}{(m-1)!} \, \Big]
	\nonumber\\
	& \leq & C \, \nuvar \, \frac{(2\tau)^{m}}{m!} 
\eeqn
for $\lambda,\tau^{-1}\leq O(\eta^2)$. This proves (\ref{eq-savprop-bd-1}).

Therefore, we conclude that
\eqn
	\eqref{eq-nesting-aux-1}
	&\leq&
	C^m \eta^2 \int du \, \frac{1}{|E(u)|+\eta^2}
	\, + \, C^m \, \eta
	\nonumber\\
	&\leq& C^m \eta^2 \log\frac1\eta
	\, + \, C^m \, \eta
	\nonumber\\
	&\leq&C^m \eta \,,
\eeqn
The gain of a factor $\eta$ is crucial, and immediately implies the bounds on nesting subgraphs
used in Section \ref{ssec-Feynman-graphs-1}.
For a more detailed discussion of nested diagrams in the context of the 
weakly disordered Anderson model, we refer to
\cite{ErdYau,Ch1}.

\newpage

\subsection{Decorated ladders and Boltzmann limit}
\label{ssec-decladders-1}

Following the list of steps explained at the beginning of Section \ref{sec-prfmainthm-1-ii},
we now carry out step 3.
Combining Lemma \ref{lm-mimmrec-1} and Proposition \ref{prp-simplad-1},
we derive the Boltzmann limit for the sum of decorated (renormalized) ladders,
and complete the proof of Theorem \ref{thm-main-1}.

For notational convenience, we introduce the multiindices
\eqn 
	\umq \, := \, (m_0,\cdots,m_q)
	\; \; \; , \; \; \; 
	\utmq \, := \, (\tm_0,\cdots,\tm_q)
\eeqn 
for fixed $q\in\N$, and 
\eqn 
	|\umq| \, := \, m_0+ \, \cdots \, + m_q
	\; \; \; , \; \; \;
	|\utmq| \, = \, \tm_0 + \, \cdots \, + \tm_q \,.
\eeqn
We use $N(\e)=O(\frac{\log\frac1\e}{\log\log\frac1\e})$ 
with $\frac1\e=t=\frac T{\eta^2}$
as in (\ref{eq-param-choice-1}), and consider
\eqn  
	\lefteqn{
	\cU^{(ren)}_t(J;f,g)  
	}
	\nonumber\\
	& := &
	\sum_{\bn=0}^{N(\e)} \; \sum_{n+\tn=2\bn} \; \sum_{\pi\subset \Gamma_{n,\tn}^{(lad)}}
	\lim_{L\rightarrow\infty} \amp_\pi(J;f,g;t;\eta)
	\\
	&=&
	\sum_{M_0,\dots,M_{N(\e)}\in\N_0\atop{\sum M_j\leq N(\e)}} 
	\, \sum_{q\in\N_0} \, 
	\sum_{\umq,\utmq \atop{|\umq|+|\utmq|+q=M_q}}
	\cU^{(ren;q;\umq,\utmq)}_t(J;f,g) 
	\nonumber
\eeqn
where
\eqn\label{eq-Utrenmain-mjtmj-0}
	\lefteqn{
	\cU^{(ren;q;\umq,\utmq)}_t(J;f,g) \, := \, (-\eta^2)^q 
	\int du_0 \cdots du_q
	\, J(u_q) \, \overline{f(u_0)} \, g(u_0) \,
	}
	\\
	&&\quad\quad\quad
	\int_0^t dt_q \cdots \int_0^{t_2} dt_1 \int_0^t d\tt_q \cdots \int_0^{\tt_2} d\tt_1
	\prod_{j=0}^q
	U_{t_{j-1},t_j}^{(m_j)}(u_j) 
	\, \overline{U_{\tt_{j-1},\tt_j}^{(\tm_j)}(u_j)} \,,
	\nonumber
\eeqn
using the convention that $t_{-1}=0=\tt_{-1}$. This is the Feynman amplitude of a 
ladder with $q$ rungs, where the two particle edges labeled by the momentum $u_j$
are decorated with $m_j$, respectively $\tm_j$, immediate recollisions.

To extract the dominant terms in this expression, we define
\eqn 
	\lefteqn{
	\cU^{(ren-main;q;\umq,\utmq)}_t(J;f,g) 
	}
	\nonumber\\
	&:=& 
	(-\eta^2)^q 
	\int_0^t dt_q \cdots \int_0^{t_2} dt_1 \int_0^t d\tt_q \cdots \int_0^{\tt_2} d\tt_1
	\int du_0 \cdots du_q
	\, J(u_q) \, \overline{f(u_0)} \, g(u_0) \,
	\nonumber\\
	&&\quad\quad\quad 
	\prod_{j=1}^q  \,
	U_{t_{j-1},t_j}^{(main;m_j)}(u_j) \, \overline{U_{\tt_{j-1},\tt_j}^{(main;\tm_j)}(u_j)}  
	\nonumber\\
	&=& 
	(-\eta^2)^q 
	\int_0^t dt_q \cdots \int_0^{t_2} dt_1 \int_0^t d\tt_q \cdots \int_0^{\tt_2} d\tt_1
	\int du_0 \cdots du_q
	\, J(u_q) \, \overline{f(u_0)} \, g(u_0) \,
	\nonumber\\
	&&\quad\quad\quad 
	\prod_{j=1}^q  \,
	\, U_{t_{j-1},t_j}^{(0)}(u_j;u_0) \,  
	\, \overline{U_{\tt_{j-1},\tt_j}^{(0)}(u_j;u_0)} \,  
	\nonumber\\
	&&\quad\quad\quad 
	\prod_{j=1}^q  \Big\{ \,
	\, \frac{1}{m_j!}
	\Big( -\eta^2 \, (t_j - t_{j-1}) \, \int du' \frac{1}{E(u')-E(u_j)-i\delt} \Big)^{m_j}
	\label{eq-Utrenmain-mjtmj-1}\\
	&&\quad\quad\quad\quad\quad\quad
	\, \frac{1}{\tm_j!}
	\Big( -\eta^2 \, (\tt_j - \tt_{j-1}) 
	\, \int du' \frac{1}{E(u')-E(u_j)+i\delt} \Big)^{\tm_j} \, \Big\} \,
	\nonumber
\eeqn 
where 
\eqn
	U_{t_{j-1},t_j}^{(0)}(u_j;u_0) \; := \,U_{t_{j-1},t_j}^{(0)}(u_j) 
	\, \overline{U_{t_{j-1},t_j}^{(0)}(u_0)} \,,
\eeqn
and where we have inserted a factor
\eqn 
	1 \, = \, 
	\overline{U_{0,t}^{(0)}(u_0)} \,  U_{0,t}^{(0)}(u_0)
	\, = \,
	\prod_{j=1}^q  \, \overline{U_{t_{j-1},t_j}^{(0)}(u_0)} \,  U_{\tt_{j-1},\tt_j}^{(0)}(u_0) \,.
\eeqn
To get (\ref{eq-Utrenmain-mjtmj-1}) from \eqref{eq-Utrenmain-mjtmj-0}, 
we have replaced the contributions from the immediate
recollisions by their dominant parts identified in Lemma \ref{lm-mimmrec-1}.

Moreover, we define
\eqn 
	\lefteqn{
	\cU^{(ren-main-0;q;\umq,\utmq)}_t(J;f,g) 
	}
	\nonumber\\
	&:=& 
	(-\eta^2)^q 
	\int_0^t dt_q \cdots \int_0^{t_2} dt_1 \int_0^t d\tt_q \cdots \int_0^{\tt_2} d\tt_1
	\int du_0 \cdots du_q
	\, J(u_q) \, \overline{f(u_0)} \, g(u_0) \,
	\nonumber\\
	&&\quad\quad\quad 
	\prod_{j=1}^q  \,
	\, U_{t_{j-1},t_j}^{(0)}(u_j;u_0) 
	\, \overline{U_{\tt_{j-1},\tt_j}^{(0)}(u_j;u_0)} \, 
	\nonumber\\
	&&\quad\quad\quad 
	\prod_{j=1}^q  \Big\{ \,
	\, \frac{1}{m_j!}
	\Big( -\eta^2 \, (t_j - t_{j-1}) \, \int du' \frac{1}{E(u')-E(u_0)-i\delt} \Big)^{m_j}
	\label{eq-Utrenmain-mjtmj-2}\\
	&&\quad\quad\quad\quad\quad\quad
	\, \frac{1}{\tm_j!}
	\Big( -\eta^2 \, (\tt_j - \tt_{j-1}) 
	\, \int du' \frac{1}{E(u')-E(u_0)+i\delt} \Big)^{\tm_j} \, \Big\} \,.
	\nonumber
\eeqn
To get this expression from (\ref{eq-Utrenmain-mjtmj-1}),
the momenta $u_j$ have been replaced by $u_0$ in the last product in  (\ref{eq-Utrenmain-mjtmj-1}).
Here, we anticipate the conservation of kinetic energy in the collision processes,
which will emerge in the kinetic scaling limit. 
We then prove the following result which immediately implies Theorem \ref{thm-main-1}.

\begin{proposition}
Let $N(\e)$ be as in Section  \ref{ssec-constants-1}. 
Then, for any fixed, finite $T>0$ and $t=\frac{T}{\eta^2}$, and $\delt=\e=\frac1t$,
\eqn
	\lefteqn{
	\lim_{\eta\rightarrow0} \sum_{M_0,\dots,M_{N(\e)}\in\N_0\atop{\sum M_j\leq N(\e)}} 
	\, \sum_{q\in\N_0} \, 
	\sum_{\umq,\utmq \atop{|\umq|+|\utmq|+q=M_q}}
	\cU^{(ren-main-0;q;\umq,\utmq)}_{T/\eta^2}(J;f,g)
	}
	\nonumber\\
	&& \quad\quad\quad\quad\quad\quad\quad\quad\quad\quad\quad\quad\quad\quad\quad
	\, = \, \int du \, \overline{f(u)} \, g(u) \, F_T(u)
	\label{eq-renladdsum-1}
\eeqn
where $F_T(u)$ satisfies the linear Boltzmann equation (\ref{eq-linBoltz-1}) with initial
condition $F_0(u)=J(u)$.
Moreover,
\eqn
	\lefteqn{
	\lim_{\eta\rightarrow0}   
	\sum_{M_0,\dots,M_{N(\e)}\in\N_0\atop{\sum M_j\leq N(\e)}} \, \sum_{q\in\N_0} \, 
	\sum_{\umq,\utmq \atop{|\umq|+|\utmq|+q=M_q}}
	\Big[ \, \cU^{(ren;q;\umq,\utmq)}_{T/\eta^2}(J;f,g)
	}
	\label{eq-renladdsum-2}\\
	&& \quad\quad\quad\quad\quad\quad\quad\quad\quad\quad\quad\quad\quad\quad\quad\quad
	\, - \, \cU^{(ren-main;q;\umq,\utmq)}_{T/\eta^2}(J;f,g) \, \Big] 
	\, = \, 0 \,,
	\nonumber
\eeqn
and
\eqn
	\lefteqn{
	\lim_{\eta\rightarrow0} 
	\sum_{M_0,\dots,M_{N(\e)}\in\N_0\atop{\sum M_j\leq N(\e)}} \, \sum_{q\in\N_0} \, 
	\sum_{\umq,\utmq \atop{|\umq|+|\utmq|+q=M_q}}
	\Big[ \, \cU^{(ren-main;q;\umq,\utmq)}_{T/\eta^2}(J;f,g)
	}
	\label{eq-renladdsum-3}\\
	&& \quad\quad\quad\quad\quad\quad\quad\quad\quad\quad\quad\quad\quad\quad\quad
	\, - \, \cU^{(ren-main-0;q;\umq,\utmq)}_{T/\eta^2}(J;f,g) \, \Big] 
	\, = \, 0 \,.
	\nonumber
\eeqn
\end{proposition}

\prf
We first verify the Boltzmann limit for the main term before proving the
error estimates.
\\

\noindent{$\bullet$ \underline{\em 1. Proof of (\ref{eq-renladdsum-1})}.}
We have
\eqn
	\lefteqn{
	\lim_{\eta\rightarrow0} \sum_{M_0,\dots,M_{N(\e)}\in\N_0\atop{\sum M_j\leq N(\e)}} 
	\, \sum_{q\in\N_0} \, 
	\sum_{\umq,\utmq \atop{|\umq|+|\utmq|+q=M_q}}
	\cU^{(ren-main-0;q;\umq,\utmq)}_{T/\eta^2}(J;f,g)
	}
	\nonumber\\
	& = & \sum_{M_0,\dots,M_{j},\dots\in\N_0 } 
	\, \sum_{q\in\N_0} \, 
	\sum_{\umq,\utmq \atop{|\umq|+|\utmq|+q=M_q}}
	\cU^{(ren-main-0;q;\umq,\utmq)}_{T/\eta^2}(J;f,g)
	\nonumber\\
	& = & 
	\, \sum_{q\in\N_0} \, 
	\sum_{\umq,\utmq\in\N_0^{q+1}}
	\cU^{(ren-main-0;q;\umq,\utmq)}_{T/\eta^2}(J;f,g) \,.
	\label{eq-renladdsum-1-1}
\eeqn
We find 
\eqn
	\lefteqn{
	\sum_{\umq,\utmq\in\N_0^{q+1}}\cU^{(ren-main-0;q;\umq,\utmq)}_t(J;f,g)
	}
	\nonumber\\
	& = &
	(-\eta^2)^q 
	\int_0^t dt_1 \cdots \int_0^{t_2} dt_1 \int_0^t d\tt_q \cdots \int_0^{\tt_2} d\tt_1
	\int du_0 \cdots du_q
	\, J(u_q) \, \overline{f(u_0)} \, g(u_0) \,
	\nonumber\\
	&&\quad\quad\quad 
	\prod_{j=1}^q  \,
	\, U_{t_{j-1},t_j}^{(0)}(u_j;u_0) \, 
	\, \overline{U_{\tt_{j-1},\tt_j}^{(0)}(u_j;u_0)} 
	\\
	&&\quad\quad\quad 
	 \prod_{j=1}^q
	e^{ -\eta^2 \, (t_j - t_{j-1}) \, \int du' \frac{1}{E(u')-E(u_0)-i\delt}}
	\, e^{-\eta^2 \, (\tt_j - \tt_{j-1}) \, \int du' \frac{1}{E(u')-E(u_0)+i\delt} } \,.
	\nonumber
\eeqn
First of all, we note that the expression obtained from setting the product 
on the last line equal to 1 is precisely 
$\cU_{t}^{(basic;q)}(J;f,g)$; that is, the amplitude of a basic ladder with $q$
rungs from Section \ref{sssec-basiclad-1}.
Clearly, the product on the last line equals
\eqn 
	e^{- \eta^2 \, t \, 2 \, {\rm Im}\int du' \frac{1}{E(u')-E(u_0)-i\delt}} \,.
\eeqn
We are choosing $\delt=\e$, and consider the limit $\delt=\e\rightarrow0$,
where
\eqn 
	\lim_{\delt\rightarrow0}{\rm Im}\int du' \frac{1}{E(u')-E(u)-i\delt}
	\, = \, \pi \int du' \, \delta( \, E(u') - E(u) \, ) \,.
\eeqn
Thus, it follows straightforwardly from this, and from Proposition \ref{prp-simplad-1}
that for $\lambda=O(\eta^2)$,  
and any fixed, finite $T>0$, 
\eqn
	F_T^{(ren-main-0)}(J;f,g) \, := \,
	\lim_{\eta\rightarrow0}\sum_{q\in\N_0}\cU_{T/\eta^2}^{(ren-main-0;q)}(J;f,g) 
\eeqn
exists, and 
\eqn
	F_T^{(ren-main-0)}(J;f,g) \, = \, \int du \,  \overline{f(u)} \, g(u) \, 
	F_T(u)
\eeqn
where
\eqn 
	F_T(u) \, = \, e^{-2\pi T \, \int du' \, \delta(E(u)-E(u'))} \, F_T^{(basic)}(u)
\eeqn
(see  Section \ref{sssec-basiclad-1} for the definition of $F_T^{(basic)}(u)$)
satisfies the linear Boltzmann equation
\eqn
	\partial_T F_T(u) \, = \,2 \pi \int du' \, \delta( \, E(u) - E(u')\, )
	\, ( \, F_T(u') \, - \, F_T(u) \, )  \, ,
\eeqn 
with initial condition  
\eqn
	F_0(u) \, = \, \lim_{L\rightarrow\infty} J(u)
	\, = \, \lim_{L\rightarrow\infty}\frac{1}{L^3} \, \omzl( \, a_{u}^+ a_{u} \, ) \,.
\eeqn
This proves (\ref{eq-renladdsum-1}).
\\

\noindent{$\bullet$ \underline{\em 2. Proof of (\ref{eq-renladdsum-2})}.}
Recalling Lemma \ref{lm-mimmrec-1}, we consider
\eqn
	\lefteqn{
	\cU^{(ren;q;\umq,\utmq)}_{T/\eta^2}(J;f,g)
	\, - \, \cU^{(ren-main;q;\umq,\utmq)}_{T/\eta^2}(J;f,g)
	}
	\label{eq-renladdsum-4}\\
	&=&
	(-\eta^2)^q \int du_0 \cdots du_q
	\, J(u_q) \, \overline{f(u_0)} \, g(u_0) \,
	\int_0^t dt_q \cdots \int_0^{t_2} dt_1 \int_0^t d\tt_q \cdots \int_0^{\tt_2} d\tt_1
	\nonumber\\
	&&\quad\quad\quad 
	\Big[ \, \prod_{j=1}^q   
	\, \big( \, U_{t_{j-1},t_j}^{(m_j)}(u_j) \, \overline{ U_{t_{j-1},t_j}^{(0)}(u_0) } \, \big)
	\, \big( \, \overline{U_{\tt_{j-1},\tt_j}^{(\tm_j)}(u_j)} \, U_{\tt_{j-1},\tt_j}^{(0)}(u_0) \, \big)
	\nonumber\\
	&&\quad\quad\quad\quad\quad
	\, - \, \prod_{j=1}^q  
	\, \big( \, U_{t_{j-1},t_j}^{(main;m_j)}(u_j) \, \overline{ U_{t_{j-1},t_j}^{(0)}(u_0) } )
	\, ( \overline{U_{\tt_{j-1},\tt_j}^{(main;\tm_j)}(u_j)} \, U_{\tt_{j-1},\tt_j}^{(0)}(u_0) \, \big) \, \Big]
	\label{eq-Utrenmain-mjtmj-10} 
	\nonumber\\
	& = & (A) \, + \, (B)
\eeqn
where
\eqn 
	\lefteqn{
	(A) \, := \, 
	(-\eta^2)^q \sum_{\ell=0}^{q}
	\int du_0 \cdots du_q
	\, J(u_q) \, \overline{f(u_0)} \, g(u_0) \,
	}
	\nonumber\\
	&&\quad\quad\quad
	\int_0^t dt_q \cdots \int_0^{t_2} dt_1 \int_0^t d\tt_q \cdots \int_0^{\tt_2} d\tt_1
	\nonumber\\
	&&\quad\quad\quad 
	\Big[ \,    
	\, \Big( \, \prod_{j=1}^{\ell-1} U_{t_{j-1},t_j}^{(m_j)}(u_j) \,\, \Big)
	\, \Delta U^{(m_j)}_{t_{\ell-1},t_{\ell}}(u_\ell) 
	\, \Big( \, \prod_{j=\ell+1}^{q} U_{t_{j-1},t_j}^{(main;m_j)}(u_j) \, \Big)
	\nonumber\\
	&&\quad\quad\quad\quad\quad\quad\quad\quad\quad   
	\, \Big[\prod_{j=1}^q\overline{ U_{t_{j-1},t_j}^{(0)}(u_0) }\Big]
	\, \prod_{j=1}^q
	\, \big( \, \overline{U_{\tt_{j-1},\tt_j}^{(\tm_j)}(u_j)} \, U_{\tt_{j-1},\tt_j}^{(0)}(u_0) \, \big) 
	\, \Big] \,.
	\label{eq-Utrenmain-mjtmj-11} 
	\nonumber\\
\eeqn
The functions $U^{(main;m)}_{t_a,t_b}(u)$ and
$\Delta U^{(m)}_{t_a,t_b}(u)$ were defined in connection with Lemma \ref{lm-mimmrec-1}.
$(B)$ is the analogous term with the roles of the variables $t_i$ and $\tt_i$ exchanged,
and with $U_{t_{j-1},t_j}^{(m)}(u_j)$ replaced by $U_{t_{j-1},t_j}^{(main;m)}(u_j)$.

To bound the integrals with respect to $\tt_j$
in (\ref{eq-Utrenmain-mjtmj-11}),
we recall from \eqref{eq-Umtatb-def-1} that 
\eqn  
  U^{(m)}_{t_a,t_b}(u) & = & (-\eta^2)^{m}
  \int_{\R_+^{2m+1}} ds_{1}\cdots ds_{2m+1}
  \, \delta( \, t_{b}-t_a-\sum s_j \, ) 
  \nonumber\\
  & &
  \int_{(\Tor^3)^{2m}} du_{2} \cdots du_{2m+1}
  \, \prod_{j=1}^{m}\delta\big( \, u_{ 2j+1} - u_{ 2j-1} \, \big)
  \nonumber\\
  & &
  \quad\quad\quad\quad\quad\quad
  \, \prod_{\ell=1}^{2m+1}
  \, e^{-i \int_{t_{\ell-1}}^{t_\ell}ds'(E(u_\ell)-\lambda\kappa_{s'}(u_\ell))} \,.
\eeqn 
Thus,
\eqn\label{eq-intttres-bd-1}
	\lefteqn{
  \int du_0\cdots du_q \,  \int_0^t d\tt_q \cdots \int_0^{\tt_2} d\tt_1 \, \prod_{j=1}^q
	\, \big( \, \overline{U_{\tt_{j-1},\tt_j}^{(m_j)}(u_j)} 
	\, U_{\tt_{j-1},\tt_j}^{(0)}(u_0) \, \big)  
  }
  \nonumber\\
  & = &
  \int du_0\cdots du_q \,  
  \prod_{j=1}^q \, \int_0^{\tt_{j}-\tt_{j-1}}d\ts_{j}
  \nonumber\\
  & &
  \prod_{j=1}^q \, (-\eta^2)^{m_j}
  \int_{(\Tor^3)^{2m_j}} du_{j,2} \cdots du_{j,2m_j+1}
  \, \prod_{i=1}^{m_j}\delta\big( \, u_{j, 2i+1} - u_{j, 2i-1} \, \big)
  \nonumber\\
  & &
  \quad\quad\quad\quad\quad\quad
  \int_{\R_+^{2m_j+1}} ds_{j,1}\cdots ds_{j,2m_j+1}
  \, \delta( \, \ts_j-\sum s_{j,i} \, ) 
  \nonumber\\
  & &
  \quad\quad\quad\quad\quad\quad 
  \, \prod_{\ell=1}^{2m_j+1}
  \, e^{-i \int_{t_{\ell-1}}^{t_\ell}ds'((E(u_{j,\ell})-E(u_0)
  -\lambda(\kappa_{s'}(u_{j,\ell})-\kappa_{s'}(u_0))}
  \,.
  \quad\quad\quad
\eeqn
We now bound this expression by performing the integrals over all time integrations
first, using Lemma \ref{lm-resolvexp-1},
where we begin with the latest time, and successively integrate out the 
preceding time variable. We obtain that this is bounded by
\eqn\label{eq-intttres-bd-1-1}
  \lefteqn{
  |\eqref{eq-intttres-bd-1}|
  }
  \nonumber\\
  & \leq &
  \int du_0\cdots du_q \,  
  \prod_{j=1}^q \, \int_{(\Tor^3)^{2m_j}} du_{j,2} \cdots du_{j,2m_j+1}
  \, \prod_{i=1}^{m_j}\delta\big( \, u_{j, 2i+1} - u_{j, 2i-1} \, \big)
  \nonumber\\
  & &
  \prod_{j=1}^q \, (-\eta^2)^{m_j} 
  \, \frac{1}{|E(u_{j})-E(u_0)|+\delt}\prod_{\ell=1}^{2m_j+1}
  \, \frac{1}{|E(u_{j,\ell})-E(u_0)|+\delt}
  \,.
\eeqn
Integrating out all delta distributions, we obtain
\eqn
  \lefteqn{
  \eqref{eq-intttres-bd-1-1}
  }
  \nonumber\\
  & \leq & \int du_0\cdots du_q \,
  \Big[ \, \prod_{j=1}^q\eta^{2m_j}\Big(\frac{1}{|E(u_{j})-E(u_0)|+\delt}\Big)^{m_j+1}
  \nonumber\\
  &&\quad\quad\quad\quad\quad\quad\quad\quad\quad\quad
  \Big(\int du \frac{1}{|E(u )-E(u_0)|+\delt}\Big)^{m_j } \, \Big]
	\nonumber\\
	&\leq&   \prod_{j=1}^q \, ( \, C \eta^2 \delt^{-1} (\log\frac1\delt )^2\, )^{m_j}  \,,
\eeqn
for $\delt=\e=O(\eta^2)$. 

From $|U_{\tt_{j-1},\tt_j}^{(0)}(u)|=1$, and \eqref{eq-Umain-m-def-1}, we  find
\eqn
  |U_{t_{j-1},t_j}^{(main;m_j)}(u_j)| \, \leq \,  \frac{(C\eta^2(t_j-t_{j-1}))^{m_j}}{m_j!} 
  \, < \, e^{C\eta^2(t_j-t_{j-1})} \,,
\eeqn
so that
\eqn
  \prod_{j=\ell+1}^{q} | U_{t_{j-1},t_j}^{(main;m_j)}(u_j)|
  \, \leq \,  e^{C\eta^2\sum_j(t_j-t_{j-1})} 
  \, < \, e^{C\eta^2t} \, < \, C' \,.
\eeqn
since $T=\eta^2 t$ is fixed, and hence of order 
$O(1)$ in $\eta$.

Moreover,  Lemma \ref{lm-mimmrec-1} implies that
\eqn
  |U_{t_{j-1},t_j}^{(m_j)}(u_j)| & \leq &  \frac{(C\eta^2(t_j-t_{j-1}))^{m_j}}{m_j!} 
  \, + \, \eta \, m_j \, (c')^{m_j}
  \nonumber\\
  & < & e^{C\eta^2(t_j-t_{j-1})} \, + \, \eta \, c^{m_j} \,,
\eeqn
(where we may, for instance, assume $c=2c'>1$) so that
\eqn\label{eq-laddresum-aux-aux-1}
  \prod_{j=1}^{\ell-1} |U_{t_{j-1},t_j}^{(m_j)}(u_j)| 
  & < & \prod_{j=1}^{q} \big( \, e^{C\eta^2(t_j-t_{j-1})} \, + \, \eta \, c^{m_j} \, \big)
  \nonumber\\
  & < & e^{C\eta^2\sum_j(t_j-t_{j-1})}\prod_{j=1}^{q}(1 \, + \, \eta \, c^{m_j} )
  \nonumber\\
  & < & e^{C\eta^2\sum_j(t_j-t_{j-1})}(1+\sum_{j=1}^q{q\choose r}\, \eta^r \, c^{\sum_j m_j} )
  \nonumber\\
  & < & e^{C\eta^2 t }(1 \, + \, 2\eta \, c^{M_q}) \, < \, C' \,.
\eeqn
For the second inequality, we used $e^{C\eta^2(t_j-t_{j-1})}\geq1$ and $c\geq1$.
For the last inequality, we have recalled that $M_q\leq N(\e) = \frac{C|\log\eta|}{\log|\log\eta|}$,
from the beginning of Section \ref{ssec-decladders-1}.
This implies that
\eqn
	\eta \, c^{M_q} \, \leq \,  \eta \, \eta^{-c \frac{1}{\log|\log\eta|}} \, < \, \eta^{\frac1{10}}
\eeqn
for $\eta$ sufficiently small. Moreover,  $T=\eta^2 t$ is fixed and of order 
$O(1)$ in $\eta$. This implies that \eqref{eq-laddresum-aux-aux-1} is bounded uniformly in $\eta$,
for $\eta$ sufficiently small.

Hence, we conclude that
\eqn  
	|(A)| & < & 
	(C\eta^2)^q \sum_{\ell=0}^{q}
	\int_0^t dt_q \cdots \int_0^{t_2} dt_1 
	\, \|J\|_\infty \, \|f\|_\infty \,\|g\|_\infty \,
	\nonumber\\
	&&\quad\quad\quad 
	\,  |\, \Delta U^{(m_\ell)}_{t_{\ell-1},t_{\ell}}(u_\ell) \,|  
	\, \prod_{j=1}^q \, ( \, C \eta^2 \delt^{-1} (\log\frac1\delt )^2 \, )^{\tm_j} \,.
	\label{eq-Utrenmain-mjtmj-12} 
\eeqn
Using the bounds (\ref{eq-DeltU-bd-lm-1}),
\eqn 
	|(A)|& < &
	(C\eta^2)^q \, \|J\|_\infty \, \|f\|_\infty \,\|g\|_\infty \,
	\sum_{\ell=0}^{q}
	\int_0^t dt_q \cdots \int_0^{t_2} dt_1  
	\nonumber\\
	&&\quad\quad\quad 
	\, \Big( 
	\, m_\ell \, \lambda \, \delt^{-1/2} 
	\, \big( \, \eta^2 \, \delt^{-1} \, \big)^{m_\ell}
	\, + \,  \delt^{1/2} \, \big( \, \eta^2 \, \delt^{-1} \, \big)^{m_\ell}  
	\, \Big) 
	\nonumber\\
	&&\quad\quad\quad
	\, \prod_{j=1}^q\ \, ( \, C \eta^2 \delt^{-1} (\log\frac1\delt)^2 \, )^{\tm_j} \,,
	\label{eq-Utrenmain-mjtmj-13} 
	\\
	& < &
	(C\log\frac1\eta)^q \,  \|J\|_\infty \, \|f\|_\infty \,\|g\|_\infty \, \
	\frac{(\eta^2 t)^q}{q!} 
	\nonumber\\
	&&\quad\quad\quad 
	\, 
	\sum_{\ell=0}^{q} \Big( 
	\, m_\ell \, \lambda \, \delt^{-1/2} 
	\, \big( \, \eta^2 \, \delt^{-1} \, \big)^{m_\ell}
	\, + \,  \delt^{1/2} \, \big( \, \eta^2 \, \delt^{-1} \, \big)^{m_\ell}
	\, \Big)  
	\nonumber\\
	&&\quad\quad\quad\quad\quad\quad
	\, \prod_{j=1}^q\ \, ( \, C \eta^2 \delt^{-1} (\log\frac1\delt)^2 \, )^{m_j} \,.
	\label{eq-Utrenmain-mjtmj-14} 
\eeqn
We notice the crucial gain of the factors $\lambda\delt^{-1/2}$ and $\delt^{1/2}$.
 
The term $(B)$ can be estimated in a similar way.

In conclusion, we arrive at
\eqn 
	\lefteqn{
	\Big| \, \sum_{M_0,\dots,M_{N(\e)}\in\N_0\atop{\sum M_j\leq N(\e)/2}} \, \sum_{q\in\N_0} \, 
	\sum_{\umq,\utmq;|\umq|+|\utmq|+q=M_q} \Big[ \, (A) \, + \, (B) \, \Big] \, \Big|
	}
	\nonumber\\
	& < & \sum_{M_0,\dots,M_{N(\e)}\in\N_0\atop{\sum M_j\leq N(\e)/2}} \, \sum_{q\leq N(\e)}  
	M_q \, (\lambda \, \delt^{-1/2} \, + \,  \delt^{1/2} ) 
	\, ( \, C \eta^2 \delt^{-1} (\log\frac1\delt)^2 \, )^{M_q} 
	\nonumber\\
	& < &  (2N(\e)+2)! \, (\lambda \, \delt^{-1/2} \, + \,  \delt^{1/2} ) 
	\, ( \, C \eta^2 \delt^{-1} (\log\frac1\delt)^2 \, )^{N(\e)}
	\nonumber\\
	& < & \eta^{1/10}
\eeqn
for the choice of parameters of Section \ref{ssec-constants-1}, and for $\delt=\e$.
This proves (\ref{eq-renladdsum-2}).
\\

\noindent{$\bullet$ \underline{\em 3. Proof of (\ref{eq-renladdsum-3})}.}
To begin with, we note that
\eqn 
	&&
	\int du' \frac{1}{E(u')-E(u_j)-i\delt}  
	\, - \, 
	\int du' \frac{1}{E(u')-E(u_0)-i\delt}   
	\\
	&&\quad\quad\quad\quad\quad\quad\quad\quad\quad\quad\quad\quad
	\, = \, (E(u_j)-E(u_0)) \, m_j \, G_1(u_0,u_j;\delt)
	\nonumber
\eeqn
where 
\eqn 
	\lefteqn{
	G_1(u_0,u_j;\delt) 
	}
	\nonumber\\
	&:=& \int du' \, \frac{1}{E(u')-E(u_0)-i\delt}
	\frac{1}{E(u')-E(u_j)-i\delt}
	\\
	&=&\int_{\R_+\times\R_+} ds_1 \, ds_2 \, 
	\int du' \, e^{-is_1(E(u')-E(u_0)-i\delt)}
	\, e^{-is_2(E(u')-E(u_j)-i\delt)} \,.
	\nonumber
\eeqn
From a stationary phase argument,
\eqn 
	| \, G_1(u_0,u_j;\delt) \, | & \leq &
	\int_{\R_+\times\R_+} ds_1 \, ds_2 \, \langle s_1+s_2\rangle^{-3/2} e^{-\delt(s_1+s_2)} 
	\nonumber\\
	&=&\int_{s\geq s'\geq0} ds'  \, ds \, \langle s \rangle^{-3/2} e^{-\delt s }
  \nonumber\\
	&\leq&\int_{s\geq 0}   \, ds \, \langle s \rangle^{-1/2} e^{-\delt s }
	\nonumber\\
	&\leq& C \, \delt^{-1/2} \,.
	\label{eq-Ediffimmrec-1}
\eeqn
More generally, for $m_j\in\N$, one can straightforwardly show along the same lines that
\eqn  
	&&
	\prod_{j=0}^q\Big( \int du' \frac{1}{E(u')-E(u_j)-i\delt} \Big)^{m_j}
	\, - \, 
	\prod_{j=0}^q\Big( \int du' \frac{1}{E(u')-E(u_0)-i\delt} \Big)^{m_j}  
	\nonumber\\
	&&
	\quad\quad\quad\quad\quad 
	\, = \, \sum_{\ell=0}^q (E(u_\ell)-E(u_0)) \,  G_{\ell;\um^{(q)}}(\uu^{(q)};\delt)
\eeqn
with $\uu^{(q)}:=(u_0,\dots,u_q)$,
for functions 
\eqn
	G_{\ell;\um^{(q)}}(\uu^{(q)};\delt) & := &
	\Big(\prod_{j=0}^{\ell-1} (R_{i\delt}(u_j))^{m_j}\Big) \, 
	\Big(\prod_{j=\ell+1}^q (R_{i\delt}(u_0))^{m_j}\Big)
	\\
	&&
	\cdot \, 
	G_1(u_0,u_\ell;\delt) \,  
	\sum_{i'=0}^{m_\ell-1}(R_{i\delt}(u_\ell))^{m_\ell-i'} (R_{i\delt}(u_0))^{i'}
	\nonumber 
\eeqn 
with $R_{i\delt}(u):=\int du' \frac{1}{E(u')-E(u)-i\delt}$.
One easily sees that, for $\ell\in\{0,\dots,q\}$,
\eqn\label{eq-Gell-aux-bd-1}
	| \, G_{\ell;\um^{(q)}}(\uu^{(q)};\delt) \, | \, < \,
	m_\ell \,  C^{\sum_{j=0}^qm_j} \, \delt^{-1/2} \,,
\eeqn
for a constant $C$ independent of $\ell$, $\delt$ and $\{m_j\}$, using
(\ref{eq-aux-immedcolldiff-2}) and (\ref{eq-Ediffimmrec-1}).

Next, we observe that
\eqn 
	\lefteqn{ 
	(E(u_j)-E(u_0)) \,
	U_{t_{j-1},t_j}^{(0)}(u_j;u_0)   
	}
	\nonumber\\
	& = &
	\big( \, i\partial_{s_j} e^{-is_j(E(u_j)-E(u_0)) } \, \big) 
	\, e^{i\lambda\int_{t_{j-1}}^{t_j}ds'( \, \kappa_{s'}(u_j)-\kappa_{s'}(u_0) \, ) }
\eeqn
where $s_j=t_{j}-t_{j-1}\geq0$. 

Therefore, one finds
\eqn 
	&&
	\cU^{(ren-main;q;\umq,\utmq)}_t(J;f,g) \, - \,  \cU^{(ren-main-0;q;\umq,\utmq)}_t(J;f,g) 
	\nonumber\\
	&&\quad\quad\quad\quad\quad
	\, = \, (I) \, + \, (II) \, + \, (III) \, + \, (IV)
\eeqn
with
\eqn  
	\lefteqn{
	(I) \, := \, i \, (-\eta^2)^q \, \sum_{\ell=0}^q 
	\int_0^t dt_q \cdots
	\int_0^{t_{\ell+2}}dt_{\ell+1}
	\int_0^{t_{\ell+1}}dt_{\ell-1}\cdots\int_0^{t_2} dt_1 
	}
	\nonumber\\
	&&\quad 
	\int_0^t d\tt_q \cdots \int_0^{\tt_2} d\tt_1
	\int du_0 \cdots du_q
	\, J(u_q) \, \overline{f(u_0)} \, g(u_0) \,  G_{\ell;\um^{(q)}}(\uu^{(q)};\delt)
	\nonumber\\ 
	&&\quad\quad
	\Big[ \, \prod_{j=1}^q  \,
	\, U_{t_{j-1},t_j}^{(0)}(u_j;u_0) \,  
	\, \overline{U_{\tt_{j-1},\tt_j}^{(0)}(u_j;u_0)} 
	\nonumber\\
	&&\quad\quad\quad 
	\,  
	\, \frac{1}{m_j!}
	\Big( -\eta^2 \, (t_j - t_{j-1}) \,  \Big)^{m_j}
	\label{eq-Utrenmain-mjtmj-4}\\
	&&\quad\quad\quad\quad\quad
	\, \frac{1}{\tm_j!}
	\Big( -\eta^2 \, (\tt_j - \tt_{j-1}) 
	\, \int du' \frac{1}{E(u')-E(u_j)+i\delt} \Big)^{\tm_j} \,  
	\Big]
	\Big|_{s_\ell=0}^{t_{\ell+1}-t_{\ell-1}} \,
	\nonumber
\eeqn
and
\eqn 
	(II)&:=& - \, i \, (-\eta^2)^q \sum_{j=0}^q
	\int_0^t dt_q \cdots \int_0^{t_{j}}ds_j
	\int_0^{t_{j-1}=t_{j}-s_j}dt_{j-2} \int_0^{t_2} dt_1 
	\nonumber\\
	&&\quad \int_0^t d\tt_q \cdots \int_0^{\tt_2} d\tt_1
	\int du_0 \cdots du_q
	\, J(u_q) \, \overline{f(u_0)} \, g(u_0) \, G_{j;\um^{(q)}}(\uu^{(q)};\delt)  
	\nonumber\\
	&&\quad\quad\quad
	\Big\{ \, \prod_{\ell=1}^q \,  \overline{U_{\tt_{\ell-1},\tt_\ell}^{(0)}(u_\ell;u_0)}  
	\nonumber\\
	&&\quad\quad\quad\quad\quad\quad
	\, \frac{1}{\tm_\ell!}
	\Big( -\eta^2 \, (\tt_\ell - \tt_{\ell-1}) 
	\, \int du' \frac{1}{E(u')-E(u_\ell)+i\delt} \Big)^{\tm_\ell} \, \Big\} 
	\nonumber\\
	&&\quad\quad\quad 
	\,  \Big\{ \, \prod_{i=1}^{j-1}  \,
	U_{t_{i-1},i_j}^{(0)}(u_i;u_0) \,  
	\frac{1}{m_i!}
	\big( -\eta^2 \, (t_i - t_{i-1}) \, \big)^{m_i} \Big\} 
	\nonumber\\
	&&\partial_{s_j}\Big[ 
	\, \frac{(\eta^2 s_j)^{m_j} }{m_j!} \,
	\, e^{i\lambda\int_{t_{j-1}}^{t_j}ds' (\kappa_{s'}(u_j)-\kappa_{s'}(u_0))}
	\label{eq-Utrenmain-mjtmj-5}
	\\
	&&\quad\quad\quad 
	\, \Big\{ \, \prod_{i=j+1}^{q}  \,
	U_{t_{i-1},t_i}^{(0)}(u_i;u_0)    
	\, \frac{1}{m_i!}
	\big( -\eta^2 \, (t_i - t_{i-1}) \, \big)^{m_i} \, \Big\} \, \Big] \,.
	\nonumber 
\eeqn 
where for each $\ell$, integration by parts has been applied to the variable $s_\ell$.
With respect to the latter, the expression (\ref{eq-Utrenmain-mjtmj-4}) 
comprises the boundary terms.

The terms $(III)$ and $(IV)$ are similar to $(I)$ and $(II)$, but in $(III)$ and $(IV)$,
integration by parts is applied to the variables  $\widetilde{s}_j=\tt_{j}-\tt_{j-1}$. 
Accordingly, the roles of $s_j$ and $t_i$ are
exchanged with those of $\widetilde{s}_j$ and $\tt_i$, respectively,
and moreover, $\prod(\int du'\frac{1}{E(u')-E(u_j)-i\delt})^{m_j}$ is exchanged with 
$\prod(\int du'\frac{1}{E(u')-E(u_0)-i\delt})^{m_j}$.
\\

\noindent{\underline{\em{Bounds on term} $(I)$}.}
Clearly,
\eqn    
	|(I)| \, \leq \,  \eta^{2q} \, \|J\|_\infty \, \|f\|_\infty \, \|g\|_\infty \, 
	\sum_{\ell=0}^q 
	\sup_{\uu^{(q)}}| G_{\ell;\um^{(q)}}(\uu^{(q)};\delt)| 
	\, A_1 \, A_2
\eeqn
where
\eqn
	\lefteqn{
	A_1 \, := \,   
	\int du_0 \cdots du_q \,  
	\Big| \,
	\, \int_0^t d\tt_q \cdots \int_0^{\tt_2} d\tt_1 \,  \prod_{j=1}^q  
	\Big[ \,\, \overline{U_{\tt_{j-1},\tt_j}^{(0)}(u_j;u_0)} 
	}
	\label{eq-Utrenmain-mjtmj-4-1} \\
	&&\quad\quad\quad\quad\quad\quad\quad 
	\, \frac{1}{\tm_j!}
	\Big( \eta^2 \, (\tt_j - \tt_{j-1}) 
	\, \int du' \frac{1}{E(u')-E(u_j)+i\delt} \Big)^{\tm_j} \, \Big]  \, \Big|
	\nonumber
\eeqn
and
\eqn 
	\lefteqn{
	A_2 \, := \, \sup_{u_0,\cdots,u_q}\int_0^t dt_1 \cdots
	\int_0^{t_{\ell+2}}dt_{\ell+1}
	\int_0^{t_{\ell}}dt_{\ell-1}\cdots\int_0^{t_2} dt_1  
	}
	\nonumber\\
	&&\quad\quad\quad\quad\quad\quad\quad\quad   
	\, \prod_{j=1}^q  \Big\{ \,
	\, \frac{1}{m_j!}
	\Big(  \eta^2 \, (t_j - t_{j-1}) \,  \Big)^{m_j} \, \Big\} 
	\Big|_{s_\ell=0}^{t_{\ell}} \, .
	\label{eq-Utrenmain-mjtmj-4-2} 
\eeqn
We use (\ref{eq-savprop-bd-1}) and bounds similar as in the
case of \eqref{eq-intttres-bd-1}, to estimate the factor (\ref{eq-Utrenmain-mjtmj-4-1})
involving the integrals in $\tt_j$ and $u_j$. Thereby, we obtain an upper bound
\eqn
	A_1 \, < \, ( \, C\log\frac1\delt \, )^{\sum\tm_j} \,.
\eeqn
Furthermore, we note that 
for every fixed index $\ell$ in the sum, there are only $q-1$ integrals 
with respect to the 
variables $t_i$, keeping in mind that there is no integration over $t_\ell$. 
Accordingly, we bound the integrals in $t_j$ in $A_2$
by $\frac{t^{q-1}}{(q-1)!}$ (the gain of a factor $\frac1t$ as compared
to $t^q$ is crucial).
Moreover, it is evident that 
\eqn
	\sum_{m_0,\dots,m_q}\prod_{j=1}^q  
	\, \frac{1}{m_j!}
	\Big(  \eta^2 \, (t_j - t_{j-1}) \,  \Big)^{m_j} 
	\, = \, e^{\eta^2\sum(t_j-t_{j-1})} \, = \, e^{\eta^2 t} \, < \, C \,
\eeqn
holds for the integrand in $A_2$.
Recalling the bound on $| G_{\ell;\um^{(q)}}(\uu^{(q)};\delt)|$ in 
(\ref{eq-Gell-aux-bd-1}), we straightforwardly obtain
\eqn 
	\lefteqn{
	\sum_{1\leq |\um^{(q)}|+|\widetilde{\um}^{(q)}|<M_q} |(I)| 
	}
	\nonumber\\
	& < & \sum_{1\leq |\um^{(q)}|+|\widetilde{\um}^{(q)}|<M_q}
	q \, m_\ell \, \eta^{2q} \, \delt^{-1/2} \, \frac{t^{q-1}}{(q-1)!}\, C^{\sum m_j } \, 
	( \, C' \,  \log\frac1\delt \, )^{1+\sum \tm_j}
	\nonumber\\
	& < & q  \, \frac{M^{2q}}{(2q-1)!}\, \eta^2 \, \delt^{-1/2}\, \frac{( \, C \, \eta^2 t \, )^{q-1}}{(q-1)!} 
	\, ( \, C' \,  \log\frac1\delt \, )^{M_q}
\eeqn
where the factor $q$ accounts for the sum with respect to $\ell=0,\dots,q$, for each fixed $\um^{(q)}$
and $\widetilde{\um}^{(q)}$. Moreover, we have used the estimate   
\eqn
	\#\{ \, (\um^{(q)},\overline{\um}^{(q)}) \, \big| \, |\um^{(q)}|+|\overline{\um}^{(q)}|<M_q\}
	\, \leq \, C \sum_{r\leq M_q} \frac{r^{2q-1}}{(2q-1)!} \, < \, \frac{C \, M_q^{2q}}{(2q-1)!}
\eeqn
from $\#\{(\um^{(q)},\overline{\um}^{(q)}) \, 
\big| \, |\um^{(q)}|+|\overline{\um}^{(q)}|=r\}\leq C \frac{r^{2q-1}}{(2q-1)!}$;
the latter bounds the number of lattice points in a simplex in $\N_0^{2q}$ of side length $r$. 

Hence, we conclude that, for any fixed $T=\eta^2 t>0$ and $\eta$ sufficiently small,
\eqn
	\sum_{1\leq |\um^{(q)}|+|\widetilde{\um}^{(q)}|<M_q} |(I)| & < & q \, \eta \, 
	\frac{( \, \eta^2 t \, )^{q-1}}{(q-1)!}  ( \, C \,  (\log\frac1\e )^3 \, )^{N(\e)}
	\nonumber\\
	&<&q \, \eta \, \e^{-3/10}
	\frac{T^{q-1}}{(q-1)!} 
	\nonumber\\
	& < & \eta^{1/2} 
	\label{eq-I-aux-bd-1}
\eeqn
using that $M_q<N(\e)=\frac{\log\frac1\e}{10\log\log\frac1\e}$ (see Section \ref{ssec-constants-1}), 
$\delt=\e=O(\eta^2)$, and $q\geq1$.
\\

\noindent{\underline{\em{Bounds on term} $(II)$}.}
For the term $\partial_{s_j}\big[ \, \cdots \, \big]$ in (\ref{eq-Utrenmain-mjtmj-5}),
We note that $\partial_{s_j}f(t_\ell)=f'(t_\ell)$ for every $\ell\geq j$,
since $t_\ell=s_0+\cdots+s_j+\cdots+s_\ell$.
For $i>j$,
\eqn 
	\lefteqn{
	\partial_{s_j} U_{t_{i-1},t_i}^{(0)}(u_i;u_0) \,  
	}
	\nonumber\\
	& = &
	\partial_{s_j} e^{-i\int_{t_{i-1}}^{t_i}ds' 
	\big( \, E(u_i)-E(u_0)-\lambda (\kappa_{s'}(u_i)-\kappa_{s'}(u_0))\, \big)}
	\nonumber\\
	& = &
	i \, \lambda \, \big( \, \kappa_{s'}(u_i)-\kappa_{s'}(u_0)\, \big)\Big|_{s'=t_{i-1}}^{t_i}
	\, U_{t_{i-1},t_i}^{(0)}(u_i;u_0) \,.
\eeqn
There is no term proportional to $(E(u_i)-E(u_0))$ on the last line
because $\partial_{s_j}t_{i}=1=\partial_{s_j}t_{i-1}$. 
Similarly, 
\eqn 
	\partial_{s_j} ( \, t_{i} - t_{i-1} \, )^{m_i} \, = \, 0
\eeqn
whenever $i>j$.
Using the a priori bound $\| \, \kappa_s \, \|_{L^\infty(\Tor^3)}<c$, uniformly in $s$,
(a consequence of the fermion statistics, as we recall), we conclude that
\eqn 
	\Big| \, \partial_{s_j}\big[ \, \cdots \, \big] \; {\rm in} \; 
	(\ref{eq-Utrenmain-mjtmj-5}) \, \Big| & < & C \, 
	\prod_{i>j}\frac{(\eta^2(t_{i}-t_{i-1}))^{m_i}}{m_i!} \, 
	\\
	&&
	\Big( \, \eta^2 \frac{(\eta^2(t_{j}-t_{j-1}))^{m_j-1}}{(m_j-1)!} 
	\, + \,  \lambda \frac{(\eta^2(t_{j}-t_{j-1}))^{m_j}}{(m_j)!}\, \Big) \,,
	\nonumber
\eeqn
where $0\leq t_{i}-t_{i-1}\leq t=O(\eta^{-2})$.
Combined with (\ref{eq-intttres-bd-1}), we arrive at
\eqn 
	\sum_{1\leq |\um^{(q)}|+|\widetilde{\um}^{(q)}|<M_q} |(II)| \, < \,
	\frac{M_q^{2q}}{(2q-1)!} \, (\eta^2+\lambda) 
	\, ( \, c \eta^2 t \, )^q \, ( \, c \,  \log\frac1\delt \, )^{M_q} \,,
	\nonumber\\
\eeqn
where the gain of a factor $(\eta^2+\lambda)$ is crucial. Using the same arguments
as above for  (\ref{eq-I-aux-bd-1}), we find 
\eqn
	\sum_{1\leq |\um^{(q)}|+|\widetilde{\um}^{(q)}|<M_q} |(II)| \, < \, \eta^{1/2}
\eeqn
for every fixed $T=\eta^2 t>0$, for $\eta$ sufficiently small, given that
$M_q<N(\e)=\frac{\log\frac1\e}{10\log\log\frac1\e}$, 
$\delt=\e=O(\eta^2)$, and $q\geq1$.
\\

\noindent{\underline{\em Concluding the proof}.}
The terms $(III)$ and $(IV)$ are estimated similarly, and yield similar bounds
as those derived for $(I)$ and $(II)$, respectively. 
In conclusion, we find that for any $T=\eta^2 t$ fixed and $\eta$ sufficiently small,
\eqn  
	\lefteqn{
	\Big| \, \sum_{M_0,\dots,M_{N(\e)}\in\N_0\atop{\sum M_j\leq N(\e)}} \, \sum_{q\in\N_0} \, 
	\sum_{\umq,\utmq \atop{|\umq|+|\utmq|+q=M_q} }
	\Big[ \, (I)  +  (II)  +  (III)  +  (IV) \, \Big] \, \Big|
	}   
	\nonumber\\
	&&\quad\quad\quad\quad\quad\quad\quad\quad< \, 
	(N(\e))^{N(\e)}\, 4 \, \eta^{1/2} \, \sum_{q\in\N_0} \, \chi(q<N(\e)) 
	\nonumber\\
	&&\quad\quad\quad\quad\quad\quad\quad\quad< \,
	2 \, \eta^{1/2} \, (N(\e))^{N(\e)+2}
	\nonumber\\
	&&\quad\quad\quad\quad\quad\quad\quad\quad< \, 
	 \eta^{1/3} \,,
	\nonumber\\
\eeqn
for the choice of parameters given in Section \ref{ssec-constants-1}; that is,
with $\lambda,\delt=O(\eta^2)$, and $N(\e)=\frac{\log\frac1\e}{10 \log\log\frac1\e}$.
Here, we have used the trivial but sufficient bound
\eqn
	\#\{ \, (M_0,\dots,M_{N(\e)})\in\N_0^{N(\e)} \, \big| \, \sum M_j\leq N(\e) \, \}
	\, < \, (N(\e))^{N(\e)}
	\,,
\eeqn
and observed that in the sum with respect to $q$, 
the conditions $\sum M_j\leq N(\e)$ and $|\umq|+|\utmq|+q=M_q$
imply that $q\leq N(\e)$.
In conclusion, (\ref{eq-renladdsum-3}) follows.
\endprf

$\;$

This completes our proof of  Theorem {\ref{thm-main-1}}.

\newpage

\section{Proof of Theorem \ref{thm-main-1-1}}

Based on our proof of  Theorem {\ref{thm-main-1}}, 
the proof of Theorem \ref{thm-main-1-1}
is straightforward.
Applying Lemma \ref{lm-resolvexp-1} with $\zeta=\lambda^{1+\delta}=\eta^{2}$,
and $\delta>0$ arbitrary but fixed, we immediately conclude 
that the estimates for crossing and nesting
diagrams used in the proof of Theorem \ref{thm-main-1} remain valid.

Adapting the proof of Theorem \ref{thm-main-1} 
to the  relative scaling of   parameters as asserted in Theorem \ref{thm-main-1-1},
yields the bound
\eqn\label{eq-apriorilambdalim-aux-1}
	\lim_{L\rightarrow\infty} |\amp_\pi(f,g;\e;\eta)| \, \leq \, C(J,f,g) \, 
	\, (\log\frac1\lambda)^4 (c\eta^2\lambda^{-1}\bn\log\frac1\e)^{\bn} \, 
\eeqn
(see (\ref{eq-crossnest-aux-1}) for comparison). Since $\lambda=\eta^{2-\delta}$,
we find
\eqn\label{eq-apriorilambdasum-bd-1}
	\lefteqn{
	\sum_{1\leq \bn \leq N} \sum_{\pi\in\Gamma_{2\bn}}
	\lim_{L\rightarrow\infty} | \, \amp_\pi( \, f,g;\e;\eta \, ) \, |
	}
	\\
	&&\quad\quad\quad
	\, < \, N! \,  ( \, \log\frac1 \lambda \, )^4 \,
  ( \, c \lambda^{\delta}N\log\frac1 \lambda \, )^{N} \,.
	\nonumber
\eeqn
In contrast to (\ref{eq-cnsum-bd-1}), we are not carrying out any classification
of Feynman graphs. The sum \eqref{eq-apriorilambdasum-bd-1} extends over all classes of
graphs, including ladder, crossing and nesting graphs.

For the given scaling of parameters, we note that instead of the overall factor
$\e^{1/5}\approx \lambda^{(1+\delta)/10}$ in  (\ref{eq-cnsum-bd-1}), we are 
now obtaining a factor $\eta^{\delta N}$
which is $\ll\lambda^{(1+\delta)/10}$ for $\lambda$ sufficiently small, 
and for parameters chosen similarly as in Section \ref{ssec-constants-1}. 

In particular, we can now substitute $\lambda^{\frac{1+\delta}{2}}$
for $\eta$, in the bounds given in Section \ref{ssec-constants-1},
so that $N=O(\frac{\log\frac1\lambda}{\log\log\frac1\lambda})$.

Likewise, the estimates on the Duhamel remainder term of Section \ref{ssec-remainder-1}
can be easily adapted to the present case, and we find that
\eqn
	\int dp \, \overline{f(p)} \, g(p) \, F_T(p)
	& = & 
	\lim_{\lambda\rightarrow0}\lim_{L\rightarrow\infty} 
	\Exp\big[ \, \rho_{T/\lambda}( \, a^+(f) \, a(g) \, ) \, \big] 
	\nonumber\\
	& = &
	\lim_{L\rightarrow\infty}\omzl(\, a^+(f) \, a(g) \, ) 
\eeqn
for any $T>0$. That is, for any initial state $\omzl$ satisfying the assumptions of the
theorem, the Boltzmann limit $F_T$ is stationary.
This proves Theorem \ref{thm-main-1-1}.
\qed

$\;$ \\  

\section{Proof of Theorem {\ref{thm-main-2}}}

Theorem {\ref{thm-main-2}} also follows as an almost immediate consequence
of our proof of Theorem {\ref{thm-main-1}}.
Since we are assuming that $F\in L^\infty(\Tor^3)$ does not depend on the time variable,
we have
\eqn
	U_{s_1,s_2}(u) \, = \, \exp\Big( \, i \big( \, (s_2-s_1)(E(u)
  	\, - \, \lambda (\widehat v* F)(u) \, \big) \, \Big) \,,
\eeqn
which yields a time-independent shift of the kinetic energy,
\eqn\label{eq-tildE-def-2}
	E(p) \, \rightarrow \, \widetilde E_\lambda(p)
  	\, := \, E(p) \, - \, \lambda (\widehat v* F)(p) \,.
\eeqn
Thus, under the assumptions that (\ref{eq-intcond-1})
and the crossing estimate (\ref{eq-intcond-2}) hold,
the derivation of the Boltzmann limit reduces to the case treated in
\cite{ChSa} for $\lambda=0$,
with $\widetilde E_\lambda(p)$ replacing the kinetic energy function $E(p)$.

In the kinetic scaling limit determined by $t=\frac{T}{\eta^2}$ and $\eta\rightarrow0$,
one accordingly obtains
\eqn
	\lim_{\eta\rightarrow0}
	\cG[ \, F ; \eta ; \lambda ; T/\eta^2 ; f , g \, ]
	\, = \, \int dp \, \overline{f(p)} \, g(p) \, \widetilde F_T(p)
\eeqn
for the given choice of $F$, where  $\widetilde F_T$ satisfies the equation 
\eqn\label{eq-auxBoltz-1}
	\partial_T \widetilde F_T(p) \, = \, 2\pi \int du \,
	\delta( \, \widetilde E_\lambda(u) - \widetilde E_\lambda(p) \, )
	\, ( \, \widetilde F_T(u) - \widetilde F_T(p) \, )
\eeqn
with initial condition $\widetilde F_0(p)=F(p)$.
While (\ref{eq-auxBoltz-1}) has the form of a linear Boltzmann equation,
$\widetilde F_T$ is an auxiliar quantity of which only the stationary solutions
are relevant for our discussion.
Here, we point out that
the renormalized energy $\widetilde E_\lambda$ in the collision kernel in (\ref{eq-auxBoltz-1}) 
is determined by $F$, not by $\widetilde F_T$; see (\ref{eq-tildE-def-2}).

Stationary solutions of (\ref{eq-auxBoltz-1}) are determined by the
condition $\partial_T \widetilde F_T(p)=0$, 
so that $\widetilde F_T(p)=\widetilde F_0(p)=F(p)$.
This holds if and only if $F(p)$ satisfies the
self-consistency condition
\eqn
	F(p) \, = \, \frac{1}{\widetilde m_\lambda(p)} \int du \,
	\delta( \, \widetilde E_\lambda(u) - \widetilde E_\lambda(p) \, )
	\, F(u)  \,,
\eeqn
where
\eqn
	\widetilde m_\lambda(p) \, := \, 2\pi \int du \,
	\delta( \, \widetilde E_\lambda(u) - \widetilde E_\lambda(p) \, ) \,.
\eeqn
This proves Theorem {\ref{thm-main-2}}.
\qed

$\;$    

\subsection*{Acknowledgements}
T.C. thanks I. Sasaki for discussions about related subjects, 
and J.L. Lebowitz for inspiring comments.
The work of T.C. was supported by NSF grant DMS-0704031 / DMS-0940145.

\parindent=0pt

\end{document}